\documentstyle[11pt,fullpage]{article}

\def\rf#1{(\ref{eq:#1})}
\def\lab#1{\label{eq:#1}}
\def\nonu{\nonumber}
\def\br{\begin{eqnarray}}
\def\er{\end{eqnarray}}
\def\be{\begin{equation}}
\def\ee{\end{equation}}

\def\foot#1{\footnotemark\footnotetext{#1}}
\def\lb{\lbrack}
\def\rb{\rbrack}

\def\llb{\left\lbrack}
\def\rrb{\right\rbrack}

\def\lcurl{\left\{}
\def\rcurl{\right\}}
\def\({\left(}
\def\){\right)}
\def\bv{\bigm\vert}               
\def\bgv{\bigg\vert}              
\def\lskip{\vskip\baselineskip\vskip-\parskip\noindent}
\def\mskp{\par\vskip 0.3cm \par\noindent}
\def\sskp{\par\vskip 0.15cm \par\noindent}
\def\bc{\begin{center}}
\def\ec{\end{center}}

\font\bfit=cmbxti10 at 10pt   

\newcommand\partder[2]{{{\partial {#1}}\over{\partial {#2}}}}

\newcommand\Sbr[2]{\Bigl\lbrack\,{#1}\, ,\,{#2}\,\Bigr\rbrack} 
 
\newcommand\pbbr[2]{\lcurl\,{#1}\, ,\,{#2}\,\rcurl}  

\def\a{\alpha}
\def\b{\beta}

\def\d{\delta}
\def\D{\Delta}
\def\eps{\epsilon}

\def\G{\Gamma}

\def\h{{1\over 2}}

\def\l{\lambda}

\def\m{\mu}

\def\o{\over}

\def\vp{\varphi}
\def\P{\Phi}
\def\pa{\partial}

\def\bpa{{\bar \partial}}
\def\pr{\prime}

\def\t{\tau}
\def\th{\theta}

\def\wti{\widetilde}

\newcommand\twomat[4]{\left(\begin{array}{cc}  
{#1} & {#2} \\ {#3} & {#4} \end{array} \right)}
\newcommand\threemat[9]{\left(\begin{array}{ccc}  
{#1} & {#2} & {#3}\\ {#4} & {#5} & {#6}\\
{#7} & {#8} & {#9} \end{array} \right)}

\def\cA{{\cal A}}
\def\cB{{\cal B}}

\def\cD{{\cal D}}
\def\cE{{\cal E}}
\def\cF{{\cal F}}
\def\cG{{\cal G}}

\def\cL{{\cal L}}
\def\cM{{\cal M}}

\def\cR{{\cal R}}

\def\cT{{\cal T}}

\def\cX{{\cal X}}
\def\cW{{\cal W}}

\def\cZ{{\cal Z}}

\font\msb=msbm10 scaled \magstep1
\newcommand{\IR}{\mbox{\msb R} }

\newcommand{\IZ}{\mbox{\msb Z} }

\newfam\euffam
\font\sixeuf=eufm6
\font\eighteuf=eufm8
\font\twelveeuf=eufm10 scaled\magstep1
 \textfont\euffam=\twelveeuf
 \scriptfont\euffam=\eighteuf
   \scriptscriptfont\euffam=\sixeuf

\def\one{\hbox{{1}\kern-.25em\hbox{l}}}
\def\0#1{\relax\ifmmode\mathaccent"7017{#1}%
        \else\accent23#1\relax\fi}
\def\Wino{{\bf W_{1+\infty}}}           

\newcommand\DB{{Darboux-B\"{a}cklund}~}
\def\bt{{\bar t}}
\def\pai{\partial^{-1}\!}

\def\Dth{\cD_\theta}
\def\sRes{{\cal R}es}
\newcommand\st[2]{\stackrel{(#1 )}{#2}}

\newcommand\cSKP{${\sl SKP}_{(R;M_B,M_F)}$~}

\newcommand{\ct}[1]{\cite{#1}}
\newcommand{\bi}[1]{\bibitem{#1}}
\newcommand\NPB[3]{{\sl Nucl. Phys.} {\bf B#1} (#2) #3}

\newcommand\CMP[3]{{\sl Commun. Math. Phys.} {\bf #1} (#2) #3}

\newcommand\PLA[3]{{\sl Phys. Lett.} {\bf #1A} (#2) #3}
\newcommand\PLB[3]{{\sl Phys. Lett.} {\bf #1B} (#2) #3}

\newcommand\JMP[3]{{\sl J. Math. Phys.} {\bf #1} (#2) #3}

\newcommand\LMP[3]{{\sl Letters in Math. Phys.} {\bf #1} (#2) #3}
\newcommand\IJMPA[3]{{\sl Int. J. Mod. Phys.} {\bf A#1} (#2) #3}
\newcommand\TMP[3]{{\sl Theor. Math. Phys.} {\bf #1} (#2) #3}
\newcommand\JPA[3]{{\sl J. Physics} {\bf A#1} (#2) #3}

\newcommand\MPLA[3]{{\sl Mod. Phys. Lett.} {\bf A#1} (#2) #3}

\newcommand\PHSA[3]{{\sl Physica} {\bf A#1} (#2) #3}
\newcommand\PHSD[3]{{\sl Physica} {\bf D#1} (#2) #3}

\newcommand\JGP[3]{{\sl J. Geom. Phys.} {\bf #1} (#2) #3}

\begin{document}
\noindent\null\hfill {nlin.SI/0102010}
 
\vskip .3in

\begin{center}
{\large {\bf Symmetries of Supersymmetric Integrable Hierarchies of
KP Type}}
\end{center}

\begin{center}
E. Nissimov${}^{1}$ and S. Pacheva${}^{1}$
\footnotetext[1]{E-mail: nissimov@inrne.bas.bg , svetlana@inrne.bas.bg} \\
{\small Institute for Nuclear Research and Nuclear Energy, 
Bulgarian Academy of Sciences} \\
{\small Boul. Tsarigradsko Chausee 72, BG-1784 ~Sofia, Bulgaria}
j\end{center}
\lskip
\begin{abstract}
This paper is devoted to the systematic study of additional (non-isospectral) 
symmetries of constrained (reduced) supersymmetric integrable hierarchies of 
KP type -- the so called \cSKP models. The latter are 
supersymmetric extensions of ordinary constrained KP hierarchies which
contain as special cases basic integrable systems such as (m)KdV, AKNS, 
Fordy-Kulish, Yajima-Oikawa etc.. As a first main result it is shown that any 
\cSKP hierarchy possesses two different mutually 
(anti-)commuting types of superloop superalgebra additional symmetries 
corresponding to the positive-grade and negative-grade parts of certain 
superloop superalgebras. The second main result is the systematic construction
of the full algebra of additional Virasoro symmetries of 
\cSKP hierarchies, which requires non-trivial 
modifications of the Virasoro flows known from the general case of unconstrained
Manin-Radul super-KP hierarchies (the latter flows {\em do not} define symmetries 
for constrained \cSKP hierarchies). As a third main 
result we provide systematic construction of the supersymmetric analogues of
multi-component (matrix) KP hierarchies and show that the latter contain
among others the supersymmetric version of Davey-Stewartson system. 
Finally, we present an explicit derivation of the general \DB solutions for the 
\cSKP super-tau functions (supersymmetric
``soliton''-like solutions) which preserve the additional (non-isospectral)
symmetries. 
\end{abstract}

\noindent
{\sl PACS numbers 05.45.Y, 11.30.Pb}

\newpage
\noindent
{\bf 1. Introduction}
\mskp
Supersymmetric generalization of integrable hierarchies of nonlinear evolution
(``soliton'' or ``soliton-like'') equations is an actively developing subject
whose main motivations come both from theoretical physics as well as
mathematics. In theoretical physics {\em supersymmetry} is a 
fundamental symmetry principle unifying bosonic and fermionic degrees of
freedom of infinite-dimensional dynamical (field-theoretic) systems which
underly modern superstring theory as an ultimate candidate for an unified
theory of all fundamental forces in Nature, including quantum gravity. In
particular, supersymmetric generalizations of Kadomtsev-Petviashvili (KP)
integrable hierarchy have been found \ct{SI-sstring,Stanciu} to be of direct 
relevance for (multi-)matrix models of non-perturbative superstring theory.
Historically, the first supersymmetric integrable
system, which appeared in the literature, is the supersymmetric
generalization of the Sine-Gordon equation \ct{Susy-SG}. Subsequently, the
subject of supersymmetrization of KP hierarchy
\ct{MR-SKP,Ueno-Yamada,SKP-other-1,SKP-other} and other basic integrable 
systems (Korteveg-de Vries, nonlinear Schr{\"o}dinger, Toda lattice {\sl etc.})
\ct{SI-KdV-NLS,SI-KdV-NLS-other,Liu-Manas,SKP-other-2,AR97,UIC-ARW-97,match,zim-ber,Sorin-Lecht}
attracted a lot of interest from purely mathematical point of view, 
especially, the supersymmetric generalizations of the inverse scattering 
method, bi-Hamiltonian structures, tau-functions and Sato Grassmannian 
approach.

An important role in the theory of integrable systems is being played by the
notion of {\em additional (non-isospectral) symmetries} whose systematic
study started with the papers \ct{Fuchs-Chen,Orlov-Schulman}. For detailed 
reviews of the latter subject we refer to \ct{addsym-review};
see also refs.\ct{addsym-models,noak-addsym}
for a systematic discussion of additional symmetries in the context of
specific integrable models. Additional (non-isospectral) symmetries, 
by definition, consist of the set of all flows on the space of the Lax 
operators of the pertinent integrable hierarchy which commute with the ordinary 
isospectral flows, the latter being generated by the complete set of
commuting integrals of motion. As shown in refs.\ct{Dickey-AvMS} (see also 
\ct{Orlov-Winternitz}), there
exists an equivalent definition of additional symmetries as vector fields
acting on the space of $\t$-functions (Sato Grassmannian) of the
corresponding integrable hierarchy. This latter formulation allows to provide a 
simple interpretation of the crucial Virasoro (and $\Wino$) constraints on
partition functions of (multi-)matrix models of string theory as
invariance of the $\t$-functions ({\sl i.e.}, the string partition functions) 
under the Borel subalgebra of the Virasoro algebra of additional non-isospectral
symmetries in the underlying integrable hierarchies of generalized $SL(r)$ 
Korteveg-de Vries (KdV) type (similarly for the $\Wino$ constraints).

Recently, a deep relationship has been uncovered in ref.\ct{Orlov-Scherbin} 
between additional symmetries of KP hierarchy and fermionic representations 
of certain basic $q$-deformed (``$q$'' staying for ``quantum group'') 
hypergeometric functions playing the role of correlation functions of
quantized integrable field theory models.
Furthermore, the notion of additional (non-isospectral) symmetries allows to:
\begin{itemize}
\item    
Provide an alternative formulation of multi-component (matrix) KP hierarchies 
\ct{matrix-KP} as ordinary one-component (scalar) KP hierarchy supplemented
with appropiate sets of mutually commuting additional symmetry flows
(see refs.\ct{multi-comp-KP,hallifax});
\item
Provide an alternative formulation of various physically relevant nonlinear
evolution equations in two- and higher-dimensional space-time
as additional-symmetry flows on ordinary (reduced) KP hierarchies, the most
interesting examples being Davey-Stewartson and $N$-wave resonant systems
(see refs.\ct{multi-comp-KP,hallifax,virflow,gauge-wz}), as well as 
Wess-Zumino-Novikov-Witten models of group-coset-valued fields (see 
refs.\ct{gauge-wz,AFGZ-00}) describing various ground states in string theory.
\end{itemize}  
The main advantage of the above mentioned reformulation over the standard
matrix pseudo-differential formulation of multi-component KP hierarchies
(and their reductions) lies in the fact that the new formulation allows to
employ the standard \DB techniques in ordinary scalar (constrained) KP
hierarchies in order to generate soliton-like solutions for the more 
complicated multi-component (matrix) KP hierarchies \ct{gauge-wz}.

The principal object of the present work is the important class of
constrained (reduced) $N=1$ supersymmetric KP hierarchies introduced in
ref.\ct{match} and called ``\cSKP models''
(see Eq.\rf{Lax-SKP-R-M} below). The latter are supersymmetric generalizations
of the class ${\sl cKP}_{R,M}$, called ``constrained KP models'', of reductions 
of ordinary (``bosonic'') KP hierarchy containing among themselves a series of
well-known integrable hierarchies such as (modified) KdV, AKNS, Fordy-Kulish,
Yajima-Oikawa {\sl etc.} 
\foot{Originally ${\sf cKP}_{R,M}$ hierarchies appeared in different
disguises from various parallel developments:
(i) symmetry reductions of the general unconstrained KP hierarchy
\ct{symm-red-cKP,oevela}; (ii) free-field realizations, in terms of finite number 
of fields, of both compatible first and second KP Hamiltonian structures
\ct{abel-cKP}; (iii) a method of extracting continuum integrable hierarchies
from generalized Toda-like lattice hierarchies underlying (multi-)matrix
models in string theory \ct{Bonora-Xiong}.}, which are collectively described 
by ordinary (``bosonic'') Sato pseudo-differential Lax operators of the form:
\be
\cL \equiv \cL_{R,M} = \pa^{R} + \sum_{i=0}^{R-2} u_i \pa^i + 
\sum_{j=1}^M \P_j \pa^{-1} \Psi_j
\lab{Lax-R-M}
\ee

We will work within the framework of Sato
super-pseudo-differential calculus in $N=1$ superspace \ct{MR-SKP,Ueno-Yamada}
(cf. Section 2 below).
For alternative treatment of supersymmetric integrable hierarchies within
the framework of the superspace generalization of Drinfeld-Sokolov
Lie-algebraic scheme we refer to \ct{susy-DS}, see also
\ct{UIC-ARW-97,flows}.
Our main task will be the systematic derivation of the full algebra of
additional non-isospectral symmetries for \cSKP
hierarchies. Our present approach is based on a superspace extension of the
approach employed in refs.\ct{virflow,gauge-wz} where the full algebra of
symmetries of the above mentioned ordinary bosonic constrained KP
hierarchies ${\sf cKP}_{R,M}$ \rf{Lax-R-M} has been explicitly constructed.
The latter turns out to be a semi-direct product of Virasoro algebra (see
also refs.\ct{noak-addsym}) with the loop algebra
$\({\widehat U}(1)\oplus{\widehat{SL}}(M)\)_{+}\oplus\({\widehat {SL}}(M+R)\)_{-}$
where the subscripts $(\pm)$ indicate taking the positive/negative-grade
parts of the corresponding loop algebras and the factor 
$\({\widehat U}(1)\)_{+}$ corresponds to the usual isospectral flows. 
For the supersymmetric constrained KP hierarchies \cSKP we find the full symmetry
algebra, {\sl i.e.}, the algebra of Manin-Radul isospectral flows together
with the additional non-isospectral symmetries to be a semi-direct product of 
Virasoro algebra with more complicated superloop superalgebras (with
half-integer loop grading) of the form given in 
\rf{loop-superalg}--\rf{loop-superalg-odd} and \rf{loop-superalg-sub} below:
\be
\({\widehat {GL}} (M_B,M_F)\)_{+} \oplus \({\widehat {GL}}^\pr (N+r+1,N+r+1)\)_{-}
\lab{plus-flow-alg-fSKP-0}
\ee
for \cSKP hierarchies which are defined by fermionic super-Lax operators with
$R \equiv 2r+1$, $M_B + M_F \equiv 2N+1$, and:
\be
\({\widehat {GL}}_{M_B,M_F}\)_{+} \oplus \({\widehat {GL}}^\pr_{N+r,N+r}\)_{-}
\lab{plus-flow-alg-bSKP-0}
\ee
for \cSKP hierarchies defined via bosonic super-Lax operators where 
$R \equiv 2r,\; M_B + M_F \equiv 2N$. Here again the subscripts $(\pm)$ indicate
taking the positive/negative-grade parts of the corresponding superloop 
superalgebras, whereas the primes in 
\rf{plus-flow-alg-fSKP-0}--\rf{plus-flow-alg-bSKP-0} indicate factoring out
the unit super-matrices in the pertinent integer-grade subspaces.

Let stress at this point that the superloop superalgebras
${\widehat {GL}} (N_1,N_2)$ (see \rf{loop-superalg}--\rf{loop-superalg-odd}
below) 
appearing in \rf{plus-flow-alg-fSKP-0} and in what follows,
are more general objects than the notion of superloop algebras introduced in
ref.\ct{Harnad-Kupershmidt} in that the former possess one more Grassmannian
grading. Also, let us point out that similarly to the
ordinary ``bosonic'' case \ct{noak-addsym,virflow}, the construction of
consistent additional Virasoro symmetries for \cSKP
hierarchies requires nontrivial modification of the straightforward
superspace extension \ct{Stanciu,susy-OS} of the well-known Orlov-Schulman 
additional symmetry flows \ct{Orlov-Schulman} for the general unconstrained 
(supersymmetric) KP hierarchy. The latter flows {\rm do not} define symmetries
for constrained (supersymmetric) KP hierarchies, since they do not 
preserve the constrained form of the pertinent (super-)Lax operators 
(see Section 9 below).

The plan of exposition in the present paper is as follows. In Section 2 we
briefly recapitulate the main ingredients of the super-pseudo-differential
operator formulation of the general unconstrained Manin-Radul $N=1$
supersymmetric KP hierarchy, including the superspace extension of such
basic objects as (adjoint) super-eigenfunctions and supersymmetric squared
eigenfunction potentials. In Section 3 we first briefly recall the main properties
of the class \cSKP of reductions of the original
unconstrained Manin-Radul super-KP hierarchy. In particular, we recall the
nontrivial modification of the original Manin-Radul fermionic isospectral
flows which is required for consistency of the latter with the constrained
form of the pertinent super-Lax operators defining \cSKP
hierarchies. Also in Section 3 we present briefly the superspace extension 
of the pseudo-differential treatment \ct{UIC-97} of inverse powers of
(super-)Lax operators. In Section 4 we describe briefly the general
formalism for studying of additional non-isospectral symmetries of
supersymmetric integrable hierarchies which is the superspace extension of
the formalism proposed in the first ref.\ct{Dickey-AvMS} in the purely 
``bosonic'' case. 

The main results of the paper are contained in Sections 
5--10. In Section 5 we provide the explicit construction of
positive-grade part of the superloop superalgebra of additional symmetries for 
constrained super-KP hierarchies defined by fermionic super-Lax operators,
and the same construction is done in Section 6 for the case of 
constrained super-KP hierarchies defined by bosonic super-Lax operators.
Section 7 presents the construction of supersymmetric analogues of
multi-component (matrix) KP hierarchies out of one-component (scalar) 
\cSKP hierarchies by adding to the latter of an infinite
subset of additional symmetry flows spanning Manin-Radul flow algebra. 
In particular, we find the supersymmetric extension of Davey-Stewartson system 
which is contained as superspace additional symmetry flow within 
\cSKP hierarchies. Section 8 is devoted to the 
construction of the negative-grade part of the superloop superalgebra of 
additional symmetries for constrained super-KP hierarchies. In Section 9 
we provide the correct Virasoro algebra of aditional symmetries for
constrained \cSKP hierarchies via a non-trivial modification of the naive 
superspace generalization of Orlov-Schulman additional symmetry flows for 
ordinary ``bosonic'' KP hierarchy. Finally, in Section 10 we identify the 
explicit form of \DB transformations for constrained super-KP hierarchies which
produce solutions to the same hierarchies, {\sl i.e.}, which are {\em auto-}\DB
transformations. Further we construct the general supersymmetric \DB solutions 
for the superspace tau-functions of \cSKP hierarchies with the additional 
property of preserving the additional non-isospectral symmetries. These solutions
are the superspace analogues of the soliton-like solutions in the ordinary 
``bosonic'' case.
\lskip
{\bf 2. Brief Account of the General Manin-Radul Supersymmetric KP Hierarchy}
\mskp
We shall use throughout the super-pseudo-differential calculus
\ct{MR-SKP} as in ref.\ct{match}
with the following notations: $\pa \equiv \partder{}{x}$ and 
$\cD = \partder{}{\th} + \th \partder{}{x}$
denote operators, whereas the symbols $\pa_x$ and $\Dth$ will indicate
application of the corresponding operators on superfield functions. As usual,
$(x,\th )$ denote $N\! =\! 1$ superspace coordinates and $\cD^2 = \pa$.
For any super-pseudo-differential operator $\cA = \sum_j a_{j/2} \cD^j$
the subscripts $(\pm )$ denote its purely differential part
($\cA_{+} = \sum_{j \geq 0} a_{j/2} \cD^j$) or its purely
pseudo-differential part ($\cA_{-} = \sum_{j \geq 1} a_{-j/2} \cD^{-j}$),
respectively. For any $\cA$ the super-residuum is defined as 
$\sRes \cA = a_{-\h}$. The rules of conjugation within the
super-pseudo-differential formalism are as follows \ct{AR97}:
$(\cA \cB )^\ast = (-1)^{|A|\, |B|} \cB^\ast \cA^\ast$
for any two elements with parities $|A|$ and $|B|$;
$\(\pa^k\)^\ast = (-1)^k \pa^k\, ,\,\(\cD^k\)^\ast = (-1)^{k(k+1)/2} \cD^k$
and $u^\ast = u$ for any coefficient superfield.
Finally, in order to avoid confusion we shall also employ the following 
notations: for any super-(pseudo-)\-differential operator $\cA$ and a 
superfield function $f$, the symbol
$\, \cA (f)\,$ will indicate application (action) of $\cA$ on $f$, whereas the
symbol $\cA f$ will denote just operator product of $\cA$ with the zero-order
(multiplication) operator $f$.

The general unconstrained Manin-Radul $N\! =\! 1$ supersymmetric KP (MR-SKP)
hierarchy \ct{MR-SKP} is given by a {\em fermionic} superspace Lax operator
$\cL$ :
\be
\cL = \cD + f_0 + \sum_{j=1}^\infty b_j \pa^{-j}\cD + 
\sum_{j=1}^\infty f_j \pa^{-j}
\lab{super-Lax}
\ee
expressed in terms of a {\em bosonic} ``dressing'' operator $\cW$ :
\be
\cL = \cW \cD \cW^{-1} \quad ,\quad
\cW = 1 + \sum_{j=1}^\infty  w_{j\o 2} \cD^{-j} \equiv
1 + \sum_{j=1}^\infty \a_j \pa^{-j}\cD + \sum_{j=1}^\infty \b_j \pa^{-j}
\lab{super-dress}
\ee
where $b_j ,\b_j$ are bosonic superfield functions whereas $f_j ,\a_j$ are
fermionic ones and where: 
\be
f_0 = 2\a_1 \quad ,\quad b_1 = - \Dth \a_1 \quad ,\quad 
f_1 = 2\a_2 - \a_1 \Dth \a_1 - 2\a_1 \b_1 - \Dth \b_1  \quad ,\quad 
b_2 = \Dth ( -\a_2 + \a_1 \b_1) +\(\Dth\a_1\)^2
\lab{L-W-rel}
\ee
and so on. The square of $\cL$ \rf{super-Lax} is a bosonic 
super-pseudo-differential operator of the form:
\be
\cL^2 = \cW \pa \cW^{-1} = \pa + \sum_{j=1}^\infty u_{j\o 2} \cD^{-j}
\lab{super-Lax-2}
\ee
\be
u_{\h} = - \pa_x \a_1 = \Dth b_1   \quad ,\quad 
u_1 = - \pa_x \b_1 + \a_1 \pa_x \a_1 = 2b_2 - b_1^2 + \Dth f_1
\lab{L2-W-rel}
\ee
and so on.

The Lax evolution equations for MR-SKP read \ct{MR-SKP}:
\br
\partder{}{t_l} \cL &=& -\Sbr{\cL^{2l}_{-}}{\cL} = \Sbr{\cL^{2l}_{+}}{\cL}
\lab{super-Lax-even} \\
D_n \cL &=& -\pbbr{\cL^{2n-1}_{-}}{\cL} =
\pbbr{\cL^{2n-1}_{+}}{\cL} - 2\cL^{2n} 
\lab{super-Lax-odd} \\
\partder{}{t_l} \cW &=& - \(\cW \pa^l \cW^{-1}\)_{-} \cW \quad ,\quad
D_n \cW = - \(\cW \cD^{2n-1}\cW^{-1}\)_{-} \cW
\lab{super-dress-eqs}
\er
with the short-hand notations for the fermionic isospectral flows $D_n$
($\partder{}{t_k}$ being the bosonic isospectral flows) :
\br
D_n = \partder{}{\th_n} - \sum_{k=1}^\infty \th_k \partder{}{t_{n+k-1}}
\quad ,\quad
\pbbr{D_k}{D_l} = - 2 \partder{}{t_{k+l-1}}
\lab{MR-D-n} \\
(t,\th ) \equiv \( t_1 \equiv x ,t_2, \ldots ; \th, \th_1 ,\th_2 ,\ldots \)
\lab{t-th-short}
\er

The super-Baker-Akhiezer (super-BA) and the adjoint super-BA wave functions
are defined as: 
\be
\psi_{BA} (t,\th ;\l ,\eta ) = \cW \( \psi^{(0)}_{BA} (t,\th ;\l ,\eta )\) 
\quad ,\quad
\psi^{\ast}_{BA} (t,\th ;\l ,\eta ) =
{\cW^\ast}^{-1} \({\psi^\ast}^{(0)}_{BA} (t,\th ;\l ,\eta )\)
\lab{super-BA-def}
\ee
(with $\eta$ being a fermionic ``spectral'' parameter), in terms of the
``free'' super-BA functions:
\br
\psi^{(0)}_{BA} (t,\th ;\l ,\eta ) \equiv e^{\xi (t,\th ;\l ,\eta )}
\quad ,\quad
{\psi^\ast}^{(0)}_{BA} (t,\th ;\l ,\eta ) \equiv e^{-\xi (t,\th ;\l ,\eta )}
\lab{free-super-BA} \\
\xi (t,\th ;\l ,\eta ) = \sum_{l=1}^\infty \l^l t_l + \eta \th +
(\eta - \l\th )\sum_{n=1}^\infty \l^{n-1} \th_n 
\lab{xi-def}
\er
For the latter it holds:
\be  
\partder{}{t_k} \psi^{(0)}_{BA} = \pa_x^k \psi^{(0)}_{BA} \quad , \quad
D_n \psi^{(0)}_{BA} = \Dth^{2n-1} \psi^{(0)}_{BA} = 
\pa_x^{n-1} \Dth \psi^{(0)}_{BA} 
\lab{dn-free-super-BA}
\ee
Because of \rf{dn-free-super-BA} (adjoint) super-BA wave functions satisfy: 
\be
\(\cL^2\)^{(\ast)} \psi^{(\ast)}_{BA} = \l \psi^{(\ast)}_{BA} \quad , \quad
\partder{}{t_l} \psi^{(\ast)}_{BA} = 
\pm \(\cL^{2l}\)^{(\ast)}_{+} (\psi^{(\ast)}_{BA}) \quad ,\quad
D_n \psi^{(\ast)}_{BA} = \pm \(\cL^{2n-1}\)^{(\ast)}_{+} (\psi^{(\ast)}_{BA})
\lab{super-BA-eqs}
\ee

Unlike the purely bosonic case \ct{Dickey-book}, the superspace tau-function
$\t (t,\th)$ is related to the super-BA function in a more complicated manner 
\ct{SKP-other}. On the other hand, $\t (t,\th)$ is simply expressed through 
super-residua of the pertinent super-Lax operator \rf{super-Lax} as follows :
\be
\sRes \cL^{2k} = \partder{}{t_k} \Dth \ln \t \quad ,\quad
\sRes \cL^{2k-1} = D_k \Dth \ln \t
\lab{tau-sres}
\ee

A basic object in our construction is the notion of (adjoint)
super-eigenfunctions $\P (t,\th)$ and $\Psi (t,\th)$, whose defining equations
read: 
\be
\partder{}{t_l} \P = \cL^{2l}_{+} (\P) \quad ,\quad
D_n \P = \cL^{2n-1}_{+} (\P) \quad ; \quad
\partder{}{t_l} \Psi = - \(\cL^{2l}\)^{\ast}_{+} (\Psi) \quad ,\quad
D_n \Psi = - \(\cL^{2n-1}\)^{\ast}_{+} (\Psi) 
\lab{super-EF-eqs}
\ee
Following the line of argument in ref.\ct{ridge} for the purely bosonic
case, one can prove that any (adjoint) super-eigenfunction possesses a
supersymmetric ``spectral'' representations: 
\be
\P (t,\th ) = \int d\l\, d\eta\,\vp (\l,\eta ) \psi_{BA} (t,\th ;\l ,\eta )
\quad ,\quad     \Psi (t,\th ) = 
\int d\l\, d\eta\,\vp^\ast (\l,\eta) \psi^\ast_{BA}(t,\th ;\l ,\eta )
\lab{super-spec}
\ee
with appropriate superspace ``spectral'' densities $\vp (\l,\eta )$ and
$\vp^\ast (\l,\eta)$. In particular, from \rf{super-BA-eqs} we note that the
(adjoint) super-BA functions are special examples of (adjoint)
super-eigenfunctions which in addition to \rf{super-EF-eqs} satisfy spectral
equations - the first Eqs.\rf{super-BA-eqs}.

In what follows we shall encounter another basic object of the form
$\Dth^{-1} (\P \Psi ) = \Dth \pa_x^{-1} (\P \Psi )$ where $\P ,\Psi$ is a 
pair of super-eigenfunction and adjoint super-eigenfunction. Similarly to 
the purely bosonic case \ct{oevela} one can
show that application of inverse derivative on such products is well-defined.
Namely, there exists a unique superfield function -- supersymmetric 
``squared eigenfunction potential'' (super-SEP) $S(\P ,\Psi )$ such that:
$\Dth S(\P ,\Psi ) = \P \Psi$. More precisely the super-SEP satisfies the
relations:
\be
\partder{}{t_k} S(\P ,\Psi ) = \sRes \(\cD^{-1}\Psi\cL^{2k}\P\cD^{-1}\)
\quad ,\quad
D_n S(\P ,\Psi ) = \sRes \(\cD^{-1}\Psi\cL^{2n-1}\P\cD^{-1}\)
\lab{super-Oevel}
\ee
In particular, eqs.\rf{super-Oevel} for $k=1$ and $n=1$ read:
\be
\pa_x S(\P ,\Psi ) = \sRes \(\cD^{-1}\Psi\cL^2\P\cD^{-1}\) = \Dth (\P \Psi )
\quad ,\quad
D_1 S(\P ,\Psi ) = \sRes \(\cD^{-1}\Psi\cL\P\cD^{-1}\) = \P \Psi
\lab{super-Oevel-1}
\ee
It is in this well-defined sense that we will be using in the sequel 
inverse superspace derivatives $\Dth^{-1} = \Dth \pai$ acting on products of 
super-eigenfunctions with adjoint super-eigenfunctions.
\lskip
{\bf 3. Constrained Supersymmetric KP Hierarchies. Inverse Powers of
Super-Lax Operators}
\mskp
{\bfit 3.1 \cSKP Hierarchies}
\sskp
In ref.\ct{match} we introduced a class of reductions of the original
MR-SKP hierarchy, called \cSKP \\
constrained super-KP models, which contain the
supersymmetric extensions of various basic bosonic integrable hierarchies such
as (modified) Korteveg-de-Vries, nonlinear Schr\"{o}dinger (AKNS hierarchy in
general), Yajima-Oikawa, coupled Boussinesq-type equations etc.. The \cSKP
hierarchies are defined by the following superspace Lax operators (we will use 
slighty different notations from \ct{match}) :
\be
\cL \equiv \cL_{(R;M_B,M_F)} = 
\cD^R + \sum_{j=0}^{R-1} v_{j\o2} \cD^j + \sum_{i=1}^M \P_i \cD^{-1} \Psi_i
\quad ,\quad  M \equiv M_B + M_F
\lab{Lax-SKP-R-M}
\ee
where $M_{B,F}$ indicate the number of bosonic/fermionic super-eigenfunctions
$\P_i$ entering the purely pseudo-differential part of $\cL_{(R;M_B,M_F)}$.
\cSKP hierarchies defined by fermionic/bosonic super-Lax operators 
\rf{Lax-SKP-R-M}, for which $R\equiv 2r+1,\; M \equiv M_B + M_F \equiv 2N+1$
and $R\equiv 2r,\; M \equiv M_B + M_F \equiv 2N$, respectively, will be called
in what follows ``fermionic''/``bosonic'' hierarchies for brevity.

One of the main results in \ct{match} was to show that the original
fermionic flows $D_n$ \rf{super-Lax-odd} for the general unconstrained MR-SKP
hierarchy do not anymore define consistent flows on the space of fermionic 
constrained \cSKP hierarchies, {\sl i.e.}, the odd flows
\rf{super-Lax-odd} do not preserve the constrained form of fermionic super-Lax
operatores $\cL \equiv \cL_{(R;M_B,M_F)}$ \rf{Lax-SKP-R-M}.
We found the following consistent modification for $D_n$ :
\be
D_n \cL = - \pbbr{\cL^{2n-1}_{-} - X_{2n-1}}{\cL} = 
\pbbr{\cL^{2n-1}_{+} + X_{2n-1}}{\cL} - 2\cL^{2n}
\lab{odd-flow-new}
\ee
where:
\be
X_{2n-1} \equiv 2\sum_{i=1}^M (-1)^{|i|} \sum_{s=0}^{n-2}
\cL^{2(n-s)-3}(\P_i) \cD^{-1} \(\cL^{2s+1}\)^\ast (\Psi_i)
\lab{X-def}
\ee
\be
D_n \P_i = \cL^{2n-1}_{+}(\P_i) - 2 \cL^{2n-1}(\P_i) + X_{2n-1}(\P_i)
\lab{P-i-flow-new}
\ee
\be
D_n \Psi_i = - \(\cL^{2n-1}\)^\ast_{+}(\Psi_i) +
2\(\cL^{2n-1}\)^\ast (\Psi_i) - \( X_{2n-1}\)^\ast (\Psi_i)
\lab{Psi-i-flow-new}
\ee
with the subscript $|i|$ in Eq.\rf{X-def} and below denoting the Grassmann
parity of the corresponding (adjoint) super-eigenfunction $\P_i,\,\Psi_i$.
The modified fermionic isospectral flows $D_n$ \rf{odd-flow-new}--\rf{X-def}
obey the anti-commutation algebra:
\be
\lcurl D_n,\, D_m\rcurl = - 2 \pa/\pa t_{R(n+m-1)}
\lab{D-n-alg}
\ee
where $\partder{}{t_l}$ are the bosonic isospectral flows for
$\cL \equiv \cL_{(R;M_B,M_F)}$ :
\be
\partder{}{t_l} \cL = \Sbr{\Bigl(\cL^{{2l}\o R}\Bigr)_{+}}{\cL}
\lab{boson-flows-R}
\ee
In checking the consistency of the new $D_n$-flows an extensive use is made
of the following superspace pseudo-differential operator identities: 
\be
\cZ_{(i,j)}\, \cZ_{(k,l)} = \cZ_{(i,j)} (\P_k) \cD^{-1} \Psi_l +
(-1)^{|j|(|k|+|l|+1)} \P_i \cD^{-1} \cZ_{(k,l)}^\ast \(\Psi_j\)
\lab{susy-pseudo-diff-id}
\ee
where $\cZ_{(k,l)} \equiv \P_k \cD^{-1} \Psi_l$. They will turn very important
in our symmetry-flows construction below. In particular, using identities
\rf{susy-pseudo-diff-id} we get the relations:
\be
\(\cL^K\)_{-} \equiv \(\cL_{(R;M_B,M_F)}^K\)_{-} =
\sum_{i=1}^M \sum_{s=0}^{K -1} 
(-1)^{s|i|} \cL^{K-1-s}(\P_i) \cD^{-1} \(\cL^s\)^\ast (\Psi_i)
\lab{SKP-Lax-plus-K}
\ee
for the purely pseudo-differential part of arbitrary positive integer power of
a fermionic super-Lax operator \rf{Lax-SKP-R-M}.

In the case of bosonic constrained \cSKP ({\sl i.e.}, $R=2r$ and 
$M\equiv M_B +M_F = 2N$ in \rf{Lax-SKP-R-M}), we can split the set of 
super-eigenfunctions entering the negative pseudo-differential part of 
\rf{Lax-SKP-R-M} in bosonic $\bigl\{\P_a,\Psi_b\bigr\}_{a,b=1}^{M_{B,F}}$
and femionic $\bigl\{{\wti \P}_b,{\wti \Psi}_a\bigr\}_{a,b=1}^{M_{B,F}}$ subsets, 
respectively, so that the expression \rf{Lax-SKP-R-M} acquires the form:
\be
L \equiv \cL_{(2r;M_B,M_F)}\bv_{M_B +M_F = 2N} = L_{+} +
\sum_{a=1}^{M_B} \P_a \cD^{-1} {\wti \Psi}_a + 
\sum_{b=1}^{M_F} {\wti \P}_b \cD^{-1}\Psi_b
\lab{Lax-SKP-R-even-M}
\ee
Henceforth we will use the short-hand notation $L$ to indicate bosonic super-Lax
operators \rf{Lax-SKP-R-even-M}. 
Before proceeding let us recall that in the case of bosonic \cSKP hierarchies 
there is no need of modification \rf{odd-flow-new}--\rf{X-def} of the original 
MR-SKP fermionic isospectral flows $D_n$ \rf{super-Lax-odd} since they preserve 
the constrained form of the bosonic super-Lax operator \rf{Lax-SKP-R-even-M} 
unlike the case with fermionic \cSKP hierarchies:
\be
\partder{}{t_l} L = \Sbr{\Bigl( L\Bigr)^{l\o r}_{+}}{L}
\quad ,\quad
D_n \cL_B = \Sbr{\( L\)^{{2n-1}\o {2r}}_{+}}{L}
\lab{super-bLax-odd}
\ee
\be
D_n \st{\sim}{\P_a} = \( L\)^{{2n-1}\o {2r}}_{+} (\st{\sim}{\P_a}) 
\quad ,\quad
D_n \st{\sim}{\Psi_a} =
- \( L^{{2n-1}\o {2r}}\)^\ast_{+} (\st{\sim}{\Psi_a})
\lab{super-EF-odd}
\ee
Note that the $2r$-th root $L^{1\o {2r}}$ of the bosonic super-Lax
operator \rf{Lax-SKP-R-even-M} (and similarly for the higher ${2n-1}\o {2r}$
powers thereof) is a {\em fermionic} super-pseudo-differential operator
of the general Manin-Radul form \rf{super-Lax} whose coefficients are determined 
recursively from the relation $\( L^{1\o {2r}}\)^{2r} = L$. 

We will also need the explicit expressions for inverse powers of
\cSKP super-Lax operators \rf{Lax-SKP-R-M}. Following
the same lines of the construction in ref.\ct{UIC-97} of inverse powers of
KP Lax operators in the purely bosonic case, we can represent the super-Lax
operator \rf{Lax-SKP-R-M} as a ratio of two purely super-differential
operators $L_{\h (R+M)}$ and $L_{\h M}$ of orders $\h (R+M)$ and $\h M$, 
respectively:
\be
\cL_{(R;M_B,M_F)} = L_{\h (R+M)} L_{\h M}^{-1} =
\left\{\begin{array}{lr} L_{N+r+1} L_{N+\h}^{-1} & M\equiv M_B +M_F =2N+1\; ,
\; R=2r+1 \\
L_{N+r} L_N^{-1} &  M\equiv M_B +M_F =2N\; ,\; R=2r   
\end{array} \right.
\lab{Lax-SKP-R-M-ratio}
\ee
where the first line refers to fermionic super-Lax operator and the second
line refers to bosonic super-Lax operator, respectively. According to
\ct{Ueno-Yamada} (see also refs.\ct{Liu-Manas,zim-ber}), any super-differential
operators $L_N$ (bosonic, of integer order) and $L_{N+\h}$ (fermionic, of
half-integer order) can be parametrized through the elements of their
respective kernels:
$$Ker (L_N) = \lcurl \vp_0, \vp_\h,\ldots ,\vp_{N-1},\vp_{N-\h} \rcurl
\quad,\quad
Ker (L_{N+\h})=\lcurl \vp_0,\vp_\h,\ldots ,\vp_{N-1},\vp_{N-\h},\vp_N \rcurl$$
as follows:
\be
L_N = \cT^{(2N-1)}_{\vp_{N-\h}} \cT^{(2N-2)}_{\vp_{N-1}} \ldots
\cT^{(1)}_{\vp_{\h}} \cT^{(0)}_{\vp_{0}}  \quad ,\quad
L_{N+\h} = \cT^{(2N)}_{\vp_{N}} \cT^{(2N-1)}_{\vp_{N-\h}} \ldots
\cT^{(1)}_{\vp_{\h}} \cT^{(0)}_{\vp_{0}}  
\lab{sdiff-op}
\ee
with the notations ($j=0,1,\ldots, 2N$) :
\be
\cT^{(j)}_{\vp_{j\o 2}} \equiv \vp^{(j)}_{j\o 2} \cD
\bigl( \vp^{(j)}_{j\o 2}\bigr)^{-1} \quad ,\quad
\vp^{(j)}_{j\o 2} \equiv \cT^{(j-1)}_{\vp_{(j-1)\o 2}} 
\cT^{(j-2)}_{\vp_{{j\o 2}-1}} \ldots 
\cT^{(1)}_{\vp_{\h}} \cT^{(0)}_{\vp_{0}} \bigl( \vp_{j\o 2}\bigr)
\lab{sdiff-notat}
\ee
Integer/half-integer indices of the corresponding elements of the kernels
indicate that the latter are bosonic/fermionic, respectively. On the other
hand all super-functions $\vp^{(j)}_{j\o 2}$ in \rf{sdiff-notat} are bosonic
for any index $j$. As shown in \ct{Liu-Manas,zim-ber}, the objects 
$\vp^{(j)}_{j\o 2}$ have explicit representations as ratios of Wronskian-like 
Berezinians (super-determinants) (see also Section 10 below).

Furthermore, as in the purely bosonic case \ct{UIC-97} one can show that the
inverse power of $L_N$ is given as:
\be
L_N^{-1} = \sum_{\a=1}^{N} \Bigl\lb \vp_{\a} \cD^{-1} {\wti \psi}_{\a} +
{\wti \vp}_{\a} \cD^{-1} \psi_{\a} \Bigr\rb
\lab{inverse-power}
\ee
where the set of super-functions 
$\lcurl \vp_{\a},{\wti \vp}_{\a}\rcurl_{\a=1}^{N}$ is spanning $Ker(L_N)$, 
whereas the super-functions 
$\lcurl \psi_{\a},{\wti \psi}_{\a}\rcurl_{\a=1}^{N}$ span $Ker(L^\ast_N)$ --
the kernel of the adjoint operator, and where we have split explicitly the
corresponding kernel elements into bosonic and fermionic (indicated by
``tilde'') subsets.
\mskp
{\bfit 3.2 Inverse Powers of Bosonic Super-Lax Operators}
\sskp
We are now ready to write the explicit expressions for inverse powers of the
super-Lax operators \rf{Lax-SKP-R-M} (cf. Eq.\rf{Lax-SKP-R-M-ratio}).
We start with the bosonic super-Lax operators $\cL_{(R;M_B,M_F)}$
\rf{Lax-SKP-R-even-M} where $R=2r$, $M\equiv M_B + M_F = 2N$. Henceforth we
will use the short-hand notation $L \equiv \cL_{(R;M_B,M_F)}$ for the latter.
Taking into account 
\rf{inverse-power} and the identities \rf{susy-pseudo-diff-id} we obtain:
\be
L^{-1} = L_N L^{-1}_{N+r} = 
\sum_{\b=1}^{N+r} \Bigl\lb L_N (\vp_{\b}) \cD^{-1} {\wti \psi}_{\b} +
L_N ({\wti \vp}_{\b}) \cD^{-1} \psi_{\b} \Bigr\rb
\lab{susy-L-minus-1}
\ee
where the sets of super-functions 
$\lcurl \vp_{\b}, {\wti \vp}_{\b}\rcurl_{\b=1}^{N+r}$ and
$\lcurl \psi_{\b}, {\wti \psi}_{\b}\rcurl_{\b=1}^{N+r}$ span the kernels
$Ker(L_{N+r})$ and $Ker(L^\ast_{N+r})$, respectively.
For later convenience it is useful to introduce the following short hand
notations:
\be
\P^{(-m)}_\b \equiv L^{-(m-1)}\bigl( L_N (\vp_{\b})\bigr) \quad,\quad
\Psi^{(-m)}_\b \equiv \( L^{-(m-1)}\)^\ast (\psi_{\b})
\lab{bEF-inverse}
\ee
\be
{\wti \P}^{(-m)}_\b \equiv L^{-(m-1)}\bigl( L_N ({\wti \vp}_{\b})\bigr) 
\quad,\quad
{\wti \Psi}^{(-m)}_\b \equiv \( L^{-(m-1)}\)^\ast ({\wti \psi}_{\b})
\lab{fEF-inverse}
\ee
where $m=1,2,\ldots$ , and $\b=1,2,\ldots ,N+r$.
Note that the superfunctions in \rf{bEF-inverse} and \rf{fEF-inverse} are
bosonic and fermionnic, respectively. In terms of the short-hand notations
\rf{bEF-inverse}--\rf{fEF-inverse}, we can write the explicit expression
for arbitrary integer $K\geq 1$ inverse power of $L$ 
generalizing \rf{susy-L-minus-1} as:
\be
L^{-K} = \sum_{\b=1}^{N+r} \sum_{s=1}^{K}
\Bigl\lb \P^{(-K-1+s)}_\b \cD^{-1} {\wti \Psi}^{(-s)}_\b +
{\wti \P}^{(-K-1+s)}_\b \cD^{-1} \Psi^{(-s)}_\b \Bigr\rb 
\lab{susy-L-minus-K}
\ee
The latter equality is completely analogous to the expression for the purely
pseudo-differential part of arbitrary positive integer powers of $L$ 
(cf. Eq.\rf{SKP-Lax-plus-K}):
\be
L^K = \sum_{s=1}^{K}\Bigl\lb \sum_{a=1}^{M_B}
\P^{(K+1-s)}_a \cD^{-1} {\wti \Psi}^{(s)}_a +
\sum_{b=1}^{M_F} {\wti \P}^{(K+1-s)}_b \cD^{-1} \Psi^{(s)}_b \Bigr\rb 
\lab{susy-L-plus-K}
\ee
where we introduced another set of convenient short-hand notations similar to
\rf{bEF-inverse}--\rf{fEF-inverse} :
\be
\P^{(m)}_a \equiv L^{m-1}(\P_a) \quad,\quad
\Psi^{(m)}_b \equiv \( L^{m-1}\)^\ast (\Psi_b) \quad,\quad
{\wti \P}^{(m)}_b \equiv L^{m-1}\bigl({\wti \P}_b \bigr) 
\quad,\quad
{\wti \Psi}^{(m)}_a \equiv \( L^{m-1}\)^\ast \bigl({\wti \Psi}_a \bigr)
\lab{EF-plus}
\ee
with $m=1,2,\ldots$ , $a=1,\ldots ,M_B$ and $b=1,\ldots ,M_F$. The derivation of 
both Eqs.\rf{susy-L-minus-K}--\rf{susy-L-plus-K} is based on systematic use of
identities \rf{susy-pseudo-diff-id}.

In what follows an essential use will be made of the following simple
consequences from the definitions of the corresponding objects above:
\be
L \Bigl( L_N \bigl(\st{\sim}{\vp}\!\!\!{}_{\b}\bigr)\Bigr) = 0 \quad ,\quad
L^\ast \Bigl(\st{\sim}{\psi}\!\!\!{}_{\b}\Bigr) = 0 \quad ,\quad
L^{-1} (\st{\sim}{\P}\!\!\!{}_a) = 0 \quad ,\quad 
\( L^{-1}\)^\ast (\st{\sim}{\Psi}\!\!\!{}_a) = 0 
\lab{susy-zero-eqs}
\ee
or, more generally, for the objects defined in 
\rf{bEF-inverse}--\rf{fEF-inverse} and \rf{EF-plus} :
\br
L^K \Bigl(\st{\sim}{\P}\!\!\!{}^{(-m)}_\b\Bigr) = 0   \quad ,\quad
\( L^\ast\)^K \Bigl(\st{\sim}{\Psi}\!\!\!{}^{(-m)}_\b\Bigr) = 0 \quad ,
\nonu \\
L^{-K} \Bigl(\st{\sim}{\P}\!\!\!{}^{(m)}_a\Bigr) = 0   \quad ,\quad
\( L^\ast\)^{-K} \Bigl(\st{\sim}{\Psi}\!\!\!{}^{(m)}_a\Bigr) = 0 \quad
{\rm for ~any} \;\; K \geq m
\lab{susy-K-zero-eqs}
\er

For later use we also observe, that the supersymmetric isospectral flow 
equations written for the inverse $L^{-1}$ of the bosonic super-Lax operator 
\rf{susy-L-minus-1} :
\be
\partder{}{t_l} L^{-1} = \Sbr{\( L^{l\o r}\)_{+}}{L^{-1}}  \quad ,\quad
D_n L^{-1} = \Sbr{\( L^{{2n-1}\o {2r}}\)_{+}}{L^{-1}}
\lab{super-bLax-inverse}
\ee
straightforwardly imply (upon using identities \rf{susy-pseudo-diff-id})
that the super-functions
$\bigl\{\st{\sim}{\P}\!\!\!{}^{(-m)}_b,\st{\sim}{\Psi}\!\!\!{}^{(-m)}_b\bigr\}$
\rf{bEF-inverse}--\rf{fEF-inverse} entering the various inverse powers
\rf{susy-L-minus-K} of $L$ \rf{Lax-SKP-R-even-M} are (adjoint) super-eigenfunctions
\rf{super-EF-eqs} of the latter:
\br
\partder{}{t_l} \st{\sim}{\P}\!\!\!{}^{(-m)}_b = 
\( L^{l\o r}\)_{+} (\st{\sim}{\P}\!\!\!{}^{(-m)}_b) \quad ,\quad
D_n \st{\sim}{\P}\!\!\!{}^{(-m)}_b = 
\( L^{{2n-1}\o {2r}}\)_{+} (\st{\sim}{\P}\!\!\!{}^{(-m)}_b) 
\nonu \\
\partder{}{t_l} \st{\sim}{\Psi}\!\!\!{}^{(-m)}_b =
- \( L^{l\o r}\)^{\ast}_{+} (\st{\sim}{\P}\!\!\!{}^{(-m)}_b) \quad ,\quad
D_n \st{\sim}{\P}\!\!\!{}^{(-m)}_b =
- \( L^{{2n-1}\o {2r}}\)^{\ast}_{+} (\st{\sim}{\P}\!\!\!{}^{(-m)}_b) 
\lab{super-EF-eqs-inverse}
\er
(cf. Eqs.\rf{SKP-EF-eqs-inverse} below for the analogous result in the case of
fermionic \cSKP hierarchies).

\mskp
{\bfit 3.3 Inverse Powers of Fermionic Super-Lax Operators}
\sskp
Repeating the same steps as in the derivation of Eqs.\rf{susy-L-minus-1} and
\rf{susy-L-minus-K}, we obtain the explicit expressions for the inverse powers
of fermionic super-Lax operators \rf{Lax-SKP-R-M}:
\br
\cL^{-1} = L_{N+\h} L^{-1}_{N+r+1} = 
\phantom{aaaaaaaaaaaaaaaaaaaaaa}
\nonu \\
= \sum_{\a=1}^{N+r+1} \Bigl\lb L_{N+\h} (\vp_{\a}) \cD^{-1} {\wti \psi}_{\a} +
L_{N+\h} ({\wti \vp}_{\a}) \cD^{-1} \psi_{\a} \Bigr\rb 
\equiv \sum_{I=1}^{2(N+r+1)} \phi_I \cD^{-1} \psi_I
\lab{SKP-Lax-minus-1}
\er
\be
\cL^{-K} = \sum_{I=1}^{2(N+r+1)} \sum_{s=0}^{K-1} (-1)^{s|I|}
\phi^{(-(K-1-s)/2)}_I \cD^{-1} \psi^{(-s/2)}_I
\lab{SKP-Lax-minus-K}
\ee
where:
\be
\lcurl \vp_{\a}, {\wti \vp}_{\a}\rcurl_{\a=1}^{N+r+1} \equiv
\lcurl \vp_I \rcurl_{I=1}^{2(N+r+1)}  \quad ,\quad
\lcurl \psi_{\a}, {\wti \psi}_{\a}\rcurl_{\a=1}^{N+r+1} \equiv
\lcurl \psi_I \rcurl_{I=1}^{2(N+r+1)} 
\lab{Ker-L}
\ee
span the kernels $Ker(L_{N+r+1})$ and $Ker(L^\ast_{N+r+1})$, respectively, and where
we have introduced further short-hand notations analogous to
\rf{bEF-inverse}--\rf{fEF-inverse} and \rf{EF-plus} :
\be
\phi_I \equiv L_{N+\h} (\vp_I)   \quad ,\quad
\phi^{(-\ell/2)}_I \equiv \cL^{-\ell} (\phi_I)   \quad,\quad
\psi^{(-\ell/2)}_I \equiv \(\cL^{-\ell}\)^\ast (\psi_I) 
\lab{EF-neg-fSKP}
\ee
\be
\P^{(\ell/2)}_i \equiv \cL^{\ell} (\P_i)   \quad,\quad
\Psi^{(\ell/2)}_i \equiv \(\cL^{\ell}\)^\ast (\Psi_i) 
\lab{EF-plus-fSKP}
\ee
Furthermore, similar to \rf{susy-zero-eqs}--\rf{susy-K-zero-eqs} the following 
relations hold for fermionic \cSKP hierarchies:
\be
\cL (\phi_I) = 0 \quad ,\quad  \cL^\ast (\psi_I) = 0   \quad ,\quad
\cL^{-1}(\P_i) = 0  \quad ,\quad  \(\cL^{-1}\)^\ast (\Psi_i) = 0
\lab{SKP-zero-eqs}
\ee

Acting with the isospectral flows $\partder{}{t_l}$ and $D_n$ on the inverse
powers of the fermionic super-Lax operator 
\rf{SKP-Lax-minus-1} (cf. \rf{odd-flow-new} and \rf{boson-flows-R}) :
\be
\partder{}{t_l} \cL^{-1} = \Sbr{\cL^{{2l}\o R}_{+}}{\cL^{-1}}  \quad ,\quad
D_n \cL^{-1} = \Bigl\{-\cL^{2n-1}_{-} + X_{2n-1},\,\cL^{-1}\Bigl\}
\lab{super-fLax-inverse}
\ee
and taking into account identities \rf{susy-pseudo-diff-id} together with 
\rf{SKP-zero-eqs}, we deduce that the sets of super-functions 
$\lcurl \phi_I \rcurl \equiv \bigl\{ L_{n+\h}(\st{\sim}{\vp}\!\!\!{}_\a)\bigr\}$
and $\lcurl \psi_I \rcurl \equiv \bigl\{ \st{\sim}{\psi}\!\!\!{}_\a \bigr\}$ 
are (adjoint) super-eigenfunctions of the fermionic \cSKP hierarchy 
(cf. Eqs.\rf{P-i-flow-new}--\rf{Psi-i-flow-new}) :
\br
\partder{}{t_l} \phi_I = \cL^{{2l}\o R}_{+}\bigl(\phi_I\bigr) \quad ,\quad
D_n \phi_I = \Bigl\lb \cL^{2n-1}_{+} + X_{2n-1}\Bigr\rb \bigl(\phi_I\bigr)
\nonu \\
\partder{}{t_l} \psi_I = - \(\cL^{{2l}\o R}\)^\ast_{+}\bigl(\psi_I\bigr) \quad ,\quad
D_n \psi_I = - \Bigl\lb \(\cL^{2n-1}\)^\ast_{+} + X^\ast_{2n-1}\Bigr\rb 
\bigl(\psi_I\bigr)
\lab{SKP-EF-eqs-inverse}
\er
Analogous result holds also for the super-functions 
$\st{\sim}{\P}\!\!\!{}^{(-m)}_b$ \rf{bEF-inverse} and 
$\st{\sim}{\Psi}\!\!\!{}^{(-m)}_b$ \rf{fEF-inverse}, connected with the bosonic
\cSKP hierarchies \rf{Lax-SKP-R-even-M} 
(see Eqs.\rf{super-EF-eqs-inverse} in Section 6 below).
\lskip
{\bf 4. Additional Symmetries for Super-KP Hierarchies: General Formalism}
\mskp
Bosonic/fermionic flows $\d_{B,F}$ on the space of Sato 
super-pseudo-differential Lax operators $\cL$ 
\rf{super-Lax} or, equivalently, on the space of Sato super-dressing
operators $\cW$ \rf{super-dress} are defined (similarly to the purely bosonic
case, see {\sl e.g.} first ref.\ct{Dickey-AvMS}) in terms of bosonic/fermionic
super-pseudo-differential operators $\cM_{B,F}$ by:
\be
\d_B \cL = \Sbr{\cM_B}{\cL} \quad ,\quad 
\d_F \cL = \Bigl\{ \cM_F,\,\cL \Bigr\}
\quad ;\quad  \d_{B,F} \cW = \cM_{B,F} \cW
\lab{super-flow-def}
\ee
where $\cM_{B,F}$ are bosonic/fermionic purely super-pseudo-differential 
operators. A flow $\d_{B,F}$ \rf{super-flow-def} is a symmetry of MR-SKP
hierarchy if and only if it (anti-)commutes with 
the isospectral Manin-Radul flows, which implies:
\be
\partder{}{t_l} \cM_{B,F} = {\Sbr{\(\cL^{2l}\)_{+}}{\cM_{B,F}}}_{-}
\lab{super-symm-def}
\ee
\be
D_n \cM_B = {\Sbr{\(\cL^{2n-1}\)_{+} + X_{2n-1}}{\cM_B}}_{-} 
+ \d_B X_{2n-1} 
\lab{super-symm-def-even}
\ee
\be
D_n \cM_F = \Bigl\{\(\cL^{2n-1}\)_{+} + X_{2n-1},\,\cM_F \Bigr\}_{-}
- \d_F X_{2n-1}
\lab{super-symm-def-odd}
\ee
In Eqs.\rf{super-symm-def-odd} we have taken into account the modification
\rf{odd-flow-new}--\rf{X-def} of the fermionic isospectral flows $D_n$ needed
in the case of fermionic constrained \cSKP hierarchies \rf{Lax-SKP-R-M}.

Extending the construction in the first ref.\ct{Dickey-AvMS} to the 
supersymmetric case, one can show that the general form of $\cM_{B,F}$ obeying 
\rf{super-symm-def}--\rf{super-symm-def-odd} can be
represented in terms of as follows:
\be
\cM_{B,F} = \int\!\! d\l d\eta_1 d\m d\eta_2\,\rho_{B,F} (\l,\eta_1;\m,\eta_2)\, 
\psi_{BA}(t,\th;\m,\eta_2) \cD^{-1} \psi^\ast_{BA}(t,\th;\l,\eta_1) =
\sum_{P,Q \in \cE} c^{(B,F)}_{PQ}\, \P_Q \cD^{-1} \Psi_P
\lab{super-M-bispec}
\ee
where $\rho_{B,F} (\l,\eta_1;\m,\eta_2)$ is arbitrary (in the case of the general
MR-SKP hierarchy \rf{super-Lax}) double Laurent series in $\l$ and $\m$.
In the second equality above the sums run in general over an infinite set 
$\cE$ of indices, and $\lcurl \P_P ,\,\Psi_P \rcurl_{P \in \cE}$ are
(adjoint) super-eigenfunctions of the super-Lax operator $\cL$, {\sl i.e.},
satisfying \rf{super-EF-eqs} for $\P = \P_P$ and $\Psi = \Psi_P$.
The second equality in \rf{super-M-bispec} arises from the general 
representation of the ``bispectral'' density:
\be
\rho_{B,F} (\l,\eta_1;\m,\eta_2) = \sum_{P,Q \in \cE} c^{(B,F)}_{PQ}\, 
\vp_Q (\m,\eta_2) \psi_P (\l,\eta_1)
\lab{bispec-repr}
\ee
with $c^{(B,F)}_{PQ}$ being constant matrices,
in terms of basis of superspace functions $\lcurl \vp_P (\l,\eta)\rcurl$ and
$\lcurl \psi_P (\l,\eta)\rcurl$ (Laurent series in $\l$), taking into account the 
spectral representation for (adjoint) super-eigenfunctions Eqs.\rf{super-spec}.

We need furthermore to define the action of $\d_{B,F}$-flows on (adjoint)
super-eigenfunctions \rf{super-EF-eqs}. First we note from 
\rf{super-BA-eqs} and the last Eqs.\rf{super-flow-def} that
$\d_{B,F}\psi^{(\ast)}_{BA} = \pm \(\cM_{B,F}\)^{(\ast)}(\psi^{(\ast)}_{BA})$.
For general (adjoint) super-eigenfunctions we have:
\be
\d_{B,F} \P = \cM_{B,F} (\P) + \cF^{B,F} \quad ,\quad
\d_{B,F} \Psi = - \(\cM_{B,F}\)^{\ast} (\Psi) + \cG^{B,F} 
\lab{flow-sEF-def}
\ee
where the inhomogeneous terms $\cF^{B,F}$ and $\cG^{B,F}$ are other (adjoint)
super-eigenfunctions (special examples are
Eqs.\rf{P-i-flow-new}--\rf{Psi-i-flow-new}). The emergence of additional
non-homogeneous terms on the r.h.s. of Eqs.\rf{flow-sEF-def} is due to the
nontrivial (in general) action of $\d_{B,F}$-flows on the pertinent spectral
densities in \rf{super-spec}. The form of the latter non-homogeneous terms
is not arbitrary in general. Namely, when $\P$ and $\Psi$ are (adjoint)
super-eigenfunctions entering the negative pseudo-differential parts of 
the super-Lax operator \rf{Lax-SKP-R-M} or its inverse powers 
\rf{SKP-Lax-minus-1}--\rf{SKP-Lax-minus-K}, then the additional terms
$\cF^{B,F}$ and $\cG^{B,F}$ in \rf{flow-sEF-def} are determined uniquely
from the consistency of the flow action \rf{super-flow-def} with the
constrained form of $\cL$ \rf{Lax-SKP-R-M}. Explicit construction of 
\rf{flow-sEF-def} with consistent non-homogeneous terms will be given in the 
next sections.

Finally, we find for the transformation of the super-tau-function 
\rf{tau-sres} under the action of bosonic/fermionic symmetry flows :
\be
\d_{B,F} \ln \t = \cD_\th^{-1} \Bigl( {\cR es}\, \cM_{B,F}\Bigr) =
\sum_{P,Q \in \cE} c^{(B,F)}_{PQ} \Dth^{-1}\(\P_Q \Psi_P\)
\lab{super-tau-flow}
\ee
\lskip
{\bf 5. Superloop Superalgebra Symmetries of Constrained SKP Hierarchies: The Case
of Fermionic Super-Lax Operators}
\mskp
We now proceed by constructing the explicit form of additional symmetry
generating super-pseudo-differential operators $\cM_{B,F}$ 
\rf{super-M-bispec} in the case of fermionic constrained \cSKP
hierarchies \rf{Lax-SKP-R-M}, {\sl i.e.}, with super-Lax operators
$\cL \equiv \cL_{(R;M_B,M_F)}$ being fermionic (recall 
$R = 2r+1,\; M\equiv M_B + M_F = 2N +1$) :
\be
\cM^{(\ell/2)}_\cA \equiv \sum_{i,j=1}^M \cA^{(\ell/2)}_{ij}
\sum_{s=0}^{\ell -1} (-1)^{s(|j| + \ell)} 
\cL^{\ell -1-s}(\P_j) \cD^{-1} \(\cL^s\)^\ast (\Psi_i)
\lab{M-A-def}
\ee
\be
\cM^{(\ell/2)}_\cF \equiv \sum_{i,j=1}^M \cF^{(\ell/2)}_{ij}
\sum_{s=0}^{\ell -1} (-1)^{s(|j| + \ell)} 
\cL^{\ell -1-s}(\P_j) \cD^{-1} \(\cL^s\)^\ast (\Psi_i)
\lab{M-F-def}
\ee
where $\ell =1,2,\ldots$ .
Here $\cA^{(\ell/2)}$ and $\cF^{(\ell/2)}$ are graded constant matrices of 
the following types:

(a) For $\ell = 2n$ the matrices $\cA^{(n)}$ and $\cF^{(n)}$ are purely 
bosonic and purely fermionic elements, respectively, belonging (as a vector
space) to the superalgebra $GL(M_B,M_F)$ of graded $(M_B,M_F)\times (M_B,M_F)$
matrices:
\be
\cA^{(n)} = \twomat{A^{(n)}}{0}{0}{D^{(n)}} \quad ,\quad  
\cF^{(n)} = \twomat{0}{B^{(n)}}{C^{(n)}}{0}
\lab{A-F-even}
\ee
Here the block matrices $A^{(n)},\, B^{(n)},\ C^{(n)}$ and $D^{(n)}$ are of
sizes $M_B\times M_B$, $M_B\times M_F$, $M_F\times M_B$ and $M_F\times M_F$,
respectively.

(b) For $\ell = 2n -1$ the matrices $\cA^{(n-\h)}$ and $\cF^{(n-\h)}$ are
purely bosonic and purely fermionic elements, respectively, belonging (as a vector
space) to ${\wti {GL}} (M_B,M_F)$ -- the superalgebra of $(M_B,M_F)\times (M_B,M_F)$
graded matrices in the ``twisted'' basis (the diagonal blocks are fermionic, 
whereas the off-diogonal blocks are bosonic; for a general discussion of
non-standard formats of matrix superalgebras, see ref.\ct{Gieres-etal}) :
\be
\cF^{(n-\h)} = \twomat{B^{(n-\h)}}{0}{0}{C^{(n-\h)}} \quad ,\quad  
\cA^{(n-\h)} = \twomat{0}{A^{(n-\h)}}{D^{(n-\h)}}{0}
\lab{A-F-odd}
\ee
In this case the sizes of the block matrices $A^{(n-\h)},\, B^{(n-\h)},\ C^{(n-\h)}$ and 
$D^{(n-\h)}$ are $M_B\times M_F$, $M_B\times M_B$, $M_F\times M_F$ and 
$M_F\times M_B$, respectively.

Thus, all graded matrices \rf{A-F-even}-\rf{A-F-odd} are special
positive-grade elements of a superloop superalgebra ${\widehat {GL}} (M_B,M_F)$
with half-integer grading $\ell = 0, \pm \h, \pm 1, \pm {3\o 2},\ldots$ .
More generally, ${\widehat {GL}} (N_1,N_2)$ here will denote an
infinite-dimensional algebra with half-integer grading:
\be
{\widehat {GL}} (N_1,N_2) = \oplus_{\ell \in \IZ} {GL}^{({\ell \o 2})}(N_1,N_2)
\lab{loop-superalg}
\ee
whose ${\ell \o 2}$-grade subspaces consist of super-matrices of the following form:
\be
{GL}^{(n)}(N_1,N_2) =\lcurl \twomat{A^{(n)}}{B^{(n)}}{C^{(n)}}{D^{(n)}}
\in GL(N_1,N_2)  \rcurl
\lab{loop-superalg-even}
\ee
\be
{GL}^{(n-\h)}(N_1,N_2) =
\lcurl \twomat{B^{(n-\h)}}{A^{(n-\h)}}{D^{(n-\h)}}{C^{(n-\h)}} \in 
{\wti GL}(N_1,N_2) \rcurl 
\lab{loop-superalg-odd}
\ee

Note from Eq.\rf{M-A-def} that for $\ell =2n$ :
\be
\cM^{(n)}_{\cA=\one} \equiv 
\sum_{j=1}^M \sum_{s=0}^{2n-1} (-1)^{s|j|} 
\cL^{2n-1-s}(\P_j) \cD^{-1} \(\cL^s\)^\ast (\Psi_j) = \(\cL^{2n}\)_{-}
\lab{Lax-2n-def}
\ee
whereas for $\ell = 2n-1$ Eq.\rf{M-F-def} implies:
\be
\cM^{(n-\h)}_{\cF=\one} \equiv
\sum_{j=1}^M \sum_{s=0}^{2n -2} (-1)^{s(|j|+1)} 
\cL^{2n-2-s}(\P_j)\cD^{-1}\(\cL^s\)^\ast (\Psi_j) =\(\cL^{2n-1}\)_{-} - X_{2n-1} 
\lab{Lax-X}
\ee
with $X_{2n-1}$ the same as in Eq.\rf{X-def}.

Now, we define the following infinite set of bosonic and fermionic flows,
respectively:
\be
\d^{(\ell/2)}_{\cA} \cL = \Sbr{\cM^{(\ell/2)}_{\cA}}{\cL} \quad ,\quad
\d^{(\ell/2)}_{\cF} \cL = \Bigl\{\cM^{(\ell/2)}_{\cF},\,\cL \Bigr\}
\lab{ghost-susy-Lax}
\ee
One can show, using the superspace identities \rf{susy-pseudo-diff-id}, that
the flows \rf{ghost-susy-Lax} are well-defined, namely, that they preserve
the specific constrained form of the superspace Lax operator
\rf{Lax-SKP-R-M} provided the action of these flows on the constituent
(adjoint) super-eigenfunctions is given by:
\be
\d^{(\ell/2)}_{\cA} \P_i = \cM^{(\ell/2)}_{\cA} (\P_i) -
\sum_{j=1}^M \cA^{(\ell/2)}_{ij}\cL^{\ell}(\P_j) 
\lab{ghost-A-EF}
\ee
\be
\d^{(\ell/2)}_{\cA} \Psi_i = - \(\cM^{(\ell/2)}_{\cA}\)^\ast (\Psi_i) +
\sum_{j=1}^M (-1)^{\ell |j|} \cA^{(\ell/2)}_{ji}\(\cL^{\ell}\)^\ast (\Psi_j)
\lab{ghost-A-adj-EF}
\ee
\be
\d^{(\ell/2)}_{\cF} \P_i = \cM^{(\ell/2)}_{\cF} (\P_i) +
\sum_{j=1}^M \cF^{(\ell/2)}_{ij}\cL^{\ell}(\P_j) 
\lab{ghost-F-EF}
\ee
\be
\d^{(\ell/2)}_{\cF} \Psi_i = - \(\cM^{(\ell/2)}_{\cF}\)^\ast (\Psi_i) -
\sum_{j=1}^M (-1)^{(\ell +1)|j|} \cF^{(\ell/2)}_{ji}
\(\cL^{\ell}\)^\ast (\Psi_j)
\lab{ghost-F-adj-EF}
\ee

Furthermore, employing again identities \rf{susy-pseudo-diff-id}, we find:
\be
\d_{\cA_1}^{(\ell/2)} \cM^{(m/2)}_{\cA_2} - 
\d_{\cA_2}^{(m/2)} \cM^{(\ell/2)}_{\cA_1}
- \Sbr{\cM^{(\ell/2)}_{\cA_1}}{\cM^{(m/2)}_{\cA_2}} = 
\cM^{((\ell +m)/2)}_{\lb \cA_1,\cA_2 \rb}
\lab{comm-A1-A2}
\ee
\be
\d_{\cA}^{(\ell/2)} \cM^{(m/2)}_{\cF} - 
\d_{\cF}^{(m/2)} \cM^{(\ell/2)}_{\cA}
- \Sbr{\cM^{(\ell/2)}_{\cA}}{\cM^{(m/2)}_{\cF}} = 
\left\{ \begin{array}{lr}
\cM^{((\ell +m)/2)}_{\lb \cA ,\cF\rb} & {\rm for} \;\; \ell=even \\
- \cM^{((\ell +m)/2)}_{\{\cA,\cF\}} & {\rm for} \;\; \ell=odd
\end{array} \right.
\lab{comm-A-F}
\ee
\br
\d_{\cF_1}^{(\ell/2)} \cM^{(m/2)}_{\cF_2} + 
\d_{\cF_2}^{(m/2)} \cM^{(\ell/2)}_{\cF_1}
- \Bigl\{ \cM^{(\ell/2)}_{\cF_1},\,\cM^{(m/2)}_{\cF_2}\Bigr\} = 
\nonu \\
= \left\{ \begin{array}{lr}
\pm \cM^{((\ell +m)/2)}_{\{\cF_1,\cF_2\}} & {\rm for} \;\; 
(\ell,m)=(odd,odd)/(even,even) \\
\pm \cM^{((\ell +m)/2)}_{\lb \cF_1,\cF_2\rb} & {\rm for} \;\; 
(\ell,m)=(odd,even)/(even,odd) \\
\end{array} \right.
\lab{comm-F1-F2}
\er
which implies the following infinite-dimensional algebra of flows:
\br
\Sbr{\d^{(\ell/2)}_{\cA_1}}{\d^{(m/2)}_{\cA_2}} = 
\d^{((\ell +m)/2)}_{\lb \cA_1,\, \cA_2\rb}
\phantom{aaaaaaaaaaaaaaaaaaaaaa}       \nonu \\
\Sbr{\d^{(\ell/2)}_{\cA}}{\d^{(m/2)}_{\cF}} =\d^{((\ell +m)/2)}_{\lb \cA,\,\cF\rb}
\;\;\; {\rm for}\;\; \ell ={\rm even}  \quad ,\quad
\Sbr{\d^{(\ell/2)}_A}{\d^{(m/2)}_F} = -\d^{((\ell +m)/2)}_{\{ A,\, F\}}
\;\;\; {\rm for}\;\; \ell ={\rm odd}
\nonu
\er
\br
\Bigl\{ \d^{(\ell/2)}_{\cF_1},\, \d^{(m/2)}_{\cF_2} \Bigr\} = 
\pm \d^{((\ell +m)/2)}_{\{ \cF_1,\cF_2\}}
\;\;\; {\rm for}\;\; (\ell,m)=(odd,odd)/(even,even)
\nonu \\
\Bigl\{ \d^{(\ell/2)}_{\cF_1},\, \d^{(m/2)}_{\cF_2} \Bigr\} = 
\pm \d^{((\ell +m)/2)}_{\lb\cF_1,\cF_2\rb}
\;\;\; {\rm for}\;\; (\ell,m)=(odd,even)/(even,odd)
\lab{super-KM-alg-flows}
\er
Recall that $\cA^{(n)}, \cA^{(n)}_{1,2}$ and $\cF^{(n)},\cF^{(n)}_{1,2}$ are 
constant graded matrices of the form \rf{A-F-even}--\rf{A-F-odd}.

From \rf{Lax-2n-def}--\rf{Lax-X} we find that:
\be
\d^{(n)}_{\cA=\one} = - \partder{}{t_n} \quad ,\quad
\d^{(n-\h)}_{\cF=\one} = - D_n
\lab{susy-isospec-flows}
\ee
are (upto an overall minus sign) the superspace isospectral flows of the
corresponding \cSKP \\
hierarchy, where the fermionic isospectral flows $D_n$ carry the relevant 
modification (see Eqs.\rf{odd-flow-new}--\rf{X-def}) found in ref.\ct{match} in
order to preserve the specific constrained form of \rf{Lax-SKP-R-M}.

Relations \rf{comm-A1-A2}--\rf{super-KM-alg-flows} show that the algebra of
symmetry flows \rf{ghost-susy-Lax} for fermionic \cSKP hierarchy 
\rf{Lax-SKP-R-M} (with $R=2r+1$, $M \equiv M_B +M_F =2N+1$), which contains also 
Manin-Radul isospectral flows according to \rf{susy-isospec-flows}, 
spans $\({\widehat {GL}}(M_B,M_F)\)_{+}$ -- the positive grade part of 
superloop superalgebra ${\widehat {GL}}(M_B,M_F)$ with half-integer grading
\rf{loop-superalg}--\rf{loop-superalg-odd}.

It is also instructive to rewrite the definitions \rf{M-A-def}--\rf{M-F-def}
and the flow Eqs.\rf{ghost-A-EF}--\rf{ghost-F-adj-EF} using the short-hand
notations \rf{EF-plus-fSKP} for the pertinent (adjoint) super-eigenfuntions:
\be
\cM^{(\ell/2)}_\cA \equiv \sum_{i,j=1}^M \cA^{(\ell/2)}_{ij}
\sum_{s=0}^{\ell-1} (-1)^{s(|j| + \ell)} 
\P^{(\ell-1-s)/2}_j \cD^{-1} \Psi^{(s/2)}_i
\lab{M-A-def-SKP}
\ee
\be
\cM^{(\ell/2)}_\cF \equiv \sum_{i,j=1}^M \cF^{(\ell/2)}_{ij}
\sum_{s=0}^{\ell-1} (-1)^{s(|j| + \ell)} 
\P^{(\ell-1-s)/2}_j \cD^{-1} \Psi^{(s/2)}_i
\lab{M-F-def-SKP}
\ee
\be
\d^{(\ell/2)}_{\cA} \P^{(m/2)}_i = \cM^{(\ell/2)}_{\cA} (\P^{(m/2)}_i) -
\sum_{j=1}^M \cA^{(\ell/2)}_{ij} \P^{((\ell + m)/2)}_j 
\lab{ghost-A-EF-m}
\ee
\be
\d^{(\ell/2)}_{\cA} \Psi^{(m/2)}_i = 
- \(\cM^{(\ell/2)}_{\cA}\)^\ast (\Psi^{(m/2)}_i) +
\sum_{j=1}^M (-1)^{\ell (|j|+m-1)} \cA^{(\ell/2)}_{ji} \Psi^{((\ell + m)/2)}_j
\lab{ghost-A-adj-EF-m}
\ee
\be
\d^{(\ell/2)}_{\cF} \P^{(m/2)}_i = \cM^{(\ell/2)}_{\cF} (\P^{(m/2)}_i) +
(-1)^{m-1} \sum_{j=1}^M \cF^{(\ell/2)}_{ij} \P^{((\ell + m)/2)}_j 
\lab{ghost-F-EF-m}
\ee
\be
\d^{(\ell/2)}_{\cF} \Psi^{(m/2)}_i = 
- \(\cM^{(\ell/2)}_{\cF}\)^\ast (\Psi^{(m/2)}_i) -
\sum_{j=1}^M (-1)^{(\ell +1)(|j|+m-1)} \cF^{(\ell/2)}_{ji} \Psi^{((\ell + m)/2)}_j
\lab{ghost-F-adj-EF-m}
\ee
Then, the construction of positive-grade superloop superalgebra symmetries of
this Section can be straightforwardly carried over to the case of the
general unconstrained MR-SKP hierarchy \rf{super-Lax}. In the latter case all 
pertinent (adjoint) super-eigenfunctions
$\lcurl \P^{\ell/2}_i,\, \Psi^{\ell/2}_i\rcurl_{i=1,\ldots ,M}^{\ell
=0,1,2,\ldots}$
are arbitrary, {\sl i.e.}, {\em not} related to a finite subset of them unlike
\rf{EF-plus-fSKP} and, moreover, their respective numbers $M_{B,F}$ ($M=M_B + M_F$)
are also arbitrary. Therefore, the general unconstrained MR-SKP hierarchy
possesses $\({\widehat {GL}} (M_B,M_F)\)_{+}$ superloop superalgebra
symmetries for {\em any} $M_{B,F}$.

Concluding this Section, let us also write down the $\d^{(\ell/2)}_{\cA,\cF}$-flow
equations for the (adjoint) super-eigenfunctions \rf{EF-neg-fSKP} entering the 
inverse powers of $\cL$, which result from consistency of the flow actions 
\rf{ghost-susy-Lax} with the specific constrained form of $\cL^{-K}$ 
\rf{SKP-Lax-minus-K} :
\be
\d^{(\ell/2)}_{\cA,\cF} \phi^{(-m/2)}_I =
\cM^{(\ell/2)}_{\cA,\cF} \bigl(\phi^{(-m/2)}_I\bigr)  \quad ,\quad
\d^{(\ell/2)}_{\cA,\cF} \psi^{(-m/2)}_I =
- \(\cM^{(\ell/2)}_{\cA,\cF}\)^\ast \bigl(\psi^{(-m/2)}_I\bigr)
\lab{ghost-A-F-EF-neg}
\ee
\lskip
{\bf 6. Superloop Superalgebra Symmetries of Constrained SKP Hierarchies: 
The Case of Bosonic Super-Lax Operators}
\mskp
Now we will extend the construction of superloop superalgebra additional
symmetries from the previous Section to the case of bosonic super-Lax operators
$\cL \equiv L$ \rf{Lax-SKP-R-even-M}. 

We find for the counterparts of \rf{M-A-def}--\rf{M-F-def} the following
expressions:
\be
\cM_\cA^{(n)} = \sum_{a,a^\pr =1}^{M_B} A^{(n)}_{aa^\pr} \sum_{k=0}^{n-1} 
L^{n-k-1}(\P_a^\pr) \cD^{-1} \( L^k\)^\ast ({\wti \Psi}_a)
+ \sum_{b,b^\pr =1}^{M_F} D^{(n)}_{bb^\pr} \sum_{k=0}^{n-1} 
L^{n-k-1}({\wti \P}_b^\pr) \cD^{-1} \( L^k\)^\ast (\Psi_b)
\lab{M-A-def-even}
\ee
\br
\cM_\cF^{(n-\h)} = - \sum_{a,b=1}^{M_B,M_F} B^{(n-\h)}_{ab} \sum_{l=0}^{n-2} 
L^{n-l-2}({\wti\P}_b)\cD^{-1}\( L^l\)^\ast ({\wti\Psi}_a) +
\nonu \\
+ \sum_{a,b=1}^{M_B,M_F} C^{(n-\h)}_{ba}
\sum_{k=0}^{n-1} L^{n-k-1}(\P_a) \cD^{-1} \( L^k\)^\ast (\Psi_b)
\lab{M-F-def-even}
\er
Here $\cA^{(n)}$ and $\cF^{(n-\h)}$ are constant supermatrices which are purely
bosonic and purely fermionic elements belonging to the superloop superalgebra
${\widehat G}(M_B,M_F)$ \rf{loop-superalg}--\rf{loop-superalg-odd} with grades 
$n$ and $n-\h$, respectively:
\be
\cA^{(n)} = \twomat{A^{(n)}}{0}{0}{D^{(n)}} \quad ,\quad  
\cF^{(n-\h)} = \twomat{0}{B^{(n-\h)}}{C^{(n-\h)}}{0}
\lab{A-even-F-odd}
\ee
(compare expressions \rf{A-even-F-odd} for bosonic super-Lax operators with
expressions \rf{A-F-even}--\rf{A-F-odd} for fermionic super-Lax operators).

In particular, we note that:
\be
\cM^{(n)}_{\cA=\one} = \sum_{a=1}^N \sum_{k=0}^{n-1} 
\Bigl\lb L^{n-k-1}({\wti \P}_a) \cD^{-1} \( L^k\)^\ast (\Psi_a)
+ L^{n-k-1}(\P_a) \cD^{-1} \( L^k\)^\ast ({\wti \Psi}_a) \Bigr\rb =
\( L^n\)_{-}
\lab{M-one}
\ee
In full analogy with Eqs.\rf{ghost-susy-Lax}--\rf{ghost-F-adj-EF} we
construct the following infinite set of bosonic and fermionic flows acting
on the bosonic constrained SKP Lax operator $\cL \equiv L$ :
\be
\d_A^{(n)} L = \Sbr{\cM_A^{(n)}}{L}  \quad ,\quad
\d_F^{(n-\h)} L = \Sbr{\cM_F^{(n-\h)}}{L} 
\lab{ghost-susy-Lax-even}
\ee
Consistency of the flow actions \rf{ghost-susy-Lax-even} with the specific
constrained form of $\cL \equiv L$ \rf{Lax-SKP-R-even-M} implies the 
following flow actions on the associated (adjoint) super-eigenfunctions:
\be
\d^{(n)}_{\cA} \P_a = \cM^{(n)}_{\cA} (\P_a) -
\sum_{a^\pr =1}^{M_B} A^{(n)}_{aa^\pr} L^n (\P_{a^\pr})   \quad ,\quad 
\d^{(n)}_{\cA} \Psi_b = - \(\cM^{(n)}_{\cA}\)^\ast (\Psi_b) +
\sum_{b^\pr=1}^{M_F}  D^{(n)}_{b^\pr b}\( L^n\)^\ast (\Psi_{b^\pr})
\lab{ghost-A-adj-bEF}
\ee
\be
\d^{(n)}_{\cA} {\wti \P}_b = \cM^{(n)}_{\cA} ({\wti \P}_b) -
\sum_{b^\pr =1}^{M_F} D^{(n)}_{bb^\pr} L^n ({\wti \P}_{b^\pr})   \quad ,\quad 
\d^{(n)}_{\cA} {\wti \Psi}_a = - \(\cM^{(n)}_{\cA}\)^\ast ({\wti \Psi}_a) +
\sum_{a^\pr =1}^{M_B} A^{(n)}_{a^\pr a}\( L^n\)^\ast ({\wti \Psi}_{a^\pr})
\lab{ghost-A-adj-fEF}
\ee
\be
\d^{(n-\h)}_{\cF} \P_a = \cM^{(n-\h)}_{\cF} (\P_a) +
\sum_{b^\pr =1}^{M_F} B^{(n-\h)}_{a b^\pr} L^{n-1}({\wti \P}_{b^\pr}) 
\lab{ghost-F-bEF}
\ee
\be
\d^{(n-\h)}_{\cF} \Psi_b = - \(\cM^{(n-\h)}_{\cF}\)^\ast (\Psi_b) -
\sum_{a^\pr =1}^{M_B} B^{(n-\h)}_{a^\pr b}\( L^{n-1}\)^\ast ({\wti \Psi}_{a^\pr})
\lab{ghost-F-adj-bEF}
\ee
\be
\d^{(n-\h)}_{\cF} {\wti \P}_b = \cM^{(n-\h)}_{\cF} ({\wti \P}_b) -
\sum_{a^\pr =1}^{M_B} C^{(n-\h)}_{b a^\pr} L^{n}(\P_{a^\pr}) 
\lab{ghost-F-fEF}
\ee
\be
\d^{(n-\h)}_{\cF} {\wti \Psi}_a = - \(\cM^{(n-\h)}_{\cF}\)^\ast ({\wti \Psi}_a) -
\sum_{b^\pr =1}^{M_F} C^{(n-\h)}_{b^\pr a}\( L^{n}\)^\ast (\Psi_{b^\pr})
\lab{ghost-F-adj-fEF}
\ee

Now, employing identities \rf{susy-pseudo-diff-id} we find that the
symmetry generating super-pseudo-differential operators 
\rf{M-A-def-even}--\rf{M-F-def-even} for bosonic constrained
super-KP hierarchies satisfy the same type of commutation relations
as relations \rf{comm-A1-A2}--\rf{comm-F1-F2} in the case of fermionic
constrained super-KP hierarchies upon replacing there the
constant supermatrices of the form \rf{A-F-even}--\rf{A-F-odd} with the
corresponding constant supermatrices \rf{A-even-F-odd} :
\be
\d^{(n)}_{\cA_1} \cM^{(m)}_{\cA_2} - \d^{(m)}_{\cA_2} \cM^{(n)}_{\cA_1}
- \Sbr{\cM^{(n)}_{\cA_1}}{\cM^{(m)}_{\cA_2}} = 
\cM^{(n+m)}_{\lb \cA_1,\cA_2 \rb}
\lab{comm-A1-A2-b}
\ee
\be
\d^{(n)}_{\cA} \cM^{(m-\h)}_{\cF} - \d^{(m-\h)}_{\cF} \cM^{(n)}_{\cA}
- \Sbr{\cM^{(n)}_{\cA}}{\cM^{(m-\h)}_{\cF}} = 
\cM^{(n+m-\h)}_{\lb \cA ,\cF \rb}
\lab{comm-A-F-b}
\ee
\be
\d^{(n-\h)}_{\cF_1} \cM^{(m-\h)}_{\cF_2} + 
\d^{(m-\h)}_{\cF_2} \cM^{(n-\h)}_{\cF_1}
- \Bigl\{\cM^{(n-\h)}_{\cF_1},\,\cM^{(m-\h)}_{\cF_2}\Bigr\} = 
\cM^{(n+m-1)}_{\{\cF_1,\cF_2\}}
\lab{comm-F1-F2-b}
\ee

Therefore, the pertinent flows $\d^{(n)}_\cA$ and $\d^{(n-\h)}_\cF$ 
\rf{ghost-susy-Lax-even}--\rf{ghost-F-adj-fEF} span the following
infinite-dimensional superalgebra: 
\be
\Sbr{\d^{(n)}_{\cA_1}}{\d^{(m)}_{\cA_2}} = \d^{(n+m)}_{\lb \cA_1,\cA_2\rb}
\quad ,\quad
\Sbr{\d^{(n)}_\cA}{\d^{(m-\h)}_\cF} = \d^{(n+m-\h)}_{\lb \cA,\,\cF\rb}
\quad ,\quad
\Bigl\{ \d^{(n-\h)}_{\cF_1},\, \d^{(m-\h)}_{\cF_2} \Bigr\} = 
\d^{(n+m-1)}_{\{ \cF_1,\cF_2\}}
\lab{super-KM-alg-flows-b}
\ee
which we denote as $\({\widehat {GL}}_{M_B,MF}\)_{+}$. Similarly to
\rf{super-KM-alg-flows} we find that $\({\widehat {GL}}_{M_B,MF}\)_{+}$ is the 
positive-grade part of an infinite-dimensional superalgebra
${\widehat {GL}}_{M_B,M_F}$ with half-integer grading 
consisting of all $(M_B,M_F)\times (M_B,M_F)$ graded matrices of the form:
\br
{\widehat {GL}}_{N_1,N_2} = \oplus_{\ell \in \IZ} {GL}^{(\ell)}_{N_1,N_2} 
\quad ,\quad N_1 \equiv M_B\;,\; N_2 \equiv M_F 
\nonu \\
{GL}^{(n)}_{N_1,N_2} =\lcurl \twomat{A^{(n)}}{0}{0}{D^{(n)}} \rcurl 
\quad ,\quad  
{GL}^{(n-\h)}_{N_1,N_2} =
\lcurl \twomat{0}{B^{(n-\h)}}{C^{(n-\h)}}{0} \rcurl
\lab{loop-superalg-sub}
\er

In the present case of bosonic constrained super-KP hierarchies 
\rf{Lax-SKP-R-even-M} the first relation in \rf{susy-isospec-flows} is again
satisfied, whereas the second relation \rf{susy-isospec-flows} holds only for
fermionic \cSKP hierarchies.

For later use let us also write down explicitly the $\d^{(n)}_\cA$ and 
$\d^{(n-\h)}_\cF$ ($\({\widehat {GL}}_{M_B,M_F}\)_{+}$ superloop superalgebra) 
flow equations for all pertinent (adjoint) super-eigenfunctions \rf{EF-plus} :
\be
\d^{(n)}_{\cA} \P^{(m)}_a = \cM^{(n)}_{\cA} (\P^{(m)}_a) -
\sum_{a^\pr =1}^{M_B} A^{(n)}_{a a^\pr} \P^{(n+m)}_{a^\pr}  
\quad ,\quad 
\d^{(n)}_{\cA} \Psi^{(m)}_b = - \(\cM^{(n)}_{\cA}\)^\ast (\Psi^{(m)}_b) +
\sum_{b^\pr=1}^{M_F} D^{(n)}_{b^\pr b} \Psi^{(n+m)}_{b^\pr}
\lab{ghost-A-EF-m-1}
\ee
\be
\d^{(n)}_{\cA} {\wti \P}^{(m)}_b = \cM^{(n)}_{\cA} ({\wti \P}^{(m)}_b) -
\sum_{b^\pr =1}^{M_F} D^{(n)}_{b b^\pr} {\wti \P}^{(n+m)}_{b^\pr}  
\quad ,\quad 
\d^{(n)}_{\cA} {\wti \Psi}^{(m)}_a =
- \(\cM^{(n)}_{\cA}\)^\ast ({\wti \Psi}^{(m)}_a) +
\sum_{b^\pr=1}^{M_B} A^{(n)}_{a^\pr a} {\wti \Psi}^{(n+m)}_{a^\pr}
\lab{ghost-A-adj-EF-m-1}
\ee
\be
\d^{(n-\h)}_{\cF} \P^{(m)}_a = \cM^{(n-\h)}_{\cF} (\P^{(m)}_a) +
\sum_{b^\pr =1}^{M_F} B^{(n-\h)}_{ab^\pr} {\wti \P}^{(n+m-1)}_{b^\pr} 
\lab{ghost-F-bEF-m}
\ee
\be
\d^{(n-\h)}_{\cF} \Psi^{(m)}_b = - \(\cM^{(n-\h)}_{\cF}\)^\ast (\Psi^{(m)}_b) -
\sum_{a^\pr =1}^{M_B} B^{(n-\h)}_{a^\pr b} {\wti \Psi}^{(n+m-1)}_{a^\pr}
\lab{ghost-F-adj-bEF-m}
\ee
\be
\d^{(n-\h)}_{\cF} {\wti \P}^{(m)}_b = \cM^{(n-\h)}_{\cF} ({\wti \P}^{(m)}_b) -
\sum_{a^\pr =1}^{M_B} C^{(n-\h)}_{b a^\pr} \P^{(n+m)}_{a^\pr} 
\lab{ghost-F-fEF-m}
\ee
\be
\d^{(n-\h)}_{\cF} {\wti \Psi}^{(m)}_a =
- \(\cM^{(n-\h)}_{\cF}\)^\ast ({\wti \Psi}^{(m)}_a) -
\sum_{b^\pr =1}^{M_F} C^{(n-\h)}_{b^\pr a} \Psi^{(n+m)}_{b^\pr}
\lab{ghost-F-adj-fEF-m}
\ee
which generalize Eqs.\rf{ghost-A-adj-bEF}--\rf{ghost-F-adj-fEF} and where
relations \rf{susy-K-zero-eqs} have been taken into account.
Accordingly, the consistency conditions for the flow
Eqs.\rf{ghost-susy-Lax-even} written in terms of $L^{-K}$ with the
specific super-pseudodifferential form of the latter \rf{susy-L-minus-K} 
imply for \rf{bEF-inverse}--\rf{fEF-inverse} the following
$\({\widehat {GL}}_{M_B,M_F}\)_{+}$ superloop superalgebra flow equations:
\br
\d^{(n)}_{\cA} \st{\sim}{\P}\!\!\!{}^{(-m)}_\b = 
\cM^{(n)}_{\cA} (\st{\sim}{\P}\!\!\!{}^{(-m)}_\b) \quad , \quad
\d^{(n)}_{\cA} \st{\sim}{\Psi}\!\!\!{}^{(-m)}_\b = 
- \(\cM^{(n)}_{\cA}\)^\ast (\st{\sim}{\Psi}\!\!\!{}^{(-m)}_\b) 
\lab{ghost-A-inverse-EF-m} \\
\d^{(n-\h)}_{\cF} \st{\sim}{\P}\!\!\!{}^{(-m)}_\b = 
\cM^{(n-\h)}_{\cF} (\st{\sim}{\P}\!\!\!{}^{(-m)}_\b) \quad , \quad
\d^{(n-\h)}_{\cF} \st{\sim}{\Psi}\!\!\!{}^{(-m)}_\b = 
- \(\cM^{(n-\h)}_{\cF}\)^\ast (\st{\sim}{\Psi}\!\!\!{}^{(-m)}_\b) 
\lab{ghost-F-inverse-EF-m}
\er
where again relations \rf{susy-K-zero-eqs} have been accounted for.
\lskip
{\bf 7. Multi-Component (Matrix) SKP Hierarchies: Supersymmetric Extension
of Davey-Stewartson System}
\mskp
Let us now consider the following subalgebra of the superloop superalgebra
symmetry flows for bosonic ${\sl SKP}_{(2r;N,N)}$ hierarchies
({\sl i.e.}, $R=2r,\; M_B = M_F =N$; cf. \rf{M-A-def-even}--\rf{M-F-def-even} 
and \rf{ghost-susy-Lax-even}) which are defined as:
\be
\d^{(n)}_{\cA=\cE_k} \equiv - \pa/\pa\!\st{k}{t}\!\!\!{}_{n} \quad ,\quad
\d^{(n-\h)}_{\cF=\cE_k} \equiv - \st{k}{D}\!\!\!{}_{n}
\lab{multi-comp-SKP-flows}
\ee
\be
\cE^{(n)}_k = \twomat{E_k}{0}{0}{E_k} \quad ,\quad
\cE^{(n-\h)}_k = \twomat{0}{E_k}{E_k}{0}
\quad 
{\rm with} \;\; E_k \equiv diag(0,\ldots ,0, \st{k}{1},0,\ldots ,0)
\ee
where $k=1,\ldots ,N$. The flows \rf{multi-comp-SKP-flows} span a direct sum 
of $N$ copies of the original Manin-Radul isospectral flow algebra 
\rf{MR-D-n} :
\be
\Bigl\{ \st{k}{D_n},\, \st{l}{D_m}\Bigr\} = 
- \d_{kl} \pa/\pa\!\st{k}{t}\!\!{}_{n+m-1}  \quad ,\quad {\rm rest} = 0 \quad ;
\;\;\; k,l=1,\ldots ,N \;\; ,\;\; n,m=1,2, \ldots
\lab{multi-comp-SKP-flows-alg}
\ee
which justifies their representation in a form similar to \rf{MR-D-n}:
\be
\st{k}{D_n} = \pa/\pa\!\st{k}{\th_n} 
- \sum_{s=1}^\infty \st{k}{\th_s} \pa/\pa\!\st{k}{t}\!\!{}_{n+s-1}
\lab{multi-comp-MR-D-n}
\ee

Now, we can construct the following supersymmetric extended integrable hierarchy
built on the original bosonic ${\sl SKP}_{(2r;N,N)}$ supersymmetric hierarchy
\rf{Lax-SKP-R-even-M} by supplementing the latter with the set of
additional superloop superalgebra supersymmetric Manin-Radul-like flows 
\rf{multi-comp-SKP-flows}--\rf{multi-comp-MR-D-n} 
(recall here $L \equiv \cL_{(2r;N,N)}$ \rf{Lax-SKP-R-even-M}) :
\be
\pa/\pa\!\st{k}{t}\!\!{}_{n} L = -\Sbr{\cM^{(n)}_k}{L} \quad ,\quad
\st{k}{D_n} L = -\Sbr{\cM^{(n-\h)}_k}{L}
\lab{N-comp-SKP}
\ee
with:
\be
\cM_k^{(n)} \equiv \sum_{s=0}^{n-1} 
\Bigl\lb L^{n-s-1}({\wti \P}_k) \cD^{-1} \( L^s\)^\ast (\Psi_k)
+ L^{n-s-1}(\P_k) \cD^{-1} \( L^s\)^\ast ({\wti \Psi}_k) \Bigr\rb
\lab{M-k-def-even}
\ee
\be
\cM_k^{(n-\h)} \equiv 
\sum_{s=0}^{n-1} L^{n-s-1}(\P_k)\cD^{-1}\( L^s\)^\ast (\Psi_k) - 
\sum_{s=0}^{n-2} L^{n-s-2}({\wti\P}_k)\cD^{-1}\( L^s\)^\ast ({\wti\Psi}_k)
\lab{M-k-def-odd}
\ee
where the flow action on the constituent (adjoint) super-eigenfunctions is
given by (cf. Eqs.\rf{ghost-A-adj-bEF}--\rf{ghost-F-adj-fEF}) :
\be
\pa/\pa\!\st{k}{t}\!\!{}_{n} \st{\sim}{\P_a} = 
-\cM^{(n)}_{k} (\st{\sim}{\P_a})   \quad ,\quad 
\pa/\pa\!\st{k}{t}\!\!{}_{n} \st{\sim}{\Psi_a} = 
\(\cM^{(n)}_{k}\)^\ast (\st{\sim}{\Psi_a})     \quad ,\quad  a \neq k
\lab{ghost-b-gen-EF}
\ee
\be
\pa/\pa\!\st{k}{t}\!\!{}_{n} \st{\sim}{\P_k} = 
-\cM^{(n)}_{k} (\st{\sim}{\P_k}) + L^n (\st{\sim}{\P_k})    \quad ,\quad 
\pa/\pa\!\st{k}{t}\!\!{}_{n} \st{\sim}{\Psi_k} = 
\(\cM^{(n)}_{k}\)^\ast (\st{\sim}{\Psi_k}) - \( L^n\)^\ast (\st{\sim}{\P_k})
\lab{ghost-b-k-EF}
\ee
\be
\st{k}{D_n} \st{\sim}{\P_a} = 
-\cM^{(n-\h)}_{k} (\st{\sim}{\P_a})   \quad ,\quad 
\st{k}{D_n} \st{\sim}{\Psi_a} = 
\(\cM^{(n-\h)}_{k}\)^\ast (\st{\sim}{\Psi_a})     \quad ,\quad  a \neq k
\lab{ghost-f-gen-EF}
\ee
\be
\st{k}{D_n} \P_k = -\cM^{(n-\h)}_{k} (\P_k) + L^{n-1}({\wti \P}_k) 
\quad ,\quad 
\st{k}{D_n} \Psi_k = \(\cM^{(n-\h)}_{k}\)^\ast (\Psi_k) +
\( L^{n-1}\)^\ast ({\wti \Psi}_k)
\lab{ghost-f-k-bEF}
\ee
\be
\st{k}{D_n} {\wti \P}_k = -\cM^{(n-\h)}_{k} ({\wti \P}_k) + L^{n}(\P_k) 
\quad ,\quad
\st{k}{D_n} {\wti \Psi}_k = \(\cM^{(n-\h)}_{k}\)^\ast ({\wti \Psi}_k) +
\( L^{n}\)^\ast (\Psi_k)
\lab{ghost-f-k-fEF}
\ee

The above construction 
is the superspace analog of our construction in refs.\ct{multi-comp-KP,hallifax},
where the corresponding ordinary bosonic scalar (one-component) KP hierarchy,
supplemented with the flows belonging to the Cartan subalgebra of additional
loop-algebra symmetries, was identified as a matrix (multi-component) KP
hierarchy. Therefore it is natural to call the supersymmetric extended KP
hierarchy defined by Eqs.\rf{N-comp-SKP}--\rf{ghost-f-k-fEF}
{\em $N$-component constrained super-KP hierarchy}.

It is well-known that ordinary bosonic multi-component KP hierarchies
contain various physically interesting nonlinear systems such as 2-dimensional
Toda lattice, Davey-Stewartson and $N$-wave resonant systems. As a
non-trivial illustration of the properties of the new $N$-component constrained
super-KP hierarchies \rf{N-comp-SKP}--\rf{ghost-f-k-fEF} we will show that the 
${\sl SKP}_{(2;2,2)}$ model
contains a supersymmetric version of ordinary Davey-Stewartson system.

Thus, we consider the special case $N=2$ in
Eqs.\rf{N-comp-SKP}--\rf{ghost-f-k-fEF}. We take for convenience the following
two mutually (anti-)commuting infinite sets of Manin-Radul-like flows:
$\Bigl\{\pa/\pa t_n, D_n\Bigr\}$ (the original MR-SKP isospectral
flows) and $\Bigl\{\pa/\pa \bt_n\! \equiv\! -\pa/\pa\!\!\!\st{1}{t}\!\!\!\!{}_{n},
\st{1}{D}\!\!{}_n\Bigr\}$
(recall
$\pa/\pa t_n = \pa/\pa\!\!\!\!\st{1}{t}\!\!{}_{1}+\pa/\pa\!\!\!\!\st{1}{t}\!\!{}_{2}$
according to first Eq.\rf{susy-isospec-flows}). In particular, we will use
the short-hand notation
$\bpa \equiv \pa/\pa \bt_1 \equiv -\pa/\pa\!\!\!\st{1}{t}\!\!{}_{1}$.
We will need the following explicit expressions 
(cf. the general notations and relations from Section 2) :
\be
L \equiv L_{(2;2,2)} = \pa + \sum_{a=1}^2 \Bigl({\wti \P}_a \cD^{-1} \Psi_a +
\P_a \cD^{-1} {\wti \Psi}_a\Bigr) \equiv \cW \pa \cW^{-1} \equiv
\pa + u_{\h} \cD^{-1} + u_1 \cD^{-1} + \ldots
\lab{Lax-SKP-1-2}
\ee
\be
u_{\h} = - \pa\a_1 \equiv 
\sum_{a=1}^2 \Bigl({\wti \P}_a \Psi_a -\P_a {\wti \Psi}_a\Bigr) 
\quad ,\quad
u_1 = -\b_1 - \a_1 \pa\a_1 
\equiv \sum_{a=1}^2 \Bigl({\wti \P}_a \Dth \Psi_a +
\P_a \Dth {\wti \Psi}_a\Bigr)
\lab{u-expr}
\ee
\be
\( L^2\)_{+} = \pa^2 + 2u_{\h} \cD + 2u_1      \quad ,\quad
\( L^2\)^\ast_{+} = \pa^2 + 2u_{\h} \cD + 2(u_1 -\Dth u_{\h})
\lab{Lax-2-plus}
\ee
\be
\cM^{(1)}_1 = {\wti \P}_1 \cD^{-1} \Psi_1 + \P_1 \cD^{-1} {\wti \Psi}_1 
\lab{ghost-1-1}
\ee
\be
\cM^{(2)}_1 = 
L ({\wti \P}_1) \cD^{-1} \Psi_1 + L (\P_1) \cD^{-1} {\wti \Psi}_1 +
{\wti \P}_1 \cD^{-1} L^\ast (\Psi_1) + 
\P_1 \cD^{-1} L^\ast ({\wti \Psi}_1)
\lab{ghost-1-2}
\ee
For the first additional symmetry flow $\bpa$ we have:
\be
\bpa \st{\sim}{\P_1} = \cM^{(1)}_1 (\st{\sim}{\P_1}) - L (\st{\sim}{\P_1})
\quad ,\quad
\bpa \st{\sim}{\Psi_1} = - \(\cM^{(1)}_1\)^\ast (\st{\sim}{\Psi_1}) 
+ L^\ast (\st{\sim}{\Psi_1})
\lab{ghost-1-EF}
\ee
\be
\bpa \a_1 = {\wti \P}_1 \Psi_1 - \P_1 {\wti \Psi}_1  \quad ,\quad
\bpa \b_1 = \P_1 \Dth{\wti \Psi}_1 + {\wti \P}_1 \Dth \Psi_1 +
\a_1 \({\wti \P}_1 \Psi_1 - \P_1 {\wti \Psi}_1\)
\lab{ghost-1-W}
\ee
where the last two Eqs.\rf{ghost-1-W} are obtained from
$\bpa \cW = \cM^{(1)}_1 \cW$ upon inserting there the expansion for $\cW$
\rf{super-dress}. Recall also that $\a_1$ and $\b_1$ are expressed in terms 
of the constituent (adjoint) super-eigenfunctions of the super-Lax operator 
\rf{Lax-SKP-1-2} through Eqs.\rf{u-expr}. Further, taking the super-residuum
of the superspace operator flow equation 
$\bpa L = \Sbr{\cM^{(1)}_1}{L}$ we get:
\be
\bpa u_{\h} = - \pa \({\wti \P}_1 \Psi_1 - \P_1 {\wti \Psi}_1\)
\lab{div-eq}
\ee

We now consider the following system of flow equations:
\be
\partder{}{t_2} \st{\sim}{\P_1} = \( L^2\)_{+} \bigl(\st{\sim}{\P_1}\bigr)
\quad ,\quad
\partder{}{t_2} \st{\sim}{\Psi_1} = 
- \( L^2\)^\ast_{+} \bigl(\st{\sim}{\Psi_1}\bigr)
\lab{EF-isoflow-2}
\ee
which are the standard (adjoint) super-eigenfunction equations
\rf{super-EF-eqs} for the second isospectral bosonic flow $\pa/\pa t_2$
where the super-differential operators on the r.h.s. are given by
\rf{Lax-2-plus}, and:
\be
\pa/\pa \bt_2 \st{\sim}{\P_1} = \cM^{(2)}_1 \bigl(\st{\sim}{\P_1}\bigr)
- L^2 \bigl(\st{\sim}{\P_1}\bigr)   \quad ,\quad
\pa/\pa \bt_2 \st{\sim}{\Psi_1} = 
- \(\cM^{(2)}_1\)^\ast \bigl(\st{\sim}{\Psi_1}\bigr)
+ \( L^2\)^\ast \bigl(\st{\sim}{\Psi_1}\bigr) 
\lab{EF-ghostflow-2}
\ee
which are the flow equations for the second bosonic
additional-symmetry flow $\pa/\pa \bt_2$ with $\cM^{(2)}_1$ as in 
\rf{ghost-1-2}. The explicit form of Eqs.\rf{EF-isoflow-2}--\rf{EF-ghostflow-2}
reads:
\be
\partder{}{t_2} \P_1 = \llb \pa^2 + 2u_{\h} \cD + 2u_1 \rrb \P_1 \quad ,\quad
\pa/\pa \bt_2 \P_1 = 
\llb - \bpa^2 + 2\bpa\(\Dth^{-1}(\P_1{\wti \Psi}_1)\)\rrb \P_1 -
2\bpa\(\Dth^{-1}(\P_1 \Psi_1)\) {\wti \P}_1
\lab{P-1-eqs}
\ee
\be
\partder{}{t_2} {\wti \P}_1 = \llb \pa^2 + 2u_{\h} \cD + 2u_1 \rrb {\wti \P}_1 
\quad ,\quad
\pa/\pa \bt_2 {\wti \P}_1 = 
\llb - \bpa^2 + 2\bpa\(\Dth^{-1}({\wti \P}_1\Psi_1)\)\rrb {\wti \P}_1 -
2\bpa\(\Dth^{-1}({\wti \P}_1 {\wti \Psi}_1)\) \P_1
\lab{ti-P-1-eqs}
\ee
\br
\partder{}{t_2} \Psi_1 = 
-\llb \pa^2 + 2u_{\h} \cD + 2(u_1 -\Dth u_\h)\rrb \Psi_1 
\nonu \\
\pa/\pa \bt_2 \Psi_1 = 
\llb \bpa^2 - 2\bpa\(\Dth^{-1}({\wti \P}_1 \Psi_1)\)\rrb \Psi_1 +
2\bpa\(\Dth^{-1}(\P_1 \Psi_1)\) {\wti \Psi}_1
\lab{Psi-1-eqs}
\er
\br
\partder{}{t_2} {\wti \Psi}_1 = 
\llb \pa^2 + 2u_{\h} \cD + 2(u_1 -\Dth u_\h)\rrb {\wti \Psi}_1 
\nonu \\
\pa/\pa \bt_2 {\wti \Psi}_1 = 
\llb \bpa^2 - 2\bpa\(\Dth^{-1}(\P_1{\wti \Psi}_1)\)\rrb {\wti \Psi}_1 -
2\bpa\(\Dth^{-1}({\wti \P}_1 {\wti \Psi}_1)\) \Psi_1
\lab{ti-Psi-1-eqs}
\er
Introducing a new time variable $T = t_2 - \bt_2$ and subtracting
$\pa/\pa \bt_2$-flow equations from the corresponding $\pa/\pa t_2$-flow
equations above,
we arrive at the following system of super-differential evolution equations
for $\P_1, {\wti \P}_1, \Psi_1, {\wti \Psi}_1$ regarded as functions of
$(T,x\equiv t_1,y\equiv \bt_1,\th)$ ({\sl i.e.}, functions on superspace
$\IR^{2|1}$ with coordinates $(x,y,\th)$) and suppressing the dependence on
the rest of the bosonic and ferminionic flow parameters:
\be
\partder{}{T} \P_1 = \Bigl\lb \h (\pa^2 + \bpa^2) + u_\h \cD +
\(\cG + 2(\P_1 \Dth {\wti \Psi}_1 + {\wti \P}_1 \Dth \Psi_1)\)\Bigr\rb \P_1
+ \cF \, {\wti \P}_1
\lab{susy-DS-dyn-1}
\ee
\be
\partder{}{T} {\wti \P}_1 = \Bigl\lb \h (\pa^2 + \bpa^2) + u_\h \cD +
\({\wti \cG} + 2(\P_1 \Dth {\wti \Psi}_1 + {\wti \P}_1 \Dth \Psi_1)\)\Bigr\rb
{\wti \P}_1  + {\wti \cF} \, \P_1
\lab{susy-DS-dyn-2}
\ee
\be
\partder{}{T} \Psi_1 = - \Bigl\lb \h (\pa^2 + \bpa^2) + u_\h \cD +
\({\wti \cG} - \Dth u_\h + 2(\P_1 \Dth {\wti \Psi}_1 + 
{\wti \P}_1 \Dth \Psi_1)\)\Bigr\rb \Psi_1 - \cF \, {\wti \Psi}_1
\lab{susy-DS-dyn-3}
\ee
\be
\partder{}{T} {\wti \Psi}_1 = - \Bigl\lb \h (\pa^2 + \bpa^2) + u_\h \cD +
\(\cG - \Dth u_\h + 2(\P_1 \Dth {\wti \Psi}_1 + 
{\wti \P}_1 \Dth \Psi_1)\)\Bigr\rb {\wti \Psi}_1 + {\wti \cF}\, \Psi_1
\lab{susy-DS-dyn-4}
\ee
The coefficient super-functions $u_\h,\cF,{\wti \cF},\cG,{\wti \cG}$ 
in Eqs.\rf{susy-DS-dyn-1}--\rf{susy-DS-dyn-4} are related with 
$\P_1, {\wti \P}_1, \Psi_1, {\wti \Psi}_1$ through non-dynamical
super-differential relations as follows:
\be
\bpa u_{\h} = - \pa \({\wti \P}_1 \Psi_1 - \P_1 {\wti \Psi}_1\)  \quad ,\quad
\Dth \cF = \bpa (\P_1\Psi_1)   \quad ,\quad
\Dth {\wti \cF} = \bpa \bigl({\wti \P}_1 {\wti \Psi_1}\bigr) 
\lab{susy-DS-nondyn-1}
\ee
\br
\pa\bpa \cG + 
(\pa +\bpa)^2 \bigl(\P_1 \Dth {\wti \Psi}_1 + {\wti \P}_1 \Dth \Psi_1\bigr) =
\nonu \\
2 \pa\Bigl( u_\h \bigl( {\wti \P}_1 \Psi_1 - \P_1 {\wti \Psi}_1\bigr)\Bigr)
+ \bpa^2 \Bigl({\wti \P}_1 \Dth\Psi_1 + {\wti \Psi}_1 \Dth\P_1 \Bigr)
\lab{susy-DS-nondyn-2}
\er
\be
\Dth (\cG - {\wti \cG}) = 
\bpa \bigl( {\wti \P}_1 \Psi_1 - \P_1 {\wti \Psi}_1\bigr)
\lab{susy-DS-nondyn-3}
\ee
The first relation in \rf{susy-DS-nondyn-1} coincides with Eq.\rf{div-eq}.
Relations \rf{susy-DS-nondyn-2} and \rf{susy-DS-nondyn-3} result from the 
definitions of $\cG$ and ${\wti \cG}$ :
\be
\cG \equiv u_1 - \bpa \bigl(\Dth^{-1}(\P_1 {\wti \Psi}_1)\bigr) -
2(\P_1 \Dth {\wti \Psi}_1 + {\wti \P}_1 \Dth \Psi_1) 
\lab{cG-def}
\ee
\be
{\wti \cG} \equiv u_1 - \bpa \bigl(\Dth^{-1}({\wti \P}_1 \Psi_1)\bigr) -
2(\P_1 \Dth {\wti \Psi}_1 + {\wti \P}_1 \Dth \Psi_1) 
\lab{cG-ti-def}
\ee
with $u_1$ as in \rf{L2-W-rel} and \rf{u-expr}, upon taking into account
Eqs.\rf{ghost-1-W}.

The system of evolution Eqs.\rf{susy-DS-dyn-1}--\rf{susy-DS-dyn-4} together
with the non-dynamical relations 
\rf{susy-DS-nondyn-1}--\rf{susy-DS-nondyn-3} is the supersymetric extension
of the ordinary bosonic Davey-Stewartson system. Indeed, let us take the 
bosonic limit in \rf{susy-DS-dyn-1}--\rf{susy-DS-nondyn-3}, meaning that we 
set all fermionic component fields in the pertinent superfields equal to zero
and in addition we put the fermionic superspace coordinate $\th=0$. Then, the
only surviving functions are:
\be
\phi_1 = \P_1 \bgv_{\th=0} \quad , \quad
\psi_1 = \Dth {\wti \Psi}_1 \bgv_{\th=0} \quad ,\quad
G = \cG \bgv_{\th=0} = {\wti \cG} \bgv_{\th=0} 
\lab{bos-limit}
\ee
and the superspace system \rf{susy-DS-dyn-1}--\rf{susy-DS-nondyn-3} reduces
to the ordinary bosonic system of nonlinear equations:
\be
\partder{}{T} \phi_1 = 
\Bigl\lb \h (\pa^2 + \bpa^2) + G + 2 \phi_1 \psi_1 \Bigr\rb \phi_1
\lab{DS-dyn-1}
\ee
\be
\partder{}{T} \psi_1 = 
- \Bigl\lb \h (\pa^2 + \bpa^2) + G + 2 \phi_1 \psi_1 \Bigr\rb \psi_1
\lab{DS-dyn-2}
\ee
\be
\pa\bpa G + (\pa +\bpa)^2 \(\phi_1 \psi_1\) = 0
\lab{DS-nondyn}
\ee
which is precisely the standard Davey-Stewartson system.
\lskip
{\bf 8. ``Negative''-Grade Superloop Superalgebra Symmetries}
\mskp
{\bfit 8.1 ``Negative''-Grade Symmetries of Bosonic \cSKP Hierarchies}
\sskp
Using the same technique as in Sections 5 and 6, we can construct a
``negative''-grade superloop superalgebra of additional symmetries for 
constrained super-KP hierarchies \rf{Lax-SKP-R-M}. First we will consider 
explicitly the case of \cSKP hierarchies with bosonic super-Lax operators 
\rf{Lax-SKP-R-even-M}.

Following the pattern in Eqs.\rf{M-A-def-even}--\rf{M-F-def-even} we consider
the following set of additional symmetry generating super-pseudo-differential
operators:
\be
\cM^{(-n)}_{{\bar \cA}} = \sum_{a,b=1}^{N+r} \sum_{s=1}^{n} \Bigl\lb
{\bar A}^{(-n)}_{ab} \P^{(-(n-s+1))}_b \cD^{-1} {\wti \Psi}^{(-s)}_a +
{\bar D}^{(-n)}_{ab} {\wti \P}^{(-(n-s+1))}_b \cD^{-1} \Psi^{(-s)}_a \Bigr\rb
\lab{susy-ghost-b-minus}
\ee
\be
\cM^{(-n+\h)}_{{\bar \cF}} = \sum_{a,b=1}^{N+r} \sum_{s=1}^{n}
\Bigl\lb {\bar C}^{(-n+\h)}_{ab} \P^{(-(n-s+1))}_b \cD^{-1} \Psi^{(-s)}_a 
- {\bar B}^{(-n+\h)}_{ab} {\wti \P}^{(-(n-s))}_b \cD^{-1} {\wti \Psi}^{(-s)}_a 
\Bigr\rb
\lab{susy-ghost-f-minus}
\ee
and their associated supersymmetric flows:
\be
\d^{(-n)}_{{\bar \cA}} L = \Sbr{\cM^{(-n)}_{{\bar \cA}}}{L}  \quad ,\quad
\d^{(-n+\h)}_{{\bar \cF}} L = \Sbr{\cM^{(-n+\h)}_{{\bar \cF}}}{L} 
\lab{ghost-susy-Lax-neg}
\ee
where the short-hand notations \rf{bEF-inverse}--\rf{fEF-inverse} have been
employed. Here ${\bar \cA}^{(-n)}$ and ${\bar F}^{(-n-\h)}$ are constant
supermatrices -- elements of superloop superalgebra ${\widehat G}_{N+r,N+r}$ 
(cf. \rf{loop-superalg-sub}) having a form similar to \rf{A-even-F-odd} :
\be
{\bar \cA}^{(-n)} = \twomat{{\bar A}^{(-n)}}{0}{0}{{\bar D}^{(-n)}} \quad ,\quad  
\cF^{(-n+\h)} = \twomat{0}{{\bar B}^{(-n+\h)}}{{\bar C}^{(-n+\h)}}{0}
\lab{A-even-F-odd-neg}
\ee
where now all matrix blocks
${\bar A}^{(-n)},\, {\bar B}^{(-n+\h)},\, {\bar C}^{(-n+\h)},\, {\bar D}^{(-n)}$
are of size $(N+r)\times (N+r)$.
Furthermore, ${\bar \cA}^{(-n)} \neq \one$ in
Eqs.\rf{susy-ghost-b-minus},\rf{ghost-susy-Lax-neg} since
$\cM^{(-n)}_{{\bar \cA} =\one} = L^{-n}$ (cf. Eq.\rf{susy-L-minus-K}), so that
$\d^{(-n)}_{{\bar \cA} =\one}$ does not generate any flow according to first 
Eq.\rf{ghost-susy-Lax-neg}. Recall also, that according to 
\rf{super-EF-eqs-inverse} all super-functions entering the 
super-pseudo-differential operators
\rf{susy-ghost-b-minus}--\rf{susy-ghost-f-minus} are again (adjoint)
super-eigenfunctions of the bosonic \cSKP hierarchy \rf{Lax-SKP-R-even-M}.

Consistency of the flows action \rf{ghost-susy-Lax-neg} with the constrained
form \rf{Lax-SKP-R-even-M} of $L$ and its inverse 
\rf{susy-L-minus-1} (or \rf{susy-L-minus-K}) requires that the
``negative-grade'' flows act on the pertinent (adjoint) super-eigenfunctions
as follows:
\br
\d^{(-n)}_{{\bar \cA}} \st{\sim}{\P}\!\!\!{}^{(m)}_a =
\cM^{(-n)}_{{\bar \cA}} (\st{\sim}{\P}\!\!\!{}^{(m)}_a) \quad ,\quad
\d^{(-n)}_{{\bar \cA}} \st{\sim}{\Psi}\!\!\!{}^{(m)}_a =
-\(\cM^{(-n)}_{{\bar \cA}}\)^\ast (\st{\sim}{\Psi}\!\!\!{}^{(m)}_a) 
\lab{ghost-A-neg-EF}\\
\d^{(-n+\h)}_{{\bar \cF}} \st{\sim}{\P}\!\!\!{}^{(m)}_a =
\cM^{(-n+\h)}_{{\bar \cF}} (\st{\sim}{\P}\!\!\!{}^{(m)}_a) \quad ,\quad
\d^{(-n+\h)}_{{\bar \cF}} \st{\sim}{\Psi}\!\!\!{}^{(m)}_a =
-\(\cM^{(-n+\h)}_{{\bar \cF}}\)^\ast (\st{\sim}{\Psi}\!\!\!{}^{(m)}_a)
\lab{ghost-F-neg-EF}
\er
\br
\d^{(-n)}_{{\bar \cA}} \st{\sim}{\P}\!\!\!{}^{(-m)}_a = 
\cM^{(-n)}_{{\bar \cA}} (\st{\sim}{\P}\!\!\!{}^{(-m)}_a) -
\sum_{b=1}^{N+r} {\bar A}^{(-n)}_{ab} \st{\sim}{\P}\!\!\!{}^{(-n-m)}_b
\nonu \\
\d^{(-n)}_{{\bar \cA}} \st{\sim}{\Psi}\!\!\!{}^{(-m)}_a = 
- \(\cM^{(-n)}_{{\bar \cA}}\)^\ast (\st{\sim}{\Psi}\!\!\!{}^{(-m)}_a) +
\sum_{b=1}^{N+r} {\bar D}^{(-n)}_{ba} \st{\sim}{\Psi}\!\!\!{}^{(-n-m)}_b
\lab{ghost-A-neg-EF-inverse}
\er
\br
\d^{(-n)}_{{\bar \cA}} \st{\sim}{\P}\!\!\!{}^{(-m)}_a = 
\cM^{(-n)}_{{\bar \cA}} (\st{\sim}{\P}\!\!\!{}^{(-m)}_a) -
\sum_{b=1}^{N+r} {\bar D}^{(-n)}_{ab} \st{\sim}{\P}\!\!\!{}^{(-n-m)}_b
\nonu \\
\d^{(-n)}_{{\bar \cA}} \st{\sim}{\Psi}\!\!\!{}^{(-m)}_a = 
- \(\cM^{(-n)}_{{\bar \cA}}\)^\ast (\st{\sim}{\Psi}\!\!\!{}^{(-m)}_a) +
\sum_{b=1}^{N+r} {\bar A}^{(-n)}_{ba} \st{\sim}{\Psi}\!\!\!{}^{(-n-m)}_b
\lab{ghost-A-neg-wti-EF-inverse}
\er
\be
\d^{(-n+\h)}_{{\bar \cF}} \P^{(-m)}_a = \cM^{(-n+\h)}_{{\bar \cF}} (\P^{(-m)}_a) +
\sum_{b=1}^{N+r} {\bar B}^{(-n+\h)}_{ab} {\wti \P}^{(-n-m+1)}_b 
\lab{ghost-F-neg-bEF-inverse}
\ee
\be
\d^{(-n+\h)}_{{\bar \cF}} \Psi^{(-m)}_a = 
- \(\cM^{(-n+\h)}_{{\bar \cF}}\)^\ast (\Psi^{(-m)}_a) -
\sum_{b=1}^{N+r} {\bar B}^{(-n+\h)}_{ba} {\wti \Psi}^{(-n-m+1)}_b
\lab{ghost-F-neg-adj-bEF-inverse}
\ee
\be
\d^{(-n+\h)}_{{\bar \cF}} {\wti \P}^{(-m)}_a = 
\cM^{(-n+\h)}_{{\bar \cF}} ({\wti \P}^{(-m)}_a) -
\sum_{b=1}^{N+r} {\bar C}^{(-n+\h)}_{ab} \P^{(-n-m)}_b 
\lab{ghost-F-neg-fEF-inverse}
\ee
\be
\d^{(-n+\h)}_{{\bar \cF}} {\wti \Psi}^{(-m)}_a =
- \(\cM^{(-n+\h)}_{{\bar \cF}}\)^\ast ({\wti \Psi}^{(-m)}_a) -
\sum_{b=1}^{N+r} {\bar C}^{(-n+\h)}_{ba} \Psi^{(-n-m)}_b
\lab{ghost-F-neg-adj-fEF-inverse}
\ee

Following the same steps as in the derivation of 
\rf{comm-A1-A2}--\rf{super-KM-alg-flows} and \rf{super-KM-alg-flows-b} we
obtain:
\be
\d^{(-n)}_{{\bar \cA}_1} \cM^{(-m)}_{{\bar \cA}_2}
- \d^{(-m)}_{{\bar \cA}_2} \cM^{(-n)}_{{\bar \cA}_1}
- \Sbr{\cM^{(-n)}_{{\bar \cA}_1}}{\cM^{(-m)}_{{\bar \cA}_2}} = 
\cM^{(-n-m)}_{\lb {\bar \cA}_1,{\bar \cA}_2 \rb}
\lab{comm-A1-A2-neg}
\ee
\be
\d^{(-n)}_{{\bar \cA}} \cM^{(-m+\h)}_{{\bar \cF}} 
- \d^{(-m+\h)}_{{\bar \cF}} \cM^{(-n)}_{{\bar \cA}}
- \Sbr{\cM^{(-n)}_{{\bar \cA}}}{\cM^{(-m+\h)}_{{\bar \cF}}} = 
\cM^{(-n-m+\h)}_{\lb {\bar \cA},{\bar \cF} \rb}
\lab{comm-A-F-neg}
\ee
\be
\d^{(-n+\h)}_{{\bar \cF}_1} \cM^{(-m+\h)}_{{\bar \cF}_2} + 
\d^{(-m+\h)}_{{\bar \cF}_2} \cM^{(-n+\h)}_{{\bar \cF}_1}
- \Bigl\{\cM^{(-n+\h)}_{{\bar \cF}_1},\,\cM^{(-m+\h)}_{{\bar \cF}_2}\Bigr\} = 
\cM^{(-n-m+1)}_{\{ {\bar \cF}_1,{\bar \cF}_2\}}
\lab{comm-F1-F2-neg}
\ee
In the present case, as explained above, the supermatrices
${\bar \cA}^{(n)},\, {\bar \cA}^{(n)}_{1,2}$ are subject to the condition 
${\bar \cA}^{(n)} \neq \one \; ,\; {\bar \cA}^{(n)}_{1,2} \neq \one$. 
Relations \rf{comm-A1-A2-neg}--\rf{comm-F1-F2-neg} imply, in complete analogy
with \rf{super-KM-alg-flows-b}, that the corresponding
infinite-dimensional algebra of the ``negative-grade'' flows
\rf{susy-ghost-b-minus}--\rf{ghost-susy-Lax-neg} :
\br
\Sbr{\d^{(-n)}_{{\bar \cA}_1}}{\d^{(-m)}_{{\bar \cA}_2}} = 
\d^{(-n-m)}_{\lb {\bar \cA}_1,\,{\bar \cA}_2\rb}
\quad ,\quad
\Sbr{\d^{(-n)}_{\bar \cA}}{\d^{(-m+\h)}_{\bar \cF}} = 
\d^{(-n-m+\h)}_{\lb {\bar \cA},\, {\bar \cF}\rb}
\nonu \\
\Bigl\{ \d^{(-n+\h)}_{{\bar \cF}_1},\, \d^{(-m+\h)}_{{\bar \cF}_2} \Bigr\} = 
\d^{(-n-m+1)}_{\{ {\bar \cF}_1,{\bar \cF}_2\}}  \phantom{aaaaaaaaaaaaaaaaaa}
\lab{super-KM-alg-flows-b-neg}
\er
is the superloop superalgebra $\({\widehat {GL}}^\pr_{N+r,N+r}\)_{-}$ -- the
negative-grade part of ${\widehat {GL}}^\pr_{N+r,N+r}$ 
(cf. \rf{loop-superalg-sub}), where the prime indicates factoring out of the 
unit matrix in any integer-grade subspace, {\sl i.e.} ${\bar \cA}^{(-n)} \neq \one$, 
whereas ${\bar \cF}^{(-n+\h)}$ is arbitrary.

From relations \rf{ghost-A-neg-EF}--\rf{ghost-F-neg-EF} and
\rf{ghost-A-inverse-EF-m}--\rf{ghost-F-inverse-EF-m} it is straighforward to
check that in bosonic constrained \cSKP supersymmetric hierarchies
\rf{Lax-SKP-R-even-M} the positive-grade $\({\widehat {GL}}_{M_B,M_F}\)_{+}$
(Eqs.\rf{super-KM-alg-flows-b} with
\rf{M-A-def-even}--\rf{ghost-susy-Lax-even}) and negative-grade 
$\({\widehat {GL}}^\pr_{N+r,N+r}\)_{-}$  (Eqs.\rf{super-KM-alg-flows-b-neg} 
with \rf{susy-ghost-b-minus}--\rf{ghost-susy-Lax-neg})
superloop superalgebra additional symmetries {\em (anti-)commute} among themselves.
\mskp
{\bfit 8.2 ``Negative''-Grade Symmetries of Fermionic \cSKP Hierarchies}
\sskp
Now we turn to the construction of ``negative-grade'' superloop superalgebra
additional symmetries for \cSKP hierarchies with fermionic
super-Lax operators \rf{Lax-SKP-R-M} (where 
$R=2r+1,\, M\equiv M_B +M_F = 2N+1$). Employing short-hand notations 
\rf{EF-neg-fSKP} we introduce the infinite set of super-pseudo-differential 
operators:
\be
\cM^{(-\ell/2)}_{{\bar \cA}} = \sum_{I,J=1}^{2(N+r+1)} {\bar \cA}^{(-\ell/2)}_{IJ}
\sum_{s=0}^{\ell-1} (-1)^{s(\ell + |J|)} 
\phi^{(-(\ell-s-1)/2)}_J \cD^{-1} \psi^{(-s/2)}_I
\lab{susy-ghost-b-minus-1}
\ee
\be
\cM^{(-\ell/2)}_{{\bar \cF}} = \sum_{I,J=1}^{2(N+r+1)} {\bar \cF}^{(-\ell/2)}_{IJ}
\sum_{s=0}^{\ell-1} (-1)^{s(\ell + |J|)}
\phi^{(-(\ell-s-1)/2)}_J \cD^{-1} \psi^{(-s/2)}_I
\lab{susy-ghost-f-minus-1}
\ee
defining the supersymmetric flows:
\be
\d^{(-\ell/2)}_{{\bar \cA}} \cL = \Sbr{\cM^{(-\ell/2)}_{{\bar \cA}}}{\cL}  \quad ,\quad
\d^{(-\ell/2)}_{{\bar \cF}} \cL = \Sbr{\cM^{(-\ell/2)}_{{\bar \cF}}}{\cL} 
\lab{ghost-susy-Lax-neg-1}
\ee
Here ${\bar \cA}^{(-\ell/2)}_{IJ}$ and ${\bar \cF}^{(-\ell/2)}_{IJ}$ are constant
graded matrices belonging to ${\widehat {GL}} (N+r+1,N+r+1)$ (cf. 
\rf{loop-superalg}--\rf{loop-superalg-odd}). Furthermore, ${\bar \cA}^{(-n)}\neq\one$ 
in Eqs.\rf{susy-ghost-b-minus-1},\rf{ghost-susy-Lax-neg-1} since
$\cM^{(-n)}_{{\bar \cA} =\one} = \cL^{-2n}$ (cf. Eq.\rf{SKP-Lax-plus-K}), so that 
the flow $\d^{(-n)}_{{\bar \cA} =\one}$ identically vanishes according to the first 
Eq.\rf{ghost-susy-Lax-neg-1}. Recall also, that according to 
\rf{SKP-EF-eqs-inverse} all super-functions entering the 
super-pseudo-differential operators
\rf{susy-ghost-b-minus-1}--\rf{susy-ghost-f-minus-1} are again (adjoint)
super-eigenfunctions of fermionic \cSKP hierachies \rf{Lax-SKP-R-M}.

Consistency of $\d^{(-\ell/2)}_{{\bar \cA},{\bar \cF}}$-flow action 
\rf{ghost-susy-Lax-neg-1} with the constrained form of $L$ \rf{Lax-SKP-R-M} 
implies:
\be
\d^{(-\ell/2)}_{{\bar \cA},{\bar \cF}} \P_i = 
\cM^{(-\ell/2)}_{{\bar \cA},{\bar \cF}} (\P_i)  \quad ,\quad
\d^{(-\ell/2)}_{{\bar \cA},{\bar \cF}} \Psi_i = 
-\(\cM^{(-\ell/2)}\)^\ast_{{\bar \cA},{\bar \cF}} (\Psi_i)
\lab{ghost-neg-1}
\ee
or, more generally using short-hand notations \rf{EF-neg-fSKP} : 
\be
\d^{(-\ell/2)}_{{\bar \cA},{\bar \cF}} \P^{(m/2)}_i = 
\cM^{(-\ell/2)}_{{\bar \cA},{\bar \cF}} (\P^{(m/2)}_i)  
\quad ,\quad
\d^{(-\ell/2)}_{{\bar \cA},{\bar \cF}} \Psi^{(m/2)}_i = 
-\(\cM^{(-\ell/2)}\)^\ast_{{\bar \cA},{\bar \cF}} (\Psi^{(m/2)}_i)
\lab{ghost-neg-m}
\ee
On the other hand, consistency of $\d^{(-\ell/2)}_{{\bar \cA},{\bar \cF}}$-flow 
action \rf{ghost-susy-Lax-neg-1} with the constrained form of $\cL^{-1}$ 
and, more generally, of $\cL^{-K}$ \rf{SKP-Lax-minus-K} yields:
\be
\d^{(-\ell/2)}_{{\bar \cA}} \phi^{(-m/2)}_I = 
\cM^{(-\ell/2)}_{{\bar \cA}} (\phi^{(-m/2)}_I)
- \sum_{J=1}^{2(N+r+1)} {\bar \cA}^{(-\ell/2)}_{IJ} \phi^{(-(m+\ell)/2)}_J
\lab{ghost-A-neg-EF-inv-1}
\ee
\be
\d^{(-\ell/2)}_{{\bar \cA}} \psi^{(-m/2)}_I = 
-\(\cM^{(-\ell/2)}_{{\bar \cA}}\)^\ast (\psi^{(-m)}_I) + \sum_{J=1}^{2(N+r+1)} 
(-1)^{\ell (|J|+m)} {\bar \cA}^{(-\ell/2)}_{JI} \psi^{(-(m+\ell)/2)}_J
\lab{ghost-A-neg-adj-EF-inv-1}
\ee
\be
\d^{(-\ell/2)}_{{\bar \cF}} \phi^{(-m/2)}_I = 
\cM^{(-\ell/2)}_{{\bar \cF}} (\phi^{(-m/2)}_I)
+ (-1)^m \sum_{J=1}^{2(N+r+1)} {\bar \cF}^{(-\ell/2)}_{IJ} \phi^{(-(m+\ell)/2)}_J
\lab{ghost-F-neg-EF-inv-1}
\ee
\be
\d^{(-\ell/2)}_{{\bar \cF}} \psi^{(-m/2)}_I = 
-\(\cM^{(-\ell/2)}_{{\bar \cF}}\)^\ast (\psi^{(-m/2)}_I) - \sum_{J=1}^{2(N+r+1)} 
(-1)^{(\ell +1)(|J|+m)} {\bar \cF}^{(-\ell/2)}_{JI} \psi^{(-(m+\ell)/2)}_J
\lab{ghost-F-neg-adj-EF-inv-1}
\ee

Using \rf{ghost-A-neg-EF-inv-1}--\rf{ghost-F-neg-adj-EF-inv-1} and repeating
the steps in the derivation of \rf{comm-A1-A2}--\rf{super-KM-alg-flows} we
find:
\be
\d_{{\bar \cA}_1}^{(-\ell/2)} \cM^{(-m/2)}_{{\bar \cA}_2} - 
\d_{{\bar \cA}_2}^{(-m/2)} \cM^{(-\ell/2)}_{{\bar \cA}_1}
- \Sbr{\cM^{(-\ell/2)}_{{\bar \cA}_1}}{\cM^{(-m/2)}_{{\bar \cA}_2}} = 
\cM^{(-(\ell +m)/2)}_{\lb {\bar \cA}_1,{\bar \cA}_2 \rb}
\lab{comm-cA1-cA2-neg}
\ee
\be
\d_{{\bar \cA}}^{(-\ell/2)} \cM^{(-m/2)}_{{\bar \cF}} - 
\d_{{\bar \cF}}^{(-m/2)} \cM^{(-\ell/2)}_{{\bar \cA}}
- \Sbr{\cM^{(-\ell/2)}_{{\bar \cA}}}{\cM^{(-m/2)}_{{\bar \cF}}} = 
\left\{ \begin{array}{lr}
\cM^{(-(\ell +m)/2)}_{\lb {\bar \cA} ,{\bar \cF}\rb} & {\rm for} \;\; \ell=even \\
- \cM^{(-(\ell +m)/2)}_{\{{\bar \cA},{\bar \cF}\}} & {\rm for} \;\; \ell=odd
\end{array} \right.
\lab{comm-cA-cF-neg}
\ee
\br
\d_{{\bar \cF}_1}^{(-\ell/2)} \cM^{(-m/2)}_{{\bar \cF}_2} + 
\d_{{\bar \cF}_2}^{(-m/2)} \cM^{(-\ell/2)}_{{\bar \cF}_1}
- \Bigl\{ \cM^{(-\ell/2)}_{{\bar \cF}_1},\,\cM^{(-m/2)}_{{\bar \cF}_2}\Bigr\} =
\nonu \\
= \left\{ \begin{array}{lr}
\pm \cM^{(-(\ell +m)/2)}_{\{{\bar \cF}_1,{\bar \cF}_2\}} & {\rm for} \;\; 
(\ell,m)=(odd,odd)/(even,even) \\
\pm \cM^{(-(\ell +m)/2)}_{\lb {\bar \cF}_1,{\bar \cF}_2\rb} & {\rm for} \;\; 
(\ell,m)=(odd,even)/(even,odd) \\
\end{array} \right.
\lab{comm-cF1-cF2-neg}
\er
which has the same form as \rf{comm-A1-A2}-\rf{comm-F1-F2}, but now
${\bar \cA},\,{\bar \cA}_{1,2},{\bar \cF},\,{\bar \cF}_{1,2}$ are graded matrices of
bigger size belonging to ${\widehat {GL}} (N+r+1,N+r+1)$. Relations 
\rf{comm-cA1-cA2-neg}--\rf{comm-cF1-cF2-neg} imply the following 
infinite-dimensional algebra of flows (cf. \rf{super-KM-alg-flows}) :
\br
\Sbr{\d^{(-\ell/2)}_{{\bar \cA}_1}}{\d^{(-m/2)}_{{\bar \cA}_2}} = 
\d^{(-(\ell +m)/2)}_{\lb {\bar \cA}_1,\, {\bar \cA}_2\rb}  
\phantom{aaaaaaaaaaaaaaaaaa}   \nonu \\
\Sbr{\d^{(-\ell/2)}_{{\bar \cA}}}{\d^{(-m/2)}_{{\bar \cF}}} =
\d^{(-(\ell +m)/2)}_{\lb {\bar \cA},\,{\bar \cF}\rb}
\;\;\; {\rm for}\;\; \ell ={\rm even}  \quad ,\quad
\Sbr{\d^{(-\ell/2)}_A}{\d^{(-m/2)}_F} = -\d^{(-(\ell +m)/2)}_{\{ A,\, F\}}
\;\;\; {\rm for}\;\; \ell ={\rm odd}
\nonu
\er
\br
\Bigl\{ \d^{(-\ell/2)}_{{\bar \cF}_1},\, \d^{(-m/2)}_{{\bar \cF}_2} \Bigr\} = 
\pm \d^{(-(\ell +m)/2)}_{\{ {\bar \cF}_1,{\bar \cF}_2\}}
\;\;\; {\rm for}\;\; (\ell,m)=(odd,odd)/(even,even)
\nonu \\
\Bigl\{ \d^{(-\ell/2)}_{{\bar \cF}_1},\, \d^{(-m/2)}_{{\bar \cF}_2} \Bigr\} = 
\pm \d^{(-(\ell +m)/2)}_{\lb{\bar \cF}_1,{\bar \cF}_2\rb}
\;\;\; {\rm for}\;\; (\ell,m)=(odd,even)/(even,odd)
\lab{super-KM-alg-flows-neg}
\er
which is isomorphic to $\({\widehat {GL}}^\pr (N+r+1,N+r+1)\)_{-}$. The
latter is the negative-grade part of ${\widehat {GL}} (N+r+1,N+r+1)$
(cf.\rf{loop-superalg}--\rf{loop-superalg-odd}), where the prime indicates
factoring out ${\bar \cA}^{(-n)}=\one$ in each integer-grade subspace (recall that
the flows $\d^{(-n)}_{{\bar \cA} =\one}$ \rf{ghost-susy-Lax-neg-1} vanish 
identically). 
  
Finally, using  \rf{ghost-A-F-EF-neg} and \rf{ghost-neg-m}, 
\rf{ghost-A-neg-EF-inv-1}--\rf{ghost-F-neg-adj-EF-inv-1}, we get:
\be
\d_{\cA}^{(\ell/2)} \cM^{(-m/2)}_{{\bar \cA}} - 
\d_{{\bar \cA}_2}^{(-m/2)} \cM^{(\ell/2)}_{\cA}
- \Sbr{\cM^{(\ell/2)}_{\cA}}{\cM^{(-m/2)}_{{\bar \cA}}} = 0
\lab{comm-A-A-plus-neg}
\ee
\be
\d_{\cA}^{(\ell/2)} \cM^{(-m/2)}_{{\bar \cF}} - 
\d_{{\bar \cF}_2}^{(-m/2)} \cM^{(\ell/2)}_{\cA}
- \Sbr{\cM^{(\ell/2)}_{\cA}}{\cM^{(-m/2)}_{{\bar \cF}}} = 0
\lab{comm-A-F-plus-neg}
\ee
\be
\d_{\cF}^{(\ell/2)} \cM^{(-m/2)}_{{\bar \cA}} - 
\d_{{\bar \cA}_2}^{(-m/2)} \cM^{(\ell/2)}_{\cF}
- \Sbr{\cM^{(\ell/2)}_{\cF}}{\cM^{(-m/2)}_{{\bar \cA}}} = 0
\lab{comm-F-A-plus-neg}
\ee
\be
\d_{\cF}^{(\ell/2)} \cM^{(-m/2)}_{{\bar \cF}} + 
\d_{{\bar \cF}_2}^{(-m/2)} \cM^{(\ell/2)}_{\cF}
- \Bigl\{ \cM^{(\ell/2)}_{\cF},\,\cM^{(-m/2)}_{{\bar \cF}}\Bigr\} = 0
\lab{comm-F-F-plus-neg}
\ee
Relations \rf{comm-A-A-plus-neg}--\rf{comm-F-F-plus-neg} imply that in fermionic
constrained \cSKP supersymmetric hierarchies \rf{Lax-SKP-R-M}
positive-grade $\({\widehat {GL}} (M_B,M_F)\)_{+}$ symmetry flows (anti-)commute
with negative-grade $\({\widehat {GL}} (N+r+1,N+r+1)\)_{-}$ symmetry flows
(recall $M \equiv M_B + M_F = 2N+1$) :
\be
\Sbr{\d^{(\ell/2)}_{\cA}}{\d^{(-m/2)}_{{\bar \cA}}} = 0 \quad ,\quad
\Sbr{\d^{(\ell/2)}_{\cA,\cF}}{\d^{(-m/2)}_{{\bar \cF},{\bar \cA}}} = 0
\quad ,\quad
\Bigl\{ \d^{(\ell/2)}_{\cF},\, \d^{(-m/2)}_{{\bar \cF}}\Bigr\} = 0
\lab{comm-plus-neg}
\ee
\sskp
{\bfit 8.3 Full Superloop Superalgebra Additional Symmetries}
\sskp
Collecting the results from Sections 5,6 and the present section we conclude
that:
\begin{itemize}
\item
Fermionic constrained \cSKP supersymmetric hierarchies \rf{Lax-SKP-R-M}
(where $R=2r+1$, $M\equiv M_B + M_F = 2N+1$)
possess the following superloop superalgebra symmetries:
\be
\({\widehat {GL}} (M_B,M_F)\)_{+} \oplus \({\widehat {GL}}^\pr (N+r+1,N+r+1)\)_{-}
\lab{plus-flow-alg-fSKP-1}
\ee
\item
Bosonic constrained \cSKP supersymmetric hierarchies \rf{Lax-SKP-R-even-M}
(where $R=2r,\, M\equiv M_B + M_F = 2N$)
possess the following superloop superalgebra symmetries:
\be
\({\widehat {GL}}_{M_B,M_F}\)_{+} \oplus \({\widehat {GL}}^\pr_{N+r,N+r}\)_{-}
\lab{plus-flow-alg-bSKP-1}
\ee
\end{itemize}
\lskip
{\bf 9. Virasoro Symmetries of Constrained SKP Hierarchies}
\mskp
The action of the operators -- (multiplication by) $\l$ and $\eta$ as well as
$\partder{}{\l}$ and $\partder{}{\eta}$ on the ``free'' BA super-function
$\psi^{(0)}_{BA}$ \rf{free-super-BA}--\rf{xi-def} can be expressed as the 
action of the following superspace operators (cf. third ref.\ct{SI-sstring}) :
\be
\l \psi^{(0)}_{BA} = \pa \psi^{(0)}_{BA} \quad ,\quad
\partder{}{\l} \psi^{(0)}_{BA} = \G_0 \psi^{(0)}_{BA}     \quad ,\quad
\eta \psi^{(0)}_{BA} = - \( Q + \G_1 \pa\) \psi^{(0)}_{BA}  \quad ,\quad
\partder{}{\eta} \psi^{(0)}_{BA} = \G_1 \psi^{(0)}_{BA}
\lab{free-super-BA-eqs}
\ee
where $Q = \partder{}{\th} - \th \pa$ is the standard super-charge operator
and:
\be
\G_0 \equiv \sum_{l=1}^\infty l t_l \pa^{l-1} + 
\sum_{n=1}^\infty (n-\h)\th_n \pa^{n-2}\cD - 
\h \sum_{n=1}^\infty \th_n \pa^{n-2} Q + 
\h \sum_{n,l=1}^\infty (n-l) \th_n \th_l \pa^{n+l-2}
\lab{G0-def}
\ee
\be
\G_1 \equiv \th + \sum_{n=1}^\infty \th_n \pa^{n-1} \quad ,\quad
Q + \G_1 \pa = \partder{}{\th} + \sum_{n=1}^\infty \th_n \pa^n
\lab{G1-def}
\ee
Dressing arbitrary products of powers of the above ``free'' superspace
operators by means of Sato superspace dressing operator $\cW$ 
\rf{super-dress} :
\be
\cM_{k,\ell,m,n} = \(\cW \G_0^k \pa^\ell \G_1^m (Q+\G_1\pa)^n \cW^{-1}\)_{-}
\lab{M-klmn}
\ee
defines via Eqs.\rf{super-flow-def} an infinite set of bosonic and fermionic
symmetry flows for the general unconstrained MR-SKP hierarchy \rf{super-Lax}
which span the supersymmetric version of $W_{1+\infty}$ algebra
\ct{SI-sstring}. For the class of reduced \cSKP
hierarchies \rf{Lax-SKP-R-M}, however, the flows constructed by \rf{M-klmn} 
{\em do not} define symmetries since they do not preserve the constrained form
of the pertinent super-Lax operators. This is a superspace analog of the
problem with the usual Orlov-Schulman operators \ct{Orlov-Schulman}, which 
do not yield symmetries in the case of constrained KP hierarchies in the purely bosonic
case. In the present Section we will follow our approach from 
refs.\ct{noak-addsym} where the latter problem has been solved via
appropriate modification of the standard additional-symmetry generating 
Orlov-Schulman operators.

In fact, we will construct here the Virasoro additional symmetries for bosonic 
constrained \cSKP hierarchies \rf{Lax-SKP-R-even-M}. This same construction 
based on the super-pseudo-differential formalism does not, however, carry over 
to the case of the fermionic part of the full super-Virasoro and the rest of 
super-$W_{1+\infty}$ symmetries, as well to the case of (super-)Virasoro and
super-$W_{1+\infty}$ symmetries for fermionic constrained \cSKP hierarchies 
\rf{Lax-SKP-R-M}.

Similarly to the purely bosonic case \ct{noak-addsym}, the action of 
Virasoro flows on super-Lax and super-dressing operators
are given by (henceforth $L \equiv \cL_{(2r;M_B,M_F)}$ and we employ notations from
Section 6 above) :
\be
\d^V_n L = \Bigl\lb -\bigl(\cW \st{0}{\cM}_n \cW^{-1}\bigr)_{-} + 
\cX_n\, ,\, L \Bigr\rb
\quad ,\quad \d^V_n \cW = 
\Bigl( -\bigl(\cW \st{0}{\cM}_n \cW^{-1}\bigr)_{-} + \cX_n\Bigr) \cW
\lab{sVir}
\ee
or, equivalently:
\be
\d^V_n L = \Bigl\lb \bigl(\cW \st{0}{\cM}_n \cW^{-1}\bigr)_{+} + 
\cX_n\, ,\,L \Bigr\rb + L^n
\lab{sVir-1}
\ee
where $\d^V_n \simeq - L_{n-1}$ (in terms of standard Virasoro notations).
Here:
\be
\st{0}{\cM}_n \equiv \G_0 \pa^n + {n\o 2}\G_1 (Q+\G_1\pa )\pa^{n-1}
\lab{M-0-Vir-n-def}
\ee
are the ``bare'' (undressed) Virasoro operators and the additional operators
$\cX_n$ are to be chosen in such a way that the flows
\rf{sVir} define a symmetry, {\sl i.e.}, they must preserve the
constrained form of $L \equiv L_{r,N}$ \rf{Lax-SKP-R-even-M}. 

For non-negative Virasoro flows ($n \geq 0$ in \rf{sVir}) we find 
the following expression for $\cX_n$ :
\be
\cX_n \equiv \sum_{s=1}^{n-1} \bigl( s - {n\o 2}\bigr)
\Bigl\lb \sum_{a=1}^{M_B} \P^{(n-s)}_a D^{-1} {\wti \Psi}^{(s)}_a +
\sum_{b=1}^{M_F} {\wti \P}^{(n-s)}_b D^{-1} \Psi^{(s)}_b \Bigr\rb
\lab{susy-X-n-def}
\ee
where the short-hand notations \rf{EF-plus} are used. Consistency of
\rf{sVir} with the constrained form of $L$
\rf{Lax-SKP-R-even-M} and its inverse powers \rf{susy-K-zero-eqs} dictates 
the specific form of the action of $\d^V_n$-flows (for $n\geq 0$) on the 
pertinent (adjoint) super-eigenfunctions 
\rf{EF-plus} and \rf{bEF-inverse}--\rf{fEF-inverse}, which reads accordingly:
\be
\d^V_n \st{\sim}{\P}\!\!\!{}^{(m)}_a = 
\Bigl\lb \bigl(\cW\st{0}{\cM}_n\cW^{-1}\bigr)_{+} + \cX_n \Bigr\rb 
(\st{\sim}{\P}\!\!\!{}^{(m)}_a) + 
\bigl({n\o 2} + m-1\bigr) \st{\sim}{\P}\!\!\!{}^{(n+m-1)}_a
\lab{sVir-flow-EF-a}
\ee
\be 
\d^V_n \st{\sim}{\Psi}\!\!\!{}^{(m)}_a = 
- \Bigl\lb \bigl(\cW\st{0}{\cM}_n\cW^{-1}\bigr)^\ast_{+} + \cX^\ast_n\Bigr\rb 
(\st{\sim}{\Psi}\!\!\!{}^{(m)}_a) + 
\bigl({n\o 2} + m-1\bigr) \st{\sim}{\Psi}\!\!\!{}^{(n+m-1)}_a
\lab{sVir-flow-adj-EF-a}
\ee
\be
\d^V_n \st{\sim}{\P}\!\!\!{}^{(-m)}_b = 
\Bigl\lb \bigl(\cW\st{0}{\cM}_n\cW^{-1}\bigr)_{+} + \cX_n \Bigr\rb 
\bigl(\st{\sim}{\P}\!\!\!{}^{(-m)}_b\bigr)  
-(m-1) \st{\sim}{\P}\!\!\!{}^{(-(m-n+1))}_b
\quad  {\rm for}\;\; m\geq n
\lab{sVir-flow-EF-b}
\ee
\be
\d^V_n \st{\sim}{\P}\!\!\!{}^{(-m)}_b = 
\Bigl\lb \bigl(\cW\st{0}{\cM}_n\cW^{-1}\bigr)_{+} + \cX_n \Bigr\rb 
\bigl(\st{\sim}{\P}\!\!\!{}^{(-m)}_b\bigr)  \quad {\rm for}\;\; m\leq n-1
\lab{sVir-flow-EF-b-0}
\ee
\be
\d^V_n \st{\sim}{\Psi}\!\!\!{}^{(-m)}_b =
- \Bigl\lb \bigl(\cW\st{0}{\cM}_n\cW^{-1}\bigr)^\ast_{+} + \cX^\ast_n\Bigr\rb 
\bigl(\st{\sim}{\Psi}\!\!\!{}^{(-m)}_b\bigr)
-(m-1) \st{\sim}{\Psi}\!\!\!{}^{-(m-n+1))}_b
\quad  {\rm for}\;\; m\geq n
\lab{sVir-flow-adj-EF-b}
\ee
\be
\d^V_n \st{\sim}{\Psi}\!\!\!{}^{(-m)}_b =
- \Bigl\lb \bigl(\cW\st{0}{\cM}_n\cW^{-1}\bigr)^\ast_{+} + \cX^\ast_n\Bigr\rb 
\bigl(\st{\sim}{\Psi}\!\!\!{}^{(-m)}_b\bigr)
\quad  {\rm for}\;\; m\leq n-1
\lab{sVir-flow-adj-EF-b-0}
\ee
where relations \rf{susy-K-zero-eqs} are taken into account.

For negative Virasoro flows ($n<0$ in \rf{sVir}) we obtain
(employing again notations \rf{bEF-inverse}--\rf{fEF-inverse}) :
\be
\cX_{(-|n|)} = \sum_{b=1}^{N+r} \sum_{j=0}^{|n|}\bigl({{|n|}\o 2}- j\bigr) 
\Bigl\lb \P^{-(|n|-j+1)}_b \cD^{-1} {\wti \Psi}^{-(j+1)}_b +
{\wti \P}^{-(|n|-j+1)}_b \cD^{-1} \Psi^{-(j+1)}_b \Bigr\rb
\lab{susy-X-n-minus-def}
\ee
Consistency of \rf{sVir} (for $n<0$) with the constrained form of 
$L$ \rf{Lax-SKP-R-even-M} and its inverse powers \rf{susy-L-minus-K} implies that
the flows $\d^V_{-|n|}$ act on the constituent (adjoint) super-eigenfunctions
\rf{EF-plus} and \rf{bEF-inverse}--\rf{fEF-inverse} as follows
(taking into account \rf{susy-K-zero-eqs}) :
\be
\d^V_{-|n|} \st{\sim}{\P}\!\!\!{}^{(m)}_a = 
\Bigl\lb \bigl(\cW\st{0}{\cM}_{-|n|}\cW^{-1}\bigr)_{+} + 
\cX_{(-|n|)}\Bigr\rb (\st{\sim}{\P}\!\!\!{}^{(m)}_a) 
+ (m-1) \st{\sim}{\P}\!\!\!{}^{(m-|n|-1)}_a 
\quad {\rm for}\;\; m \geq |n|+2
\lab{sVir-minus-EF}
\ee
\be
\d^V_{-|n|} \st{\sim}{\P}\!\!\!{}^{(m)}_a = 
\Bigl\lb \bigl(\cW\st{0}{\cM}_{-|n|}\cW^{-1}\bigr)_{+} + 
\cX_{(-|n|)}\Bigr\rb (\st{\sim}{\P}\!\!\!{}^{(m)}_a) 
\quad {\rm for}\;\; m \leq |n|+1
\lab{sVir-minus-EF-0}
\ee
\be
\d^V_{-|n|} \st{\sim}{\Psi}\!\!\!{}^{(m)}_a  = - 
\Bigl\lb \bigl(\cW\st{0}{\cM}_{-|n|}\cW^{-1}\bigr)^\ast_{+} + 
\cX_{(-|n|)}^\ast \Bigr\rb (\st{\sim}{\P}\!\!\!{}^{(m)}_a)
+ (m-1) \st{\sim}{\Psi}\!\!\!{}^{(m-|n|-1)}_a 
\quad {\rm for}\;\; m \geq |n|+2
\lab{sVir-minus-adj-EF}
\ee
\be
\d^V_{-|n|} \st{\sim}{\Psi}\!\!\!{}^{(m)}_a  = - 
\Bigl\lb \bigl(\cW\st{0}{\cM}_{-|n|}\cW^{-1}\bigr)^\ast_{+} + 
\cX_{(-|n|)}^\ast \Bigr\rb (\st{\sim}{\P}\!\!\!{}^{(m)}_a)
\quad {\rm for}\;\; m \leq |n|+1
\lab{sVir-minus-adj-EF-0}
\ee
\be
\d^V_{-|n|} \st{\sim}{\P}\!\!\!{}^{(-m)}_b =
\Bigl\lb \bigl(\cW\st{0}{\cM}_{-|n|}\cW^{-1}\bigr)_{+} + 
\cX_{(-|n|)} \Bigr\rb \bigl(\st{\sim}{\P}\!\!\!{}^{(-m)}_b\bigr)
-\bigl({{|n|}\o 2}+m\bigr) \st{\sim}{\P}\!\!\!{}^{(-(m+|n|+1))}_b
\lab{sVir-minus-EF-inverse}
\ee
\be
\d^V_{-|n|} \st{\sim}{\Psi}\!\!\!{}^{(-m)}_b = -
\Bigl\lb \bigl(\cW\st{0}{\cM}_{-|n|}\cW^{-1}\bigr)^\ast_{+} + 
\cX_{(-|n|)}^\ast \Bigr\rb \bigl(\st{\sim}{\Psi}\!\!\!{}^{(-m)}_b\bigr)
-\bigl({{|n|}\o 2}+m\bigr) \st{\sim}{\Psi}\!\!\!{}^{(-(m+|n|+1))}_b
\lab{sVir-minus-adj-EF-inverse}
\ee

The consistency of the negative flow definitions \rf{sVir}
(or \rf{sVir-1}) with $n<0$, where
$\cX_{(-|n|)}$ is as in Eq.\rf{susy-X-n-minus-def}, crucially depends on
relations \rf{susy-K-zero-eqs}. Also, in the process of derivation of 
\rf{susy-X-n-def}--\rf{sVir-minus-adj-EF-inverse} essential use is made of the
super-pseudo-differential operator identities \rf{susy-pseudo-diff-id}.

What is left is to check that the flows \rf{sVir} indeed
satisfy the commutation relations of the standard Virasoro algebra. To this
end let us consider the commutator of the Virasoro flows
$\d^V_{n} \simeq - L_{n-1}$ and $\d^V_{m} \simeq - L_{m-1}$ (with $L_n$ being
the standard notations for the basis of Virasoro algebra) acting on the
bosonic super-Lax operator $L$
(where $(n,m)$ are arbitrary non-negative or negative indices) which yields:
\br
\d^V_{n} \Bigl( -\bigl(\cW\st{0}{\cM}_m \cW^{-1}\bigr)_{-} + \cX_{m}\Bigr) -
\d^V_{m} \Bigl( -\bigl(\cW\st{0}{\cM}_n \cW^{-1}\bigr)_{-} + \cX_{n}\Bigr)
\nonu \\
- \Bigl\lb -\bigl(\cW\st{0}{\cM}_n \cW^{-1}\bigr)_{-} + \cX_{n}\, ,\,
-\bigl(\cW\st{0}{\cM}_m \cW^{-1}\bigr)_{-} + \cX_{m} \Bigr\rb
\lab{susy-commut-n-m}
\er
Using the identity:
\br
\d^V_n \bigl(\cW\st{0}{\cM}_m \cW^{-1}\bigr)_{-} - 
\d^V_m \bigl(\cW\st{0}{\cM}_n \cW^{-1}\bigr)_{-} =
-(n-m) \bigl(\cW\st{0}{\cM}_{n+m-1}\cW^{-1}\bigr)_{-} \nonu \\
-\Bigl\lb \bigl(\cW\st{0}{\cM}_n \cW^{-1}\bigr)_{-}\, ,\, 
\bigl(\cW\st{0}{\cM}_m \cW^{-1}\bigr)_{-} \Bigr\rb 
+ \Bigl\lb \cX_n \, ,\,\cW\st{0}{\cM}_m \cW^{-1}\Bigr\rb_{-} -
\Bigl\lb \cX_m \, ,\,\cW\st{0}{\cM}_n \cW^{-1}\Bigr\rb_{-}
\lab{susy-rel-n-m}
\er
the r.h.s. of Eq.\rf{susy-commut-n-m} can be rewritten in the form:
\br
(n-m) \bigl(\cW\st{0}{\cM}_{n+m-1}\cW^{-1}\bigr)_{-} + \d^V_n \cX_m - 
\Bigl\lb \bigl(\cW\st{0}{\cM}_n \cW^{-1}\bigr)_{+}\, ,\, \cX_m \Bigr\rb_{-}
\nonu \\
- \d^V_m \cX_n + \Bigl\lb \bigl(\cW\st{0}{\cM}_m \cW^{-1}\bigr)_{+}
\, ,\, \cX_n \Bigr\rb_{-} - \Sbr{\cX_n}{\cX_m}
\lab{susy-commut-n-m-rhs}
\er
Now, employing again the super-pseudo-differential identities 
\rf{susy-pseudo-diff-id} we find, taking into account 
\rf{sVir-flow-EF-a}--\rf{sVir-flow-adj-EF-b-0}
and \rf{sVir-minus-EF}--\rf{sVir-minus-adj-EF-inverse},
that the sum of all terms in \rf{susy-commut-n-m-rhs} involving $\cX_{n,m}$
yield: 
\be
\d^V_n \cX_m - 
\Bigl\lb \bigl(\cW\st{0}{\cM}_n \cW^{-1}\bigr)_{+}\, ,\, \cX_m \Bigr\rb_{-}
- \d^V_m \cX_n + \Bigl\lb \bigl(\cW\st{0}{\cM}_m \cW^{-1}\bigr)_{+}
\, ,\, \cX_n \Bigr\rb_{-} - \Sbr{\cX_n}{\cX_m} = -(n-m) \cX_{n+m-1}
\lab{X-terms}
\ee
Thus, we verify the closure of the full Virasoro algebra of additional
symmetries without central extension:
\be
\Sbr{\d^V_{n}}{\d^V_{m}} = -(n-m) \d^V_{n+m-1}
\lab{full-Vir-alg}
\ee
\lskip
{\bf 10. Superspace \DB Transformations and Wronskian-like
Super-Determinant Solutions}
\mskp
{\bfit 10.1 \DB Transformations for Constrained Super-KP Hierarchies}
\sskp
In what follows we shall consider \DB (DB) transformations for
the whole class \cSKP of constrained (reduced)
supersymmetric KP integrable hierarchies \rf{Lax-SKP-R-M}. 
For definiteness we shall explicitly discuss the case of \cSKP
hierarchies defined by fermionic super-Lax operators. 
DB and adjoint-DB transformed objects will be indicated by tilde and hat, 
respectively, on top of the corresponding symbol.

In analogy with the ordinary ``bosonic'' case, DB transformations within the Sato 
super-pseudo-differential operator approach are defined as ``gauge'' 
transformations of special kind on the pertinent super-Lax operator of the 
supersymmetric integrable hierarchy:
\be
\cL \;\;\; \to\;\;\; {\wti \cL} = \cT_\phi \cL \cT_\phi^{-1} \quad ,\quad  
\cT_\phi \equiv \phi \cD \phi^{-1}
\lab{super-DB-def}
\ee
with $\phi$ being a bosonic superfunction, which obey the following requirements:
\mskp

(A) Super-DB transformations \rf{super-DB-def} have to preserve the specific
constrained form \rf{Lax-SKP-R-M}  of $\cL$ (or \rf{Lax-SKP-R-even-M} for
bosonic \cSKP hierarchies), {\sl i.e.}, 
the transformed super-Lax operator ${\wti \cL}$ \rf{super-DB-def} must be again
of the form:
\be
{\wti \cL} \equiv {\wti \cL}_{(R;{\wti M}_B,{\wti M}_F)} =
\cD^R + \sum_{j=0}^{R-1} {\wti v}_{j\o2} \cD^j + 
\sum_{i=1}^M {\wti \P}_i \cD^{-1} {\wti \Psi}_i
\quad ,\quad  M = {\wti M}_B + {\wti M}_F
\lab{DB-transf-Lax}
\ee
where ${\wti M}_{B,F}$ are the numbers of DB-transformed bosonic/fermionic
(adjoint) super-eigenfunctions ${\wti \P}_i$, ${\wti \Psi}_i$. Let us stress
that we require the total number $M$ of negative super-pseudo-differential
terms in ${\wti \cL}$ to be the same as in the initial super-Lax operator $\cL$
\rf{Lax-SKP-R-M}. Let us also note that, using the superspace pseudo-differential
operator identities \rf{susy-pseudo-diff-id}, the DB-transformed fermionic \cSKP
super-Lax operator \rf{super-DB-def} acquires the form:
\be
{\wti \cL} = \({\wti \cL}\)_{+} + 
\Bigl(\cT_\phi \cL (\phi)\Bigr)\cD^{-1}\phi^{-1} +
\sum_{i=1}^M (-1)^{|i|} \cT_\phi (\P_i) \cD^{-1} \(\cT_\phi^{-1}\)^\ast (\Psi_i)
\lab{DB-transf-Lax-1}
\ee
Therefore, one of the $M+1$ negative super-pseudo-differential terms on
the r.h.s. of \rf{DB-transf-Lax-1} has to vanish.
\sskp

(B) Super-DB transformations \rf{super-DB-def} have to preserve the bosonic 
\rf{boson-flows-R} and fermionic \rf{odd-flow-new} isospectral evolution equations
(in the case of fermionic \cSKP hierarchies) or Eqs.\rf{super-bLax-odd} (for 
bosonic \cSKP hierarchies). As we will see below, the fermionic isospectral flows
\rf{odd-flow-new} can be strictly preserved under super-DB transformations only 
for the subclass ${\sl SKP}_{(R;1,0)}$ of constrained super-KP hierarchies \cSKP. 
In the more general case we will require preservation of fermionic isospectral 
flows under super-DB transformations up to an overall sign change.
\mskp

Similarly, we can define adjoint-DB transformations:
\be
\cL \;\;\; \to\;\;\; {\widehat \cL} = 
\Bigl(-\cT_\psi^{-1}\Bigr)^\ast \cL \cT_\psi^\ast \quad ,\quad  
\cT_\psi \equiv \psi \cD \psi^{-1}
\lab{super-adj-DB-def}
\ee
obeying the same requirements (A) and (B). In this case the counterpart of 
Eq.\rf{DB-transf-Lax-1} now reads:
\be
{\widehat \cL} = \({\widehat \cL}\)_{+} + 
\psi^{-1}\cD^{-1} \Bigl(\cT_\psi \cL^\ast (\psi)\Bigr) +
\sum_{i=1}^M (-1)^{|i|}\Bigl( -\cT_\psi^{-1}\Bigr)^\ast (\P_i) \cD^{-1}
\cT_\psi (\Psi_i)
\lab{adj-DB-transf-Lax-1}
\ee
As in \rf{DB-transf-Lax-1}, one of the $M+1$ negative super-pseudo-differential 
terms on the r.h.s. of \rf{adj-DB-transf-Lax-1} has to vanish.

Comparing \rf{DB-transf-Lax-1} with \rf{DB-transf-Lax} (and similarly for the
adjoint-DB transformations \rf{adj-DB-transf-Lax-1}), and taking into account
relations \rf{SKP-zero-eqs} (for fermionic super-Lax operators) or 
\rf{susy-zero-eqs} (for bosonic super-Lax operators), we find that condition (A) 
above can be satisfied for two different choices of the (adjoint-)DB generating 
superfunctions $\phi$ and $\psi$ :
\mskp

(i) First choice: $\phi = \P_{i_0}$ where $\P_{i_0}$ is some fixed {\em bosonic} 
super-eigenfunction entering the negative pseudo-differential part of the 
original super-Lax operator \rf{Lax-SKP-R-M}. In this case we obtain:
\be
{\wti \P}_{i_0} = \cT_\phi \cL (\phi) \quad ,\quad 
{\wti \Psi}_{i_0} = \phi^{-1} \quad ,\quad  \phi \equiv \P_{i_0}
\lab{DB-EF-1-1}
\ee
\be
{\wti \P}_i = \cT_\phi (\P_i) \quad ,\quad 
{\wti \Psi}_i = (-1)^{|i|}\Bigl(\cT_\phi^{-1}\Bigr)^\ast (\Psi_i) \quad,\quad
i \neq i_0
\lab{DB-EF-1-2}
\ee
Similarly, the first choice for adjoint-DB transformations is
$\psi = \Psi_{i_0}$ where $\Psi_{i_0}$ is some fixed {\em bosonic} adjoint
super-eigenfunction entering the negative pseudo-differential part of the 
original super-Lax operator \rf{Lax-SKP-R-M}. Accordingly, for the
adjoint-DB transformed (adjoint) super-eigenfunctions we have:
\be
{\widehat \P}_{i_0} = - \psi^{-1} \quad ,\quad 
{\widehat \Psi}_{i_0} = - \cT_\psi \cL^\ast (\psi) \quad ,\quad
\psi \equiv \Psi_{i_0}
\lab{adj-DB-EF-1-1}
\ee
\be
{\widehat \P}_i = (-1)^{|i|} \(\cT_\psi^{-1}\)^\ast (\P_i) \quad ,\quad
{\widehat \Psi}_i = - \cT_\psi (\Psi_i) \quad ,\quad i \neq i_0
\lab{adj-DB-EF-1-2}
\ee
Let us note that the Grassmann parity of the DB transformed (adjoint)
super-eigenfunctions ${\wti \P}_i$ and ${\wti \Psi}_i$ \rf{DB-EF-1-2} 
for $i \neq i_0$ changes from $|i|$ to $|i|+1$, and similarly for the adjoint-DB
transformed ones \rf{adj-DB-EF-1-2}.
\sskp

(ii) Second choice: $\phi = L_{N+\h}({\wti \vp}_{\a_0})$ (for super-DB
transformations) and $\psi = \psi_{\a_0}$ (for adjoint super-DB transformations)
where $L_{N+\h}({\wti \vp}_{\a_0})$ and $\psi_{\a_0}$ are some fixed bosonic 
(adjoint) super-eigenfunctions \rf{Ker-L}--\rf{EF-neg-fSKP} entering the 
expression \rf{SKP-Lax-minus-1} for the inverse power of the super-Lax operator
$\cL$. Since according to \rf{SKP-zero-eqs} the defined above $\phi$ and $\psi$
obey the relations $\cL (\phi) = 0$ and $\cL^\ast (\psi) = 0$, we get for the 
(adjoint) DB-transformed (adjoint) super-eigenfunctions in \rf{DB-transf-Lax-1}
and \rf{adj-DB-transf-Lax-1} :
\be
{\wti \P}_i = \cT_\phi (\P_i) \quad ,\quad 
{\wti \Psi}_i = (-1)^{|i|}\Bigl(\cT_\phi^{-1}\Bigr)^\ast (\Psi_i) \quad,\quad
\phi \equiv L_{N+\h}({\wti \vp}_{\a_0}) 
\lab{DB-EF-2-1}
\ee
\be
{\widehat \P}_i = (-1)^{|i|} \(\cT_\psi^{-1}\)^\ast (\P_i) \quad ,\quad
{\widehat \Psi}_i = - \cT_\psi (\Psi_i) \quad ,\quad \psi = \psi_{\a_0}
\lab{adj-DB-EF-2-2}
\ee
for all $i=1,\ldots ,M$ .
\mskp

For later use let us write down the (adjoint) super-DB transfomations for
the whole series of (adjoint) super-eigenfunctions 
\rf{EF-plus-fSKP}--\rf{EF-neg-fSKP} entering in the definition of additional 
non-isospectral symmetry flow generating operators 
\rf{M-A-def}--\rf{M-F-def},\rf{ghost-susy-Lax} and 
\rf{susy-ghost-b-minus-1}--\rf{ghost-susy-Lax-neg-1} :
\begin{itemize}
\item
For the first choice (i) of DB-generating (adjoint) eigenfunctions 
(cf. \rf{DB-EF-1-1}--\rf{adj-DB-EF-1-2}) we have:
\be
{\wti \P}^{(\ell/2)}_{i_0} = \cT_\phi (\P^{((\ell +1)/2)}_{i_0})  \quad, \quad
{\wti \Psi}^{(\ell/2)}_{i_0} = 
(-1)^{\ell -1} \( \cT_\phi^{-1}\)^\ast (\Psi^{((\ell -1)/2)}_{i_0})
\quad {\rm for}\;\; n \geq 2
\lab{DB-EF-1-1-all}
\ee
\vspace{-.1in}
\br
{\wti \Psi}_{i_0} = {1\o \phi} \equiv {1\o {\P_{i_0}}} 
\nonu 
\er
\be
{\wti \P}^{(\ell/2)}_i = \cT_\phi (\P^{(\ell/2)}_i) \quad, \quad
{\wti \Psi}^{(\ell/2)}_i = (-1)^{|i|} \(\cT^{-1}_\phi\)^\ast (\Psi^{(\ell/2)}_i) 
\quad {\rm for} \;\; i \neq i_0
\lab{DB-EF-1-2-all}
\ee
\be
{\wti \phi}^{(-\ell/2)}_I = \cT_\phi (\phi^{(-\ell/2)}_I) \quad, \quad
{\wti \psi}^{(-\ell/2)}_I = \(\cT^{-1}_\phi\)^\ast (\psi^{(-\ell/2)}_I)
\lab{DB-EF-neg-1-all}
\ee
\be
{\widehat \P}_{i_0} = - {1\o \psi} \equiv - {1\o {\Psi_{i_0}}} \quad ,\quad
{\widehat \P}^{(\ell/2)}_{i_0} = 
- \(\cT^{-1}_\psi\)^\ast (\P^{((\ell -1)/2)}_{i_0}) \quad 
{\rm for}\;\; \ell \geq 1
\lab{adj-DB-EF-1-1-all}
\ee
\vspace{-.1in}
\br
{\widehat \Psi}^{(\ell)}_{i_0} = - \cT_\psi (\Psi^{((\ell +1)/2)}_{i_0}) 
\nonu
\er
\be
{\widehat \P}^{(\ell/2)}_i = - \(\cT^{-1}_\psi\)^\ast (\P^{(\ell/2)}_i)  
\quad ,\quad
{\widehat \Psi}^{(\ell/2)}_i = - \cT_\psi (\Psi^{(\ell/2)}_i)  
\quad ,\quad i \neq i_0
\lab{adj-DB-EF-1-2-all}
\ee
\be
{\widehat \phi}^{(-\ell/2)}_I = - \(\cT^{-1}_\psi\)^\ast (\phi^{(-\ell/2)}_I)  
\quad ,\quad
{\widehat \psi}^{(-\ell/2)}_I = - \cT_\psi (\psi^{(-\ell/2)}_I)  
\lab{adj-DB-EF-neg-1-all}
\ee
\item
For the second choice (ii) of DB-generating (adjoint) eigenfunctions 
(cf. \rf{DB-EF-2-1}--\rf{adj-DB-EF-2-2}) we obtain:
\be
{\wti \P}^{(\ell/2)}_i = \cT_\phi (\P^{(\ell/2)}_i) \quad ,\quad 
{\wti \Psi}^{(\ell/2)}_i = (-1)^{|i|} \(\cT^{-1}_\phi\)^\ast (\Psi^{(\ell/2)}_i) 
\quad , \quad 
\phi \equiv L_{N+\h} ({\wti \vp}_{\a_0}) \equiv \phi^{(0)}_{\a_0}
\lab{DB-EF-2-all}
\ee
\be
{\wti \phi}^{(-\ell/2)}_{\a_0} = \cT_\phi \bigl(\phi^{(-(\ell +1)/2)}_{\a_0}\bigr)
\;\; ,\;\; 
{\wti \psi}^{(-\ell/2)}_{\a_0} = 
\(\cT_\phi^{-1}\)^\ast \bigl(\psi^{(-(\ell -1)/2)}_{\a_0}\bigr)
\;\; {\rm for}\;\; \ell \geq 1 \;\; ,\;\;
{\wti \psi}^{(0)}_{\a_0} = {1\o {\phi}} \equiv {1\o {\phi^{(0)}_{\a_0}}}
\lab{DB-EF-neg-2-1-all}
\ee
\be
{\wti \phi}^{(-\ell/2)}_I = \cT_\phi \bigl(\phi^{(-\ell/2)}_I\bigr) \quad ,\quad
{\wti \psi}^{(-\ell/2)}_I = 
(-1)^{|I|} \(\cT^{-1}_\phi\)^\ast \bigl(\psi^{(-\ell/2)}_I\bigr) 
\quad  {\rm for} \;\; I \neq \a_0
\lab{DB-EF-neg-2-2-all}
\ee
\be
{\widehat \P}^{(\ell/2)}_i = 
(-1)^{|i|} \(\cT^{-1}_\psi\)^\ast (\P^{(\ell/2)}_i) \quad ,\quad 
{\widehat \Psi}^{(\ell/2)}_i = - \cT_\psi (\Psi^{(\ell/2)}_i) \quad , \quad 
\psi \equiv \psi_{\a_0} \equiv \psi^{(0)}_{\a_0}
\lab{adj-DB-EF-2-all}
\ee
\be
{\widehat \psi}^{(0)}_{\a_0} = - {1\o \psi} \equiv
- {1\o {\psi^{(0)}_{\a_0}}} \;\; ,\;\;
{\widehat \phi}^{(-\ell/2)}_{\a_0} = 
- \(\cT^{-1}_\psi\)^\ast (\phi^{(-(\ell -1)/2}_{\a_0}) 
\;\; {\rm for} \;\; \ell \geq 1
\lab{adj-DB-EF-neg-2-1-all}
\ee
\vspace{-.1in}
\br
{\widehat \psi}^{(-\ell/2)}_{\a_0} = 
- \cT_\psi \bigl(\Psi^{(-(\ell +1)/2)}_{\a_0}\bigr)
\nonu
\er
\be
{\widehat \phi}^{(-\ell/2)}_I = 
(-1)^{|I|} \(\cT^{-1}_\psi\)^\ast \bigl(\P^{(-\ell/2)}_a\bigr) 
\quad, \quad
{\widehat \psi}^{(-\ell/2)}_I = - \cT_\psi \bigl(\psi^{(-\ell/2)}_I\bigr) 
\quad {\rm for} \;\; I \neq \a_0
\lab{adj-DB-EF-neg-2-2-all}
\ee
\end{itemize}

Let us now study the fulfillment of condition (B) above by the (adjoint) DB
transformations \rf{DB-EF-1-1}--\rf{DB-EF-1-2} and
\rf{adj-DB-EF-1-1}--\rf{adj-DB-EF-1-2}. It is straightforward to check,
using the super-pseudo-differential identities \rf{susy-pseudo-diff-id}, that
the latter preserve the bosonic isospectral flow Eqs.\rf{boson-flows-R} :
\be
\partder{}{t_l} {\wti \cL} = \Sbr{\partder{}{t_l}\cT_\phi\, \cT^{-1}_\phi
+ \cT_\phi \Bigl(\cL^{{2l}\o R}\Bigr)_{+}\cT^{-1}_\phi}{{\wti \cL}} =
\Sbr{\Bigl({\wti \cL}^{{2l}\o R}\Bigr)_{+}}{{\wti \cL}}
\lab{DB-boson-flows-R}
\ee
for fermionic super-Lax operators, and similarly for bosonic super-Lax
operators (first Eqs.\rf{super-bLax-odd}). Next, we compute the action of the 
fermionic isospectral flows $D_n$ on the DB-transformed bosonic super-Lax 
operator ${\wti L} = \cT_\phi L \cT^{-1}_\phi$ \rf{super-DB-def} taking into 
account the second Eq.\rf{super-bLax-odd} and using the identities 
\rf{susy-pseudo-diff-id} to obtain:
\be
D_n {\wti L} = \Sbr{D_n \cT_\phi\, \cT^{-1}_\phi - 
\cT_\phi \Bigl( L^{{2n-1}\o{2r}}\Bigr)_{+} \cT^{-1}_\phi  }{{\wti L}} = 
- \Sbr{\Bigl({\wti L}^{{2n-1}\o{2r}}\Bigr)_{+}}{{\wti L}}
\lab{DB-super-bLax-odd}
\ee
Comparing \rf{DB-super-bLax-odd} with second Eqs.\rf{super-bLax-odd} we note
that $D_n$ fermionic flows for bosonic \cSKP hierarchies are preserved under
(adjoit-) DB transformations up to an overall sign.

Let us now discuss the case of fermionic \cSKP hierarchies.
The action of the modified fermionic isospectral flows $D_n$ \rf{odd-flow-new} 
on the DB-transformed fermionic super-Lax operators \rf{super-DB-def} reads:
\be
D_n {\wti \cL} = \lcurl D_n \cT_\phi \, \cT^{-1}_\phi +
\cT_\phi \(\cL^{2n-1}_{-} - X_{2n-1}\) \cT^{-1}_\phi\,\,{\wti \cL}\rcurl
\lab{DB-odd-flow-new} 
\ee
where, using the identities \rf{susy-pseudo-diff-id}, we have:
\br
D_n \cT_\phi \, \cT^{-1}_\phi +
\cT_\phi \(\cL^{2n-1}_{-} - X_{2n-1}\) \cT^{-1}_\phi =
\cT_\phi \Bigl( D_n \phi + \bigl(\cL^{2n-1}_{-} - X_{2n-1}\bigr)(\phi)\Bigr)
\cD^{-1} \phi^{-1} +
\nonu \\
+ \sum_{i=1}^M \sum_{s=0}^{2n-2} (-1)^{s(|i|+2n-1)}
\cT_\phi \(\cL^{2n-2-s}(\P_i)\) \cD^{-1} (-1)^{|i|+s} 
\bigl(\cT_\phi^{-1}\bigr)^\ast \(\bigl(\cL^s\bigr)^\ast (\Psi_i)\)
\lab{DB-odd-flow-new-a}
\er
For the second choice \rf{DB-EF-2-all} of super-DB transformations
we find, using \rf{Lax-X} and \rf{ghost-A-F-EF-neg}, that the r.h.s. of
Eq.\rf{DB-odd-flow-new-a} becomes equal to
${\wti \cL}^{2n-1}_{-} - {\wti X}_{2n-1}$ where ${\wti X}_{2n-1}$ is of the
same form as $X_{2n-1}$ \rf{X-def} with all (adjoint) super-eigenfunctions
replaced by their DB-transformed counterparts. Thus, comparing with 
\rf{odd-flow-new} we conclude that under the second type \rf{DB-EF-2-all} of 
DB transformations on fermionic \cSKP super-Lax operators the fermionic 
isospectral flows are preserved up to an overall minus sign:
\be
D_n {\wti \cL} = 
+ \lcurl {\wti \cL}^{2n-1}_{-} - {\wti X}_{2n-1}\, ,\,{\wti \cL}\rcurl
\lab{DB-odd-flow-new-2}
\ee

The situation with the first type of DB transformations 
\rf{DB-EF-1-1-all}--\rf{DB-EF-1-2-all} on fermionic \cSKP \\ 
hierarchies is slightly more complicated. First, let us consider the subclass of 
${\sl SKP}_{(R;1,0)}$ hierarchies defined by fermionic super-Lax operators 
($R=2r+1$) :
\be
\cL \equiv \cL_{(R;1,0)} = 
\cD^R + \sum_{j=0}^{R-1} v_{j\o2} \cD^j + \P \cD^{-1} \Psi
\lab{Lax-SKP-R-1}
\ee
As already shown in ref.\ct{match}, the r.h.s. of Eq.\rf{DB-odd-flow-new-a}
becomes in this case:
\be
- \sum_{s=0}^{2n -2} (-1)^s 
{\wti \cL}^{2n-2-s}({\wti \P})\cD^{-1}\({\wti \cL}^s\)^\ast ({\wti \Psi}) =
- \(\bigl({\wti \cL}^{2n-1}\bigr)_{-} - {\wti X}_{2n-1}\) 
\lab{DB-Lax-X-R-1}
\ee
where we have used $\phi \equiv \P$, ${\wti \P}=\cT_\P \bigl(\cL (\P)\bigr)$, 
${\wti \Psi} = \P^{-1}$ (cf. \rf{DB-EF-1-1}) and also the identities from
\ct{match} :
\be
{\wti \cL}^s ({\wti \P}) = \cT_\P \bigl(\cL^{s+1} (\P)\bigr)  \quad ,\quad
\Bigl({\wti \cL}^{s+1}\Bigr)^\ast ({\wti \Psi}) = 
(-1)^s \(\cT_\P^{-1}\)^\ast \(\bigl(\cL^s\bigr)^\ast (\Psi)\)
\lab{match-id}
\ee
Therefore, substituting the first term in the anti-commutator in 
\rf{DB-odd-flow-new} with the expression \rf{DB-Lax-X-R-1} we conclude that
for fermionic ${\sl SKP}_{(R;1,0)}$ hierarchies \rf{Lax-SKP-R-1} the
fermionic isospectral flows $D_n$ \rf{odd-flow-new} are strictly preserved
(no overall sign change) under first type of super-DB transformations
\rf{DB-EF-1-1}.

In the more general case of fermionic \cSKP hierarchies with 
$M = M_B + M_F \geq 2$ the r.h.s. of Eq.\rf{DB-odd-flow-new-a} becomes under
the first type of DB transformations \rf{DB-EF-1-1} :
\be
- \sum_{s=0}^{2n -2} (-1)^s {\wti \cL}^{2n-2-s}({\wti \P}_{i_0})\cD^{-1}
\({\wti \cL}^s\)^\ast ({\wti \Psi}_{i_0})
+ \sum_{i=1\, ,i\neq i_0}^M \sum_{s=0}^{2n -2} (-1)^{s|i|} 
{\wti \cL}^{2n-2-s}({\wti \P}_i)\cD^{-1} \({\wti \cL}^s\)^\ast ({\wti \Psi}_i)
\lab{DB-odd-flow-new-b}
\ee
which is {\em not} equal to
$\pm \(\bigl({\wti \cL}^{2n-1}\bigr)_{-} - {\wti X}_{2n-1}\)$  due to the
opposite signs in front of both sums in \rf{DB-odd-flow-new-b}. Therefore,
fermionic isospectral flows $D_n$ are preserved under first type of super-DB
transformations only for the subclass ${\sl SKP}_{(R;1,0)}$ \rf{Lax-SKP-R-1} 
of fermionic constrained super-KP hierarchies.

Finally, let us recall that according to \ct{match} the super-tau function 
\rf{tau-sres} undergoes the following (adjoint-)DB transformations:
\be
\t \longrightarrow {\wti \t} = \frac{\phi}{\t} \quad, \quad
\t \longrightarrow {\widehat \t} = -\frac{1}{\psi\,\t}
\lab{DB-supertau}
\ee
The latter relations are to be constrasted with their counterparts in the
ordinary ``bosonic''case where
${\wti \t} = \phi\,\t\; ,\; {\widehat \t} = -\psi\,\t$.
\mskp
{\bfit 10.2 Superspace \DB Transformations Preserving Additional Symmetries}
\sskp
We are now interested in consistency of super-DB transformations of \cSKP
constrained super-KP hierarchies with the whole algebra of the additional
non-isospectral symmetries (Sections 5,6 and 8 above). Acting with the
pertinent additional symmetry flows of positive grades 
(\rf{ghost-susy-Lax}, \rf{M-A-def}--\rf{M-F-def}) and of negative grades 
(\rf{susy-ghost-b-minus-1}--\rf{ghost-susy-Lax-neg-1}) on the DB-transformed 
super-Lax operator \rf{super-DB-def} (we take fermionic super-Lax operator for
definiteness) we have:
\be
\d^{(\pm \ell/2)}_{\cA,{\bar \cA}} {\wti \cL} = 
\Sbr{{\wti \cM}^{(\pm \ell/2)}_{\cA,{\bar \cA}}}{{\wti \cL}}
\quad ,\quad
\d^{(\pm \ell/2)}_{\cF,{\bar \cF}} {\wti \cL} = 
\lcurl {\wti \cM}^{(\pm \ell/2)}_{\cF,{\bar \cF}},\, {\wti \cL}\rcurl
\lab{DB-ghost-susy-Lax}
\ee
Here:
\br
{\wti \cM}^{(\ell/2)}_{\cA} \equiv 
\d^{(\ell/2)}_{\cA}\cT_\phi\, \cT^{-1}_\phi +
\cT_\phi \cM^{(\ell/2)}_{\cA} \cT^{-1}_\phi = 
\cT_\phi \(\cM^{(\ell/2)}_{\cA}(\phi) - 
\d^{(\ell/2)}_{\cA}\phi\) \cD^{-1}\phi^{-1} + 
\nonu \\
+ \sum_{i,j=1}^M \cA^{(\ell/2)}_{ij} \sum_{s=0}^{\ell -1}
(-1)^{s(|j|+\ell)} \cT_\phi (\P^{((\ell -1-s)/2)}_j) \cD^{-1}
(-1)^{|i|+s}\(\cT^{-1}_\phi\)^\ast (\Psi^{(s/2)}_i)
\lab{DB-M-A-def}
\er
\br
{\wti \cM}^{(\ell/2)}_{\cF} \equiv 
\d^{(\ell/2)}_{\cF}\cT_\phi\, \cT^{-1}_\phi -
\cT_\phi \cM^{(\ell/2)}_{\cF} \cT^{-1}_\phi =
\cT_\phi \(\d^{(\ell/2)}_{\cF}\phi -\cM^{(\ell/2)}_{\cF}(\phi)\)
\cD^{-1}\phi^{-1} + 
\nonu \\
+ \sum_{i,j=1}^M \cF^{(\ell/2)}_{ij} \sum_{s=0}^{\ell -1}
(-1)^{s(|j|+\ell)} \cT_\phi (\P^{((\ell -1-s)/2)}_j) \cD^{-1}
(-1)^{|i|+s}\(\cT^{-1}_\phi\)^\ast (\Psi^{(s/2)}_i)
\lab{DB-M-F-def}
\er
for positive-grade additional symmetries, and:
\br
{\wti \cM}^{(-\ell/2)}_{{\bar \cA}} \equiv 
\d^{(-\ell/2)}_{{\bar \cA}}\cT_\phi\, \cT^{-1}_\phi +
\cT_\phi \cM^{(-\ell/2)}_{{\bar \cA}} \cT^{-1}_\phi =
\cT_\phi \(\cM^{(-\ell/2)}_{{\bar \cA}}(\phi) - \d^{(-\ell/2)}_{{\bar\cA}}\phi\)
\cD^{-1}\phi^{-1} +
\nonu \\
+ \sum_{I,J=1}^{2(N+r+1)} {\bar \cA}^{(-\ell/2)}_{IJ} \sum_{s=0}^{\ell -1}
(-1)^{s(|J|+\ell)} \cT_\phi (\phi^{(-(\ell -1-s)/2)}_J) \cD^{-1}
(-1)^{|I|+s}\(\cT^{-1}_\phi\)^\ast (\psi^{(-s/2)}_I)
\lab{DB-M-A-neg-def}
\er
\br
{\wti \cM}^{(-\ell/2)}_{{\bar \cF}} \equiv 
\d^{(-\ell/2)}_{{\bar \cF}}\cT_\phi\, \cT^{-1}_\phi -
\cT_\phi \cM^{(-\ell/2)}_{{\bar \cF}} \cT^{-1}_\phi =
\cT_\phi \(\d^{(-\ell/2)}_{{\bar\cF}}\phi - \cM^{(-\ell/2)}_{{\bar \cF}}(\phi)\)
\cD^{-1}\phi^{-1} +
\nonu \\
+ \sum_{I,J=1}^{2(N+r+1)} {\bar \cF}^{(-\ell/2)}_{IJ} \sum_{s=0}^{\ell -1}
(-1)^{s(|J|+\ell)} \cT_\phi (\phi^{(-(\ell -1-s)/2)}_J) \cD^{-1}
(-1)^{|I|+s}\(\cT^{-1}_\phi\)^\ast (\psi^{(-s/2)}_I)
\lab{DB-M-F-neg-def}
\er
for negative-grade additional symmetries 
(recall $R = 2r+1$, $M \equiv M_B +M_F = 2N+1$).

Now, we can repeat the same steps as in the analysis in the previous
subsection 10.1 of the consistency of 
super-DB transformations with the bosonic and fermionic isospectral flows
in order to find the conditions under which the (adjoint-)DB transformations 
(relations \rf{DB-EF-1-1}--\rf{DB-EF-1-2} and 
\rf{adj-DB-EF-1-1}--\rf{adj-DB-EF-1-2})
preserve also the additional non-isospectral symmetries of \cSKP integrable
hierarchies. In other words, we have to find the conditions under which the
super-pseudo-differential operators \rf{DB-M-A-def}--\rf{DB-M-F-neg-def},
generating the additional symmetries of the DB-transformed \cSKP
hierarchy, can be represented in the same form as \rf{M-A-def}--\rf{M-F-def} and
\rf{susy-ghost-b-minus-1}--\rf{susy-ghost-f-minus-1}, respectively,
with all pertinent (adjoint) super-eigenfunctions replaced with their
(adjoint) DB-transformed counterparts. We obtain the following results
for fermionic \cSKP hierarchies \rf{Lax-SKP-R-M} :

\begin{itemize}
\item
Super-DB transformations of the first type (i) 
(\rf{DB-EF-1-1-all}--\rf{DB-EF-neg-1-all}) preserve (up to an overall sign change 
of the fermionic flows) the following subalgebra of additional non-isospectral
symmetries:
\be
\({\widehat {GL}}(M_B -1,M_F)\)_{+} \oplus \({\widehat{GL}}^\pr (N+r+1,N+r+1)\)_{-}
\lab{DB-alg-fSKP-1}
\ee
\item
Super-DB transformations of the second type (ii) 
(\rf{DB-EF-2-all}--\rf{DB-EF-neg-2-2-all}) preserve (up to an overall sign change 
of the fermionic flows) the following subalgebra of additional symmetries:
\be
\({\widehat {GL}}(M_B,M_F)\)_{+} \oplus \({\widehat {GL}}(N+r,N+r+1)\)_{-}
\lab{DB-alg-fSKP-2}
\ee
(here the positive-grade part includes the Manin-Radul isospectral flows).
\end{itemize}

For bosonic \cSKP hierarchies \rf{Lax-SKP-R-even-M} we obtain similar results with
\rf{DB-alg-fSKP-1} and \rf{DB-alg-fSKP-2} replaced by:
\be
\({\widehat {GL}}_{M_B -1,M_F}\)_{+} \oplus \({\widehat{GL}}^\pr_{N+r,N+r}\)_{-}
\lab{DB-alg-bSKP-1}
\ee
and 
\be
\({\widehat {GL}}_{M_B,M_F}\)_{+} \oplus \({\widehat {GL}}_{N+r-1,N+r}\)_{-}
\lab{DB-alg-bSKP-2}
\ee
respectively.  
\mskp
{\bfit 10.3 Iterations of Superspace \DB Transformations and 
Wronskian-like Super-Determinant Solutions}
\sskp
The general super-\DB orbit consists of successive applications of the allowed
DB \rf{super-DB-def} and adjoint-DB \rf{super-adj-DB-def} transformations as
defined in subsection 10.1 (see Eqs.\rf{DB-EF-1-1}--\rf{adj-DB-EF-neg-2-2-all}).
In particular, pairs of successive DB and adjoint-DB transformations are
called {\em binary} DB transformations. Let us consider an iteration of $n$
successive binary DB transformations followed by $2m$ successive DB
transformations applied on arbitrary initial bosonic super-eigenfunction $\P$ :

\br
\P^{(n+2m;n)} = 
\st{n+2m-1;n}{\cT_{\vp_{m-\h}}} \st{n+2m-2;n}{\cT_{\vp_{n+m-1}}} 
\ldots \st{n+3;n}{\cT_{\vp_{3/2}}} \st{n+2;n}{\cT_{\vp_{n+1}}} 
\st{n+1;n}{\cT_{\vp_\h}} \st{n;n}{\cT_{\vp_n}} \times
\nonu \\
\times \bigl( -\st{n;n-1}{\cT_{\psi_{n-\h}}}\!{}^{-1}\bigr)^\ast
\st{n-1;n-1}{\cT_{\vp_{n-1}}} \ldots
\bigl( -\st{2;1}{\cT_{\psi_{3/2}}}\!{}^{-1}\bigr)^\ast \st{1;1}{\cT_{\vp_1}}
\bigl( -\st{1;0}{\cT_{\psi_{\h}}}\!{}^{-1}\bigr)^\ast \st{0;0}{\cT_{\vp_0}} (\P)
\lab{DB-Phi-gen}
\er
Recall that each (adjoint) DB transformation flips the Grassmann parity of
the transformed object.
Similarly, let us consider an iteration of $n$ successive binary DB 
transformations followed by $2m+1$ successive DB transformations applied on 
arbitrary initial fermionic super-eigenfunction $F$ :
\br
F^{(n+2m+1;n)} = \st{n+2m;n}{\cT_{\vp_{n+m}}}
\st{n+2m-1;n}{\cT_{\vp_{m-\h}}} \st{n+2m-2;n}{\cT_{\vp_{n+m-1}}} 
\ldots \st{n+3;n}{\cT_{\vp_{3/2}}} \st{n+2;n}{\cT_{\vp_{n+1}}} 
\st{n+1;n}{\cT_{\vp_\h}} \st{n;n}{\cT_{\vp_n}} \times
\nonu \\
\times \bigl( -\st{n;n-1}{\cT_{\psi_{n-\h}}}\!{}^{-1}\bigr)^\ast
\st{n-1;n-1}{\cT_{\vp_{n-1}}} \ldots
\bigl( -\st{2;1}{\cT_{\psi_{3/2}}}\!{}^{-1}\bigr)^\ast \st{1;1}{\cT_{\vp_1}}
\bigl( -\st{1;0}{\cT_{\psi_{\h}}}\!{}^{-1}\bigr)^\ast \st{0;0}{\cT_{\vp_0}} (F)
\lab{DB-F-gen}
\er
The upper indices $(k;l)$ in \rf{DB-Phi-gen}--\rf{DB-F-gen} and below indicate 
iteration of (adjoint) DB transformations consisting of $k$ DB steps and $l$ 
adjoint-DB steps. The objects entering each (adjoint) DB step in 
\rf{DB-Phi-gen}--\rf{DB-F-gen} are recurrsively defined as follows:
\be
\st{k;k}{\cT_{\vp_{k}}} \equiv
\vp_{k}^{(k;k)} \cD \frac{1}{\vp_{k}^{(k;k)}}  \quad ,\quad
\st{k+1;k}{\cT_{\psi_{k+\h}}} \equiv
\psi_{k+\h}^{(k+1;k)} \cD \frac{1}{\psi_{k+\h}^{(k+1;k)}} 
\lab{bin-DB-steps}
\ee
\be
\vp_{k}^{(k;k)} = \bigl( -\st{k;k-1}{\cT_{\psi_{k-\h}}}\!{}^{-1}\bigr)^\ast
\st{k-1;k-1}{\cT_{\vp_{k-1}}} \ldots
\bigl( -\st{2;1}{\cT_{\psi_{3/2}}}\!{}^{-1}\bigr)^\ast \st{1;1}{\cT_{\vp_1}}
\bigl( -\st{1;0}{\cT_{\psi_{\h}}}\!{}^{-1}\bigr)^\ast \st{0;0}{\cT_{\vp_0}}
(\vp_k)
\lab{bin-DB-vp}
\ee
\be
\psi_{k+\h}^{(k+1;k)} = -
\bigl(\st{k;k}{\cT_{\vp_{k}}}\!{}^{-1}\bigr)^\ast
\st{k;k-1}{\cT_{\psi_{k-\h}}} 
\bigl(\st{k-1;k-1}{\cT_{\vp_{k-1}}}\!{}^{-1}\bigr)^\ast \ldots
\st{1;0}{\cT_{\psi_{\h}}} \bigl(\st{0;0}{\cT_{\vp_{0}}}\!{}^{-1}\bigr)^\ast 
(\psi_{k+\h})
\lab{bin-DB-psi}
\ee
for $k=1,\ldots ,n-1$, and:
\be
\st{n+2l;n}{\cT_{\vp_{n+l}}} \equiv
\vp_{n+l}^{(n+2l;n)} \cD \frac{1}{\vp_{n+l}^{(n+2l;n)}}
\quad ,\quad
\st{n+2l+1;n}{\cT_{\vp_{l+\h}}} \equiv
\vp_{l+\h}^{(n+2l+1;n)} \cD \frac{1}{\vp_{l+\h}^{(n+2l+1;n)}}
\lab{bin-DB+DB-steps}
\ee
\br
\vp_{n+l}^{(n+2l;n)} =
\st{n+2l-1;n}{\cT_{\vp_{l-\h}}} \st{n+2l-2;n}{\cT_{\vp_{n+l-1}}} 
\ldots \st{n+3;n}{\cT_{\vp_{3/2}}} \st{n+2;n}{\cT_{\vp_{n+1}}} 
\st{n+1;n}{\cT_{\vp_\h}} \st{n;n}{\cT_{\vp_n}} \times
\nonu \\
\times \bigl( -\st{n;n-1}{\cT_{\psi_{n-\h}}}\!{}^{-1}\bigr)^\ast
\st{n-1;n-1}{\cT_{\vp_{n-1}}} \ldots
\bigl( -\st{2;1}{\cT_{\psi_{3/2}}}\!{}^{-1}\bigr)^\ast \st{1;1}{\cT_{\vp_1}}
\bigl( -\st{1;0}{\cT_{\psi_{\h}}}\!{}^{-1}\bigr)^\ast \st{0;0}{\cT_{\vp_0}}
(\vp_{n+l})
\lab{bin-DB+DB-b}
\er
\br
\vp_{l+\h}^{(n+2l+1;n)} =
\st{n+2l;n}{\cT_{\vp_{n+l}}}
\st{n+2l-1;n}{\cT_{\vp_{l-\h}}} \st{n+2l-2;n}{\cT_{\vp_{n+l-1}}} 
\ldots \st{n+3;n}{\cT_{\vp_{3/2}}} \st{n+2;n}{\cT_{\vp_{n+1}}} 
\st{n+1;n}{\cT_{\vp_\h}} \st{n;n}{\cT_{\vp_n}} \times
\nonu \\
\times \bigl( -\st{n;n-1}{\cT_{\psi_{n-\h}}}\!{}^{-1}\bigr)^\ast
\st{n-1;n-1}{\cT_{\vp_{n-1}}} \ldots
\bigl( -\st{2;1}{\cT_{\psi_{3/2}}}\!{}^{-1}\bigr)^\ast \st{1;1}{\cT_{\vp_1}}
\bigl( -\st{1;0}{\cT_{\psi_{\h}}}\!{}^{-1}\bigr)^\ast \st{0;0}{\cT_{\vp_0}}
(\vp_{l+\h})
\lab{bin-DB+DB-f}
\er
where $l=0,1,\ldots ,m-1$. In \rf{DB-Phi-gen}--\rf{bin-DB+DB-f} the sets
$\lcurl \vp_k\rcurl_{k=0}^{n+m}$ and $\lcurl \vp_{l-\h}\rcurl_{l=1}^{m}$
are bosonic/fermionic super-eigenfunctions, whereas
$\lcurl \psi_{k-\h}\rcurl_{k=1}^n$ are fermionic adjoint super-eigenfunctions.
Let us also stress that DB-transformed superfunctions $F^{(n+2m-1;n)}$
\rf{DB-F-gen}, $\vp_{l+\h}^{(n+2l+1;n)}$ \rf{bin-DB+DB-f} and
$\psi_{k+\h}^{(k+1;k)}$ \rf{bin-DB-psi} are bosonic although the initial
$F$, $\vp_{l+\h}$ and $\psi_{k+\h}$ are fermionic.

During iteration of (adjoint) super-DB transformations we encounter
Berezinians (super-deter\-mi\-nants) whose matrix blocks possess the following
special generalized Wronskian-like $k \times (m+n)$ matrix form:
\br
{\wti W}^{(k;n)}_{k,m+n} \llb \{\vp\} \bv \{\psi\} \rrb \equiv 
{\wti W}^{(k;n)}_{k,m+n} 
\llb \vp_0,\ldots ,\vp_{k-1} \bv \psi_{\h},\ldots ,\psi_{n-\h}\rrb =
\nonu \\
= \left(
\begin{array}{cccc}
\vp_0 & \cdots & \cdots & \vp_{k-1} \\
\vdots & \ddots & \ddots & \vdots   \\
\pa^{m-1}\vp_0 & \cdots & \cdots & \pa^{m-1}\vp_{k-1} \\
\Dth^{-1}\! (\vp_0\psi_\h)& \cdots &\cdots &\Dth^{-1}\! (\vp_{k-1}\psi_\h) \\
\vdots & \ddots & \ddots & \vdots   \\
\Dth^{-1}\! (\vp_0\psi_{n-\h})& \cdots &\cdots
&\Dth^{-1}\! (\vp_{k-1}\psi_{n-\h})    
\end{array} \right)    
\lab{susy-wti-Wronski}
\er
where $\{\vp\} \equiv \{\vp_0,\ldots ,\vp_{k-1}\}$ is a set of $k$
bosonic or fermionic superfunctions whereas
$\{\psi\} \equiv \bigl\{\psi_\h,\ldots ,\psi_{n-\h}\bigr\}$ is a set of $n$
fermionic superfunctions. The generalized Wronskian-like matrix 
\rf{susy-wti-Wronski} is the supersymmetric generalization of the
Wronskian-like block matrices entering the general \DB determinant solutions
for the tau-functions of ordinary ``bosonic'' constrained KP hierarchies
\ct{hallifax,gauge-wz}.
In the special case of $n=0$ \rf{susy-wti-Wronski} reduces to the rectangular
$k \times m$ Wronskian matrix:
\be
W_{k,m} \llb \vp_0,\ldots ,\vp_{k-1} \rrb =
\left(
\begin{array}{cccc}
\vp_0 & \cdots & \cdots & \vp_{k-1} \\
\pa \vp_0 & \cdots & \cdots & \pa \vp_{k-1} \\
\vdots & \ddots & \ddots & \vdots   \\
\pa^{m-1}\vp_0 & \cdots & \cdots & \pa^{m-1}\vp_{k-1}
\end{array} \right)    
\lab{Wronski-matrix}
\ee

In ref.\ct{zim-ber} the explicit form of iterations of super-DB
transformations {\em not} accompanied by adjoint-DB transformations,
{\sl i.e.}, with $n=0$ in \rf{DB-Phi-gen}--\rf{DB-F-gen}, has been derived:
\br
\P^{(2m;0)} \equiv \st{2m-1;0}{\cT_{\vp_{m-\h}}} \st{2m-2;0}{\cT_{\vp_{m-1}}} 
\ldots \st{3;0}{\cT_{\vp_{3/2}}} \st{2;0}{\cT_{\vp_1}} 
\st{1;0}{\cT_{\vp_\h}} \st{0;0}{\cT_{\vp_0}} (\P) = \phantom{aaaaaaaaaaa} 
\lab{Phi-Ber-2m} \\ 
= 
\frac{{\rm Ber} \threemat{W_{m+1,m+1} \lb \vp_0,\ldots ,\vp_{m-1},\P \rb}{|}{
W_{m,m+1} \lb \vp_\h ,\ldots ,\vp_{m-\h}\rb}{--------------}{|}{------------}{ 
W_{m+1,m} \lb \Dth\vp_0 ,\ldots ,\Dth \vp_{m-1},\Dth \P \rb}{|}{
W_{m,m} \lb \Dth\vp_\h ,\ldots ,\Dth\vp_{m-\h}\rb}
}{
{\rm Ber} \threemat{W_{m,m} \lb \vp_0,\ldots ,\vp_{m-1}\rb}{|}{
W_{m,m} \lb \vp_\h ,\ldots ,\vp_{m-\h}\rb}{------------}{|}{------------}{ 
W_{m,m} \lb \Dth\vp_0 ,\ldots ,\Dth \vp_{m-1}\rb}{|}{
W_{m,m} \lb \Dth\vp_\h ,\ldots ,\Dth\vp_{m-\h}\rb}
}
\nonu
\er
\vspace{.1in}
\br
F^{(2m+1;0)} \equiv \st{2m;0}{\cT_{\vp_m}} \st{2m-1;0}{\cT_{\vp_{n-\h}}} 
\ldots \st{3;0}{\cT_{\vp_{3/2}}} \st{2;0}{\cT_{\vp_1}} 
\st{1;0}{\cT_{\vp_\h}} \st{0;0}{\cT_{\vp_0}} (F) = \phantom{aaaaaaaaaaa}
\lab{F-Ber-2m+1} \\
= \frac{{\rm Ber} \threemat{W_{m+1,m+1} \lb \vp_0,\ldots ,\vp_{m-1},\vp_m \rb}{|}{
W_{m,m+1} \lb \vp_\h ,\ldots ,\vp_{m-\h}\rb}{--------------}{|}{------------}{ 
W_{m+1,m} \lb \Dth\vp_0 ,\ldots ,\Dth \vp_{m-1},\Dth \vp_m \rb}{|}{
W_{m,m} \lb \Dth\vp_\h ,\ldots ,\Dth\vp_{m-\h}\rb}
}{
{\rm Ber} \threemat{W_{m+1,m+1} \lb \vp_0,\ldots ,\vp_{m}\rb}{|}{
W_{m+1,m+1} \lb \vp_\h ,\ldots ,\vp_{m-\h},F\rb}{------------}{|}{---------------}{ 
W_{m+1,m+1} \lb \Dth\vp_0 ,\ldots ,\Dth \vp_{m}\rb}{|}{
W_{m+1,m+1} \lb \Dth\vp_\h ,\ldots ,\Dth\vp_{m-\h},\Dth F\rb}
}
\nonu 
\er
Following similar techniques as in ref.\ct{zim-ber} we can similarly express
the general (adjoint) DB iterations \rf{DB-Phi-gen}--\rf{DB-F-gen}
in the form of ratios of Berezinians containing generalized Wronskian-like
matrix blocks \rf{susy-wti-Wronski}. The results are as follows:
\be
\P^{(n+2m;n)} =
\frac{{\rm Ber} \threemat{
{\wti W}^{(n+m+1;n)}_{n+m+1,n+m+1} \llb \{\vp\},\P \bv \{\psi\}\rrb}{|}{
{\wti W}^{(m;n)}_{m,n+m+1} \llb \{\vp_{(\h)}\}\bv \{\psi\} \rrb
}{---------------}{|}{------------}{ 
W_{n+m+1,m} \llb \Dth\vp_0,\ldots ,\Dth\vp_{n+m-1},\Dth \P \rrb}{|}{
W_{m,m} \llb \Dth\vp_\h ,\ldots ,\Dth\vp_{m-\h}\rrb}
}{
{\rm Ber} \threemat{
{\wti W}^{(n+m;m)}_{n+m,n+m} \llb \{\vp\} \bv \{\psi\}\rrb}{|}{
{\wti W}^{(m;n)}_{m,n+m} \llb \{\vp_{(\h)}\}\bv \{\psi\}\rrb
}{-------------}{|}{------------}{ 
W_{n+m,m} \lb \Dth\vp_0 ,\ldots ,\Dth \vp_{n+m-1}\rb}{|}{
W_{m,m} \lb \Dth\vp_\h ,\ldots ,\Dth\vp_{m-\h}\rb}
}
\lab{Phi-Ber-2m+n}
\ee
\vspace{.1in}
\br
F^{(n+2m+1;n)} = (-1)^n  \times \phantom{aaaaaaaaaaaaaaaaaaaaaaaaaaaa}
\nonu \\
\phantom{aaaaaaaaaaaaaaaaaaaaaaaaaaaa}
\nonu \\
\frac{{\rm Ber} \threemat{
{\wti W}^{(n+m+1;n)}_{n+m+1,n+m+1} \llb \{\vp\},\vp_{n+m} \bv \{\psi\}\rrb}{|}{
{\wti W}^{(m;n)}_{m,n+m+1} \llb \{\vp_{(\h)}\}\bv \{\psi\} \rrb
}{----------------}{|}{------------}{ 
W_{n+m+1,m} \llb \Dth\vp_0,\ldots ,\Dth\vp_{n+m-1},\Dth\vp_{n+m}\rrb}{|}{
W_{m,m} \llb \Dth\vp_\h ,\ldots ,\Dth\vp_{m-\h}\rrb}
}{
{\rm Ber} \threemat{
{\wti W}^{(n+m+1;m)}_{n+m+1,n+m+1} \llb \{\vp\},\vp_{n+m} \bv \{\psi\}\rrb}{|}{
{\wti W}^{(m+1;n)}_{m+1,n+m+1} \llb \{\vp_{(\h)}\},F \bv \{\psi\}\rrb
}{--------------}{|}{---------------}{ 
W_{n+m+1,m+1} \lb \Dth\vp_0 ,\ldots ,\Dth \vp_{n+m}\rb}{|}{
W_{m+1,m+1} \lb \Dth\vp_\h ,\ldots ,\Dth\vp_{m-\h},\Dth F\rb}
}
\lab{F-Ber-2m+n}
\er
with the notations:
\be
\{\vp\} \equiv \{\vp_0,\ldots ,\vp_{n+m-1}\} \quad ,\quad
\{\vp_{(\h)}\} \equiv \{\vp_\h,\ldots ,\vp_{m-\h}\} \quad ,\quad
\{\psi\} \equiv \{\psi_\h,\ldots ,\psi_{n-\h}\}
\lab{Ber-notat}
\ee
As above, $\{\vp\}$ and $\{\vp_{(\h)}\}$ are sets of bosonic/fermionic
super-eigenfunctions whereas $\{\psi\}$ is a set of fermionic adjoint
super-eigenfunctions of the constrained super-KP hierarchy \cSKP 
\rf{Lax-SKP-R-M}. Let us recall that all pertinent (adjoint) 
super-eigenfunctions are of the form \rf{EF-plus-fSKP}--\rf{EF-neg-fSKP}.

While calculating super-tau functions on the general super-DB orbit we will
also need the following iteration of (adjoint) DB transformations on
fermionic adjoint super-eigenfunctions $\Psi_F$ :
\br
\st{k+1;k}{\Psi_F} = \bigl( -\st{k;k}{\cT_{\vp_k}}\!{}^{-1}\bigr)^\ast
\st{k;k-1}{\cT_{\psi_{k-\h}}}
\bigl(\st{k-1;k-1}{\cT_{\vp_{k-1}}}\!{}^{-1}\bigr)^\ast \ldots
\st{2;1}{\cT_{\psi_{3/2}}} 
\bigl(\st{1;1}{\cT_{\vp_1}}\!{}^{-1}\bigr)^\ast
\st{1;0}{\cT_{\psi_{\h}}} 
\bigl(\st{0;0}{\cT_{\vp_0}}\!{}^{-1}\bigr)^\ast (\Psi_F) =
\nonu
\er
\be
= (-1)^{k+1}
\frac{\det\left\Vert
\begin{array}{cccc}
\Dth^{-1}\! (\vp_0 \psi_\h) & \cdots & \Dth^{-1}\! (\vp_{k-1} \psi_\h) &
\Dth^{-1}\! (\vp_k \psi_\h) \\
\vdots & \ddots & \vdots & \vdots \\
\Dth^{-1}\! (\vp_0 \psi_{k-\h}) & \cdots & \Dth^{-1}\! (\vp_{k-1} \psi_{k-\h}) &
\Dth^{-1}\! (\vp_k \psi_{k-\h}) \\
\Dth^{-1}\! (\vp_0 \Psi_F) & \cdots & \Dth^{-1}\! (\vp_{k-1} \Psi_F) &
\Dth^{-1}\! (\vp_k \Psi_F)
\end{array} \right\Vert
}{\det\left\Vert
\begin{array}{cccc}
\vp_0 & \cdots & \vp_{k-1} & \vp_k \\
\Dth^{-1}\! (\vp_0 \psi_\h) & \cdots & \Dth^{-1}\! (\vp_{k-1} \psi_\h) &
\Dth^{-1}\! (\vp_k \psi_\h) \\
\vdots & \ddots & \vdots & \vdots \\
\Dth^{-1}\! (\vp_0 \psi_{k-\h}) & \cdots & \Dth^{-1}\! (\vp_{k-1} \psi_{k-\h}) &
\Dth^{-1}\! (\vp_k \psi_{k-\h})
\end{array} \right\Vert}
\lab{Psi-Det-k}
\ee
where $k=0,1,\ldots ,n-1$.

Now, using relations \rf{DB-supertau} and taking into account 
\rf{Phi-Ber-2m+n}--\rf{F-Ber-2m+n} and \rf{Psi-Det-k}, we derive the explicit 
expressions for the super-tau functions of constrained super-KP hierarchies 
on the general super-DB orbit:
\br
\frac{\t^{(0;0)}}{\t^{(n+2m;n)}} = (-1)^{mn + n(n-1)/2}
\times \phantom{aaaaaaaaaaaaaaaaaa}
\nonu \\
\phantom{aaaaaaaaaaaaaaaaaa}
\nonu \\
{\rm Ber} \threemat{
{\wti W}^{(n+m;m)}_{n+m,n+m} \llb \{\vp\} \bv \{\psi\}\rrb}{|}{
{\wti W}^{(m;n)}_{m,n+m} \llb \{\vp_{(\h)}\}\bv \{\psi\}\rrb
}{-------------}{|}{------------}{ 
W_{n+m,m} \lb \Dth\vp_0 ,\ldots ,\Dth \vp_{n+m-1}\rb}{|}{
W_{m,m} \lb \Dth\vp_\h ,\ldots ,\Dth\vp_{m-\h}\rb}
\lab{DB-supertau-2m+n}
\er
\vspace{.1in}
\br
\t^{(n+2m+1;n)}\, \t^{(0;0)} = (-1)^{mn + n(n-1)/2}  
\times \phantom{aaaaaaaaaaaaaaaaaa}
\nonu \\
\phantom{aaaaaaaaaaaaaaaaaa}
\nonu \\
{\rm Ber} \threemat{
{\wti W}^{(n+m+1;n)}_{n+m+1,n+m+1} \llb \{\vp\},\vp_{n+m}\bv \{\psi\}\rrb}{|}{
{\wti W}^{(m;n)}_{m,n+m+1} \llb \{\vp_{(\h)}\}\bv \{\psi\} \rrb
}{------------------}{|}{------------}{ 
W_{n+m+1,m} \llb \Dth\vp_0,\ldots ,\Dth\vp_{n+m-1},\Dth\vp_{n+m} \rrb}{|}{
W_{m,m} \llb \Dth\vp_\h ,\ldots ,\Dth\vp_{m-\h}\rrb}
\lab{DB-supertau-2m+1+n}
\er
where again notaions \rf{Ber-notat} have been used.
\mskp
{\bfit 10.4 Examples: ``Super-Soliton'' Solutions}
\sskp
Now, let us write down some explicit examples of Wronskian-like Berezinian
solutions for the superspace tau-function 
\rf{DB-supertau-2m+n}--\rf{DB-supertau-2m+1+n}. We shall
consider the simplest case of a constrained super-KP hierarchy -- the
${\sl SKP}_{(1;1,0)}$ hierarchy defined by the super-Lax operator:
\be
\cL \equiv \cL_{(1;1,0)} = \cD + f_0 + \P \cD^{-1}\Psi
\lab{SKP-1-1-0}
\ee
where $\P,\,\Psi$ are bosonic (adjoint) super-eigenfunctions. We take the
initial $\t^{(0;0)}\! =\! const$, {\sl i.e.}, the initial super-Lax operator
is the ``free'' one $\cL^{(0;0)} \equiv \cL_{\h ,\h}^{(0;0)} = \cD$. 
The initial ``free'' super-eigenfunction $\P^{(0;0)} \equiv \P_0$ satisfies 
according to \rf{P-i-flow-new} :
\br
\partder{}{t_k} \P_0 &=& \pa_x^k \P_0 \quad ,\quad
\cD_n \P_0 = - \Dth^{2n-1} \P_0
\lab{free-sEF-0} \\
\P_0 (t,\th ) &=& 
\int d\l\, \Bigl\lb \vp_B (\l) + 
\Bigl(\th - \sum_{n\geq 1} \l^{n-1} \th_n \Bigr) \vp_F (\l) \Bigr\rb
e^{\sum_{l\geq 1} \l^l (t_l + \th \th_l )}  
\lab{free-sEF}
\er
where $\vp_B (\l),\, \vp_F (\l)$ are arbitrary bosonic (fermionic) 
``spectral'' densities. 

Let us consider iterations of pure DB transformations ({\sl i.e.}, no mixed
binary DB transformations). For the simplest ${\sl SKP}_{(1;1,0)}$ case this
means substituting in the Berezinian expressions 
\rf{Phi-Ber-2m}--\rf{F-Ber-2m+1}:
\be
\vp_k = \pa_x^{k-1} \P_0 \;\; ,\;\; \vp_{k-\h} = \Dth^k \P_0 \;\; 
{\rm for}\;\; k=0,1,\ldots ,m 
\lab{SKP-1-1-0-subst}
\ee
It is easy to show \ct{match,zim-ber} that in this case
\rf{Phi-Ber-2m}--\rf{F-Ber-2m+1} reduce to the following ratios of
ordinary Wronskian determinants:
\be
\t^{(2m;0)} = \frac{\cW_m \llb \pa_x\P_0,\ldots ,\pa_x^m \P_0\rrb}{
\cW_m \llb\P_0 ,\ldots ,\pa_x^{(m-1)}\P_0 \rrb }   \quad , \quad
\t^{(2m+1;0)} = \frac{\cW_{m+1}\llb\P_0 ,\ldots ,\pa_x^m \P_0 \rrb}{
\cW_m \llb \pa_x \P_0,\ldots ,\pa_x^m \P_0\rrb}
\lab{tau-match}
\ee
\br
\cW_m \lb \vp_0,\ldots ,\vp_{m-1}\rb \equiv 
\det W_{m,m} \lb \vp_0,\ldots ,\vp_{m-1}\rb
\nonu
\er
where $\P_0$ is given by \rf{free-sEF}. In particular, choosing for the
bosonic (fermionic) ``spectral'' densities in Eq.\rf{free-sEF} 
~$\vp_B (\l) = \sum_{i=1}^N c_i \d (\l - \l_i )$ ,
$\vp_F (\l) = \sum_{i=1}^N \eps_i \d (\l - \l_i )$ ,
where $c_i,\l_i$ and $\eps_i$ are Grassmann-even and Grassmann-odd
constants, respectively, we have for $\P_0$ :
\be
\P_0 = \sum_{i=1}^N \Bigl\lb c_i + 
\Bigl(\th - \sum_{n\geq 1} \l_i^{n-1} \th_n \Bigr)\eps_i \Bigr\rb
e^{\sum_{l\geq 1} \l_i^l (t_l + \th \th_l )}  
\lab{free-sEF-00}
\ee
Substituting \rf{free-sEF-00} into \rf{tau-match} we obtain the following
``super-soliton'' solutions for the super-tau function of simplest constrained
super-KP hierarchy ${\sl SKP}_{(1;1,0)}$ \rf{SKP-1-1-0} :
\br
&&\t^{(2m+1;0)} = \frac{\sum_{1 \leq i_1 < \ldots < i_{m+1} \leq N}
{N \choose m+1} {\wti c}_{i_1} \ldots {\wti c}_{i_{m+1}}
E_{i_1} \ldots E_{i_{m+1}} \D_{m+1}^2 (\l_{i_1},\ldots ,\l_{i_{m+1}})}{
\sum_{1 \leq j_1 < \ldots < j_m \leq N}
{N \choose m} {\wti c}_{j_1} \ldots {\wti c}_{j_m} E_{j_1} \ldots E_{j_m} 
\l_{j_1} \ldots \l_{j_m} \D_{m}^2 (\l_{j_1},\ldots ,\l_{j_m})}  \nonu \\
&&\phantom{aaaaaaaaaaaaaaaaaaaa} \lab{tau-super-sol} \\
&&{\wti c}_i \equiv c_i + 
\Bigl(\th - \sum_{n\geq 1} \l_i^{n-1} \th_n \Bigr) \eps_i   \quad ,\quad
E_i \equiv e^{\sum_{l\geq 1} \l_i^l (t_l + \th \th_l )}   \nonu \\
&&\D_{m} (\l_{i_1},\ldots ,\l_{i_m}) \equiv 
\det\left\Vert \l_{i_a}^{b-1}\right\Vert_{a,b=1,\ldots ,m}
\lab{super-sol-notation}
\er
\lskip
{\bf 11. Outlook}
\sskp
In the present paper we have provided a systematic derivation of the full
algebra of additional non-isospectral symmetries of constrained (reduced)
supersymmetric KP hierarchies of integrable (``super-soliton'') nonlinear
evolution equations in $N=1$ superspace, which turns out to be a semi-direct 
product of Virasoro algebra with a superloop superalgebra of the form given in 
\rf{plus-flow-alg-fSKP-1}--\rf{plus-flow-alg-bSKP-1} above. We also explicitly
constructed the superspace analogues of (constrained) multi-component KP
hierarchies where the multi-component set of Manin-Radul-type isospectral
evolution ``times'' can be viewed as special subsets of additional symmetry
non-isospectral flows of ordinary one-component supersymmetric KP
hierarchies. We also showed that the (constrained) multi-component
supersymmetric KP hierarchies contain the supersymmetric generalization
of Davey-Stewartson higher-dimensional nonlinear evolution equations.
We studied in detail the conditions for (adjoint) super-\DB transformations
to preserve both Manin-Radul isospectral flows as well as the algebra of
additional non-isospectral symmeries of constrained super-KP hierarchies,
and we presented the explicit \DB solutions for the pertinent super-tau
functions (``super-soliton'' solutions).

The results of the present work suggest a number of interesting problems for
further research:
\begin{itemize}
\item
Systematic study of the supersymmetric extended hierarchies (multi-component
constrained supersymmetric KP hierarchies) introduced in Section 7 above,
which are obtained from scalar (one-component) \cSKP
hierarchies enhanced by Manin-Radul-like subsets 
of additional symmetry flows -- of both ``positive'' and ``negative'' grades.
This implies providing an explicit super-Lax and superspace tau-function 
description of the Manin-Radul-like subsets of additional symmetry flows, as
it has been done in ref.\ct{virflow,gauge-wz} for the ordinary ``bosonic'' 
case.
\item
Revealing other physically interesting nonlinear systems contained within the
multi-component constrained supersymmetric KP hierarchies besides the
supersymmetric Davey-Stewartson system (Section 7), such as
supersymmetric extensions of the $N$-wave resonant wave system,
supersymmetric Toda lattice etc. 
\item
Systematic reformulation of the results of the present paper about
additional non-isospectral symmetries, obtained in the framework of
Sato super-pseudo-differential operator formalism, within the supersymmetric
generalization of the Drinfeld-Sokolov algebraic ``dressing'' approach
(for initial steps in this direction, see ref.\ct{flows})
\item
Systematic study of the physical properties and significance of the new very
broad class of super-soliton-like solutions obtained in Section 10 above.
\end{itemize}
{\underline{\bf Acknowledgements}}
We gratefully acknowledge support from U.S. National Science Foundation 
grant {\sl INT-9724747}. We thank Prof. H. Aratyn and the Physics Department
of University of Illinois at Chicago for hospitality during NATO Advanced
Workshop on {\sl ``Integrable Hierarchies and Modern Physical Theories''} 
(July 2000) where the results of the present paper were first reported. 
We are grateful to H. Aratyn for collaboration in the initial stage of the 
project. This work is also partially supported by Bulgarian NSF grant 
{\sl F-904/99}. 


\end{document}